\documentclass[aps,preprint,tightenlines,superscriptaddress,showpacs,byrevtex,amsmath,amssymb]{revtex4}
\usepackage{graphicx} 
\usepackage{multirow}
\usepackage{xspace}
\usepackage{subfigure}
\usepackage{dcolumn}
\usepackage{color}

\newcommand{\overbar}[1]{\mkern 5mu\overline{\mkern-5mu#1\mkern-1.5mu}\mkern 1.5mu}

\def\bpiplnu{\ensuremath{\bar{B}^{0}\!\rightarrow\!\pi^+\ell^{-}\bar{\nu}_\ell}\xspace}
\def\bpizlnu{\ensuremath{B^{-}\!\rightarrow\!\pi^0\ell^{-}\bar{\nu}_\ell}\xspace}
\def\brhoplnu{\ensuremath{\bar{B}^{0}\!\rightarrow\!\rho^+\ell^{-}\bar{\nu}_\ell}\xspace}
\def\brhozlnu{\ensuremath{B^{-}\!\rightarrow\!\rho^0\ell^{-}\bar{\nu}_\ell}\xspace}
\def\brholnu{\ensuremath{\bar{B}\rightarrow\rho\ell^{-}\bar{\nu}_\ell}\xspace}

\def\bomegalnu{\ensuremath{B^{-}\rightarrow\omega\ell^{-}\bar{\nu}_\ell}\xspace}
\def\bomegappplnu{\ensuremath{B^{-}\!\!\rightarrow\!\omega(3\pi)\ell^{-}\bar\nu_\ell}\xspace}
\def\bomegapglnu{\ensuremath{B^{-}\!\!\rightarrow\!\omega(\pi^0\gamma)\ell^{-}\bar\nu_\ell}\xspace}

\def\bxulnu{\ensuremath{\bar{B}\rightarrow X_u\ell^{-}\bar{\nu}_\ell}\xspace}
\def\bxclnu{\ensuremath{\bar{B}\rightarrow X_c\ell^{-}\bar{\nu}_\ell}\xspace}

\def\xpulnu{\ensuremath{X_u^{+}\ell^{-}\bar{\nu}_\ell}\xspace}
\def\xzulnu{\ensuremath{X_u^{0}\ell^{-}\bar{\nu}_\ell}\xspace}
\def\bxclnu{\ensuremath{\bar{B}\rightarrow X_c\ell^{-}\bar{\nu}_\ell}\xspace}
\def\bpilnu{\ensuremath{\bar{B}\rightarrow \pi\ell^{-}\bar{\nu}_\ell}\xspace}
\def\brholnu{\ensuremath{\bar{B}\rightarrow \rho\ell^{-}\bar{\nu}_\ell}\xspace}
\def\bfzlnu{\ensuremath{B^{-}\rightarrow f_0\ell^{-}\bar{\nu}_\ell}\xspace}
\def\bftwolnu{\ensuremath{B^{-}\rightarrow f_2\ell^{-}\bar{\nu}_\ell}\xspace}

\def\bpdslnu{\ensuremath{B^{-}\rightarrow D^{(*)0}\ell^{-}\bar{\nu}_\ell}\xspace}
\def\bzdslnu{\ensuremath{\bar{B}^{0}\rightarrow D^{(*)+}\ell^{-}\bar{\nu}_\ell}\xspace}
\def\bdkplnu{\ensuremath{B^{-}\rightarrow D^{(*)0}(K^-\pi^+)\ell^{-}\bar{\nu}_\ell}\xspace}
\def\bpdpplnu{\ensuremath{B^{-}\rightarrow D^{(*)0}(\pi^+\pi^-)\ell^{-}\bar{\nu}_\ell}\xspace}
\def\bzdpplnu{\ensuremath{\bar{B}^{0}\rightarrow D^{(*)+}(\pi^+\pi^0)\ell^{-}\bar{\nu}_\ell}\xspace}

\def\bomegapplnu{\ensuremath{B^{-}\!\!\rightarrow\!\omega(\pi\pi)\ell^{-}\bar{\nu}_\ell}\xspace}
\def\bpipilnu{\ensuremath{\bar{B}\rightarrow X_u(\pi\pi)\ell^{-}\bar{\nu}_\ell}\xspace}

\def\bb{\ensuremath{{B\!\overbar{B}}}\xspace}
\def\qq{\ensuremath{{q\bar{q}}}\xspace}
\def\bz{\ensuremath{{B^0}}\xspace}
\def\bp{\ensuremath{{B^\pm}}\xspace}
\def\bzbz{\ensuremath{{B^0\overbar{B}^0}}\xspace}
\def\bpbm{\ensuremath{{B^+B^-}}\xspace}

\def\GeV{\ensuremath{\mathrm{GeV}}\xspace}
\def\GeVc{\ensuremath{\mathrm{GeV}\!/c}\xspace}
\def\GeVcc{\ensuremath{\mathrm{GeV}\!/c^2}\xspace}

\def\GeVGeVcc{\ensuremath{\mathrm{GeV}^2\!/c^2}\xspace}

\def\MeV{\ensuremath{\mathrm{MeV}}\xspace}
\def\MeVc{\ensuremath{\mathrm{MeV}\!/c}\xspace}
\def\MeVcc{\ensuremath{\mathrm{MeV}\!/c^2}\xspace}
\def\lnNBc{\ensuremath{\ln{o_\mathrm{tag}^\mathrm{cs}}}\xspace}

\def\NBc{\ensuremath{o_\mathrm{tag}^\mathrm{cs}}\xspace}
\def\NB{\ensuremath{o_\mathrm{tag}}\xspace}
\def\mm2{\ensuremath{M^2_\mathrm{miss}}}

\def\mbc{\ensuremath{M_\mathrm{bc}}\xspace}

\def\eres{\ensuremath{E_\mathrm{ECL}}\xspace}

\def\npi0{\ensuremath{N_{\pi^0}^\mathrm{Extra}}\xspace}
\def\Br{{\cal B}}
\def\misq{\ensuremath{M_\mathrm{miss}^2}\xspace}

\def\IntLumShort{\ensuremath{711\ \mathrm{fb}^{-1}}\!\xspace}
\def\Nbb{\ensuremath{772 \times 10^6} \bb pairs\xspace}
\def\NbbSVDOne{\ensuremath{152 \times 10^6} \bb pairs\xspace}
\def\NbbSVDTwo{\ensuremath{620 \times 10^6} \bb pairs\xspace}

\def\Nbbwithshorterr{\ensuremath{\left( 772 \pm 11 \right) \times 10^6}\xspace}
\def\vub{\ensuremath{|V_{ub}|}\xspace}
\def\d{\ensuremath{\mathrm{d}}}
\def\pitot{\ensuremath{|\vec{p}_\pi|}\xspace}
\def\pitotp{\ensuremath{|\vec{p}_\pi\kern-0.3em{'}|}\xspace}
\usepackage{relsize}
\def\babar{\mbox{\slshape B\kern-0.1em{\smaller A}\kern-0.1em
    B\kern-0.1em{\smaller A\kern-0.2em R}}\xspace}
\def\frange{\mbox{\large\ensuremath{{}^\text{\kern0.4em{full}}_\text{range}}}\xspace}
\def\nfit{\ensuremath{N^\text{fit}}\xspace}
\def\nmc{\ensuremath{N^\text{MC}}\xspace}

\begin{document}

\vspace*{-3\baselineskip}
\resizebox{!}{3cm}{\includegraphics{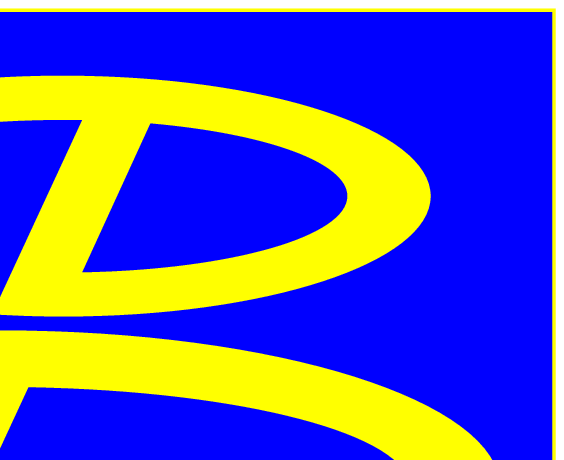}}

\preprint{
\vbox{\hbox{}
   \hbox{Belle Preprint 2013-9}
   \hbox{KEK Preprint 2013-8}
}}

\title{\boldmath Study of Exclusive $B\rightarrow X_u\ell\nu$ Decays and Extraction of $|V_{ub}|$
using Full Reconstruction Tagging at the Belle Experiment}

\begin{abstract}
  We report the results of a study of the exclusive semileptonic
  decays \bpizlnu, \bpiplnu, \brhozlnu, \brhoplnu\ and \bomegalnu,
  where $\ell$ represents an electron or a muon.  The events are
  tagged by fully reconstructing a second $B$ meson in the event in
  a hadronic decay mode.  The measured branching fractions are $
  \Br(\bpizlnu) =(0.80 \pm 0.08 \pm 0.04)\times10^{-4} $, $
  \Br(\bpiplnu) =(1.49 \pm 0.09 \pm 0.07)\times10^{-4} $, $
  \Br(\brhozlnu) =(1.83 \pm 0.10 \pm 0.10)\times10^{-4} $, $
  \Br(\brhoplnu) =(3.22 \pm 0.27 \pm 0.24)\times10^{-4} $, and
  $\Br(\bomegalnu) =(1.07 \pm 0.16 \pm 0.07)\times10^{-4} $, where the
  first error is statistical and the second one is systematic. The
  obtained branching fractions are inclusive of soft photon
  emission. We also determine the branching fractions as a function of
  the 4-momentum transfer squared to the leptonic system
  $q^2=(p_\ell+p_\nu)^2$, where $p_\ell$ and $p_\nu$ are the lepton
  and neutrino 4-momenta, respectively. Using the pion modes, a
  recent LCSR calculation, lattice QCD results and a model-independent
  description of the hadronic form factor, a value of the
  CKM matrix element $|V_{ub}|=(3.52\pm0.29)\times10^{-3}$ is
  extracted. For the first time, a
  charmless state with invariant mass greater than 1~\GeVcc, which
  might be dominated by the decay \bftwolnu, is observed. These results
  are obtained from a \IntLumShort data sample that contains \Nbb,
  collected near the $\Upsilon(4S)$ resonance with the Belle
  detector at the KEKB asymmetric-energy $e^+ e^-$ collider.
\end{abstract}


\noaffiliation
\affiliation{University of the Basque Country UPV/EHU, 48080 Bilbao}
\affiliation{University of Bonn, 53115 Bonn}
\affiliation{Budker Institute of Nuclear Physics SB RAS and Novosibirsk State University, Novosibirsk 630090}
\affiliation{Faculty of Mathematics and Physics, Charles University, 121 16 Prague}
\affiliation{University of Cincinnati, Cincinnati, Ohio 45221}
\affiliation{Deutsches Elektronen--Synchrotron, 22607 Hamburg}
\affiliation{Justus-Liebig-Universit\"at Gie\ss{}en, 35392 Gie\ss{}en}
\affiliation{Gifu University, Gifu 501-1193}
\affiliation{Hanyang University, Seoul 133-791}
\affiliation{University of Hawaii, Honolulu, Hawaii 96822}
\affiliation{High Energy Accelerator Research Organization (KEK), Tsukuba 305-0801}
\affiliation{Hiroshima Institute of Technology, Hiroshima 731-5193}
\affiliation{Ikerbasque, 48011 Bilbao}
\affiliation{Indian Institute of Technology Guwahati, Assam 781039}
\affiliation{Indian Institute of Technology Madras, Chennai 600036}
\affiliation{Institute of High Energy Physics, Chinese Academy of Sciences, Beijing 100049}
\affiliation{Institute of High Energy Physics, Vienna 1050}
\affiliation{Institute for High Energy Physics, Protvino 142281}
\affiliation{INFN - Sezione di Torino, 10125 Torino}
\affiliation{Institute for Theoretical and Experimental Physics, Moscow 117218}
\affiliation{J. Stefan Institute, 1000 Ljubljana}
\affiliation{Kanagawa University, Yokohama 221-8686}
\affiliation{Institut f\"ur Experimentelle Kernphysik, Karlsruher Institut f\"ur Technologie, 76131 Karlsruhe}
\affiliation{Korea Institute of Science and Technology Information, Daejeon 305-806}
\affiliation{Korea University, Seoul 136-713}
\affiliation{Kyungpook National University, Daegu 702-701}
\affiliation{\'Ecole Polytechnique F\'ed\'erale de Lausanne (EPFL), Lausanne 1015}
\affiliation{Faculty of Mathematics and Physics, University of Ljubljana, 1000 Ljubljana}
\affiliation{Luther College, Decorah, Iowa 52101}
\affiliation{University of Maribor, 2000 Maribor}
\affiliation{Max-Planck-Institut f\"ur Physik, 80805 M\"unchen}
\affiliation{School of Physics, University of Melbourne, Victoria 3010}
\affiliation{Moscow Physical Engineering Institute, Moscow 115409}
\affiliation{Moscow Institute of Physics and Technology, Moscow Region 141700}
\affiliation{Graduate School of Science, Nagoya University, Nagoya 464-8602}
\affiliation{Kobayashi-Maskawa Institute, Nagoya University, Nagoya 464-8602}
\affiliation{Nara Women's University, Nara 630-8506}
\affiliation{National United University, Miao Li 36003}
\affiliation{Department of Physics, National Taiwan University, Taipei 10617}
\affiliation{H. Niewodniczanski Institute of Nuclear Physics, Krakow 31-342}
\affiliation{Nippon Dental University, Niigata 951-8580}
\affiliation{Niigata University, Niigata 950-2181}
\affiliation{University of Nova Gorica, 5000 Nova Gorica}
\affiliation{Osaka City University, Osaka 558-8585}
\affiliation{Pacific Northwest National Laboratory, Richland, Washington 99352}
\affiliation{Panjab University, Chandigarh 160014}
\affiliation{University of Pittsburgh, Pittsburgh, Pennsylvania 15260}
\affiliation{University of Science and Technology of China, Hefei 230026}
\affiliation{Seoul National University, Seoul 151-742}
\affiliation{Soongsil University, Seoul 156-743}
\affiliation{Sungkyunkwan University, Suwon 440-746}
\affiliation{School of Physics, University of Sydney, NSW 2006}
\affiliation{Tata Institute of Fundamental Research, Mumbai 400005}
\affiliation{Excellence Cluster Universe, Technische Universit\"at M\"unchen, 85748 Garching}
\affiliation{Toho University, Funabashi 274-8510}
\affiliation{Tohoku Gakuin University, Tagajo 985-8537}
\affiliation{Tohoku University, Sendai 980-8578}
\affiliation{Department of Physics, University of Tokyo, Tokyo 113-0033}
\affiliation{Tokyo Institute of Technology, Tokyo 152-8550}
\affiliation{Tokyo Metropolitan University, Tokyo 192-0397}
\affiliation{Tokyo University of Agriculture and Technology, Tokyo 184-8588}
\affiliation{CNP, Virginia Polytechnic Institute and State University, Blacksburg, Virginia 24061}
\affiliation{Wayne State University, Detroit, Michigan 48202}
\affiliation{Yamagata University, Yamagata 990-8560}
\affiliation{Yonsei University, Seoul 120-749}
  \author{A.~Sibidanov}\affiliation{School of Physics, University of Sydney, NSW 2006} 
  \author{K.~E.~Varvell}\affiliation{School of Physics, University of Sydney, NSW 2006} 
  \author{I.~Adachi}\affiliation{High Energy Accelerator Research Organization (KEK), Tsukuba 305-0801} 
  \author{H.~Aihara}\affiliation{Department of Physics, University of Tokyo, Tokyo 113-0033} 
  \author{D.~M.~Asner}\affiliation{Pacific Northwest National Laboratory, Richland, Washington 99352} 
  \author{V.~Aulchenko}\affiliation{Budker Institute of Nuclear Physics SB RAS and Novosibirsk State University, Novosibirsk 630090} 
  \author{T.~Aushev}\affiliation{Institute for Theoretical and Experimental Physics, Moscow 117218} 
  \author{A.~M.~Bakich}\affiliation{School of Physics, University of Sydney, NSW 2006} 
  \author{A.~Bala}\affiliation{Panjab University, Chandigarh 160014} 
  \author{A.~Bozek}\affiliation{H. Niewodniczanski Institute of Nuclear Physics, Krakow 31-342} 
% ...
\collaboration{The Belle Collaboration}

\maketitle

\section{Introduction}

The Standard Model (SM) of particle physics contains a number of
parameters whose values are not predicted by theory and must therefore
be measured by experiment. In the quark sector, the elements of the
Cabibbo-Kobayashi-Maskawa (CKM) matrix \cite{CKM} determine the rates of the weak
transitions between quark flavours, and precision measurements of
their values are desirable. In particular, in the context of $B$-meson
decays, there is currently much experimental and theoretical effort to
test the consistency of the well-known CKM unitarity triangle (UT).

The UT angle $\phi_1$~\cite{beta}, characterising indirect $CP$ violation
in $b \to c \bar{c} s$ transitions, was first observed to be non-zero
in 2001~\cite{firstCPV}, and $\sin 2\phi_1$ is now known to a
precision of better than 3\%~\cite{lastCPV}. This makes a
corresponding precision measurement of the length of the side of the
unitarity triangle opposite $\phi_1$ particularly important as a
consistency check of the SM picture. The length of this side is
determined to good approximation by the ratio of the magnitudes of two
CKM matrix elements, $|V_{ub}/V_{cb}|$. Both can be
measured using exclusive semileptonic $B$-meson decays. Using charmed
semileptonic decays, the precision to which $|V_{cb}|$ has been
determined is 2-3\%~\cite{PDG}. In comparison, $|V_{ub}|$,
which can be measured using charmless semileptonic decays, is poorly
known. Both inclusive and exclusive methods of measuring
$|V_{ub}|$ have been pursued, with the results of the two approaches
being in some tension~\cite{HFAG12}. It is the aim of an ongoing
programme at the $B$ factories to improve the
precision of these measurements, in order to provide a more stringent
comparison of exclusive and inclusive results, which have somewhat
different experimental and theoretical uncertainties, and to provide a
sharp consistency test with the value of $\sin 2\phi_1$.

Measurements of branching fractions for exclusive \bxulnu decays,
where $X_u$ denotes a light meson containing a $u$ quark and $\ell$ an
electron or muon, have been reported by the CLEO \cite{CLEOUntagged},
\babar~\cite{BaBarUntaggedPiRho, BaBarUntaggedPiEta, BaBarTaggedPi, 
BaBarSemiTaggedPiEtaEtap, BaBarUntaggedPiOmegaEta} 
and Belle~\cite{BelleUntaggedPi, BelleSLTagPiRho, BelleUntaggedOmega}
collaborations.  Three methods of identifying signal candidates have
been employed in these studies. In untagged analyses, the missing
energy and momentum of the whole event are used to reconstruct the
neutrino from the signal semileptonic decay. Semileptonic tagging
involves partial reconstruction of a $B \to D^{(*)} \ell \nu$ decay as
the tagging mode. In this case, two neutrinos are present in the event
and the kinematics cannot be fully constrained.  In full
reconstruction tagging, a hadronically decaying $B$ meson is
reconstructed, against which the signal decay recoils.

In this article, we present measurements of the total and partial
branching fractions for the exclusive semileptonic decays \bpiplnu,
\bpizlnu, \brhoplnu, \brhozlnu and \bomegalnu~\cite{conj} using the full
reconstruction tagging technique. The measurement is based on a
\IntLumShort data sample that contains \Nbbwithshorterr \bb pairs,
collected with the Belle detector at the KEKB asymmetric-energy
$e^+e^-$ (3.5 on 8~GeV) collider~\cite{KEKB} operating at the
$\Upsilon(4S)$ resonance.

The Belle detector is a large-solid-angle magnetic spectrometer that
consists of a silicon vertex detector (SVD), a 50-layer central drift
chamber (CDC), an array of aerogel threshold Cherenkov counters (ACC),
a barrel-like arrangement of time-of-flight scintillation counters
(TOF) and an electromagnetic calorimeter comprised of CsI(Tl)
crystals (ECL) located inside a superconducting solenoid coil that
provides a 1.5~T magnetic field.  An iron flux-return yoke located outside
of the coil is instrumented to detect $K_L^0$ mesons and to identify
muons (KLM).  The detector is described in detail
elsewhere~\cite{Belle}.  Two inner detector configurations were
used. A 2.0 cm beampipe and a 3-layer silicon vertex detector were
used for the first sample of \NbbSVDOne, while a 1.5 cm beampipe, a
4-layer silicon detector and a small-cell inner drift chamber were
used to record the remaining \NbbSVDTwo~\cite{svd2}.

Recently, a new reconstruction procedure for $B$-meson hadronic decays based
on the NeuroBayes neural network package~\cite{NeuroBayes} has been
introduced in Belle.  This procedure reconstructs $B$ mesons in more
than 1100 exclusive hadronic decay channels. Compared to the previous
cut-based algorithm, it offers roughly a factor of two efficiency gain
and about $2.1\times10^{6}$ ($1.4\times10^{6}$) fully reconstructed
charged (neutral) $B$-meson decays within the data sample collected at
the $\Upsilon(4S)$ resonance.

\section{Differential decay rates}
The decay rate for the process $B\to f_1f_2...$, where the $f_i$
represent final state particles, is given by
\begin{equation}
 \d \Gamma(B\to f_1f_2 ...) = \dfrac{1}{2m_B}|{\cal
M}(B\to f_1f_2...)|^2\d\Pi,
\end{equation}
where $m_B$ is the mass of the $B$ meson, ${\cal M}$ is the matrix
element for the decay,
\begin{equation}
 \d\Pi = (2\pi)^4\delta^{(4)}(p_B-\sum_i p_i)\prod_i\dfrac{\d^3\vec{p}_i}{(2\pi)^32E_i}
\end{equation}
is the total decay phase space element, $p_B$ is the 4-vector of the
parent $B$ meson and $p_i = (E_i, \vec{p}_i)$ are the 4-vectors of the
final state particles $f_i$.

The matrix element for weak semileptonic $B$-meson decays at first
order can be written as
\begin{equation}
{\cal M}(\bar{B}\to X_q\ell^{-}\bar{\nu}_\ell) = \dfrac{G_F}{\sqrt{2}}V_{qb}L^\mu H_\mu,
\end{equation}
where $G_F$ is the Fermi constant, $V_{qb}$ is the element of the
CKM matrix corresponding to the $b\to q$ transition, $L^\mu =
\bar{u}_\ell\gamma^\mu(1-\gamma^5)v_\nu$ is the leptonic current and
$H_\mu$ is the hadronic current, which depends on the
particular hadronic final state. More details about the subsequent
formulae can be found elsewhere~\cite{Richman:1995wm}.

\subsection{The \bpilnu decay}
We can parametrise the hadronic current for the \bpilnu decay as
\begin{equation}
H_\mu   = \langle \pi(p_\pi)|V_\mu|B(p_B)\rangle
        =  f_+(q^2) \left(p_{B}+p_{\pi}-q\dfrac{m_B^2-m_\pi^2}{q^2}\right)_\mu + 
           f_0(q^2)\dfrac{m_B^2-m_\pi^2}{q^2}q_\mu, 
\end{equation}
where $q=p_B-p_\pi = p_W = p_\ell + p_\nu$ is the 4-momentum transferred to the
leptonic system, $f_+(q^2)$ is a vector form factor, and $f_0(q^2)$ is a scalar
form factor; we use $f_+(0) \equiv f_0(0)$ to avoid a kinematic singularity at
$q^2 = 0$.

The differential decay rate for the process involving pseudoscalar mesons is written as
\begin{multline}
\frac{\d\Gamma}{\d q^2}\left(\bar{B}\to\pi\ell^{-}\bar{\nu}_\ell\right)
 =  \frac{G_F^2|V_{ub}|^2}{24\pi^3m_{B}^2q^4}\left(q^2-m_\ell^2\right)^2\pitot \times \\ 
    \left[\left(1+\frac{m_\ell^2}{2q^2}\right)m_{B}^2\pitot^2\left|f_+\!\left(q^2\right)\!\right|^2 +
    \frac{3m_\ell^2}{8q^2}\left(m_{B}^{2}-m_{\pi}^{2}\right)^2\left|f_0\!\left(q^2\right)\!\right|^2\right],
\end{multline}
where \pitot is the magnitude of the pion momentum in the $B$
rest frame. For light leptons ($e$ and $\mu$), we can neglect terms
proportional to $m_\ell^2$ so that only $f_+(q^2)$ is relevant:
\begin{equation}
\dfrac{\d\Gamma}{\d q^2} = \dfrac{G_F^2}{24\pi^3}|V_{ub}|^2\left|f_+\!\left(q^2\right)\!\right|^2\pitot^3.
\end{equation}

\subsection{The \brholnu and \bomegalnu decays}
For semileptonic decays with vector mesons in the final state,
{\it i.e.,} $\rho$ or $\omega$, we can define the hadronic current with four
dimensionless form factors:
\begin{multline}
H_\mu = \langle V(p)|(V-A)_\mu|B(p_B)\rangle =
-i\epsilon^*_\mu(m_B+m_V)A_1^V(q^2) + i(p_B+p)_\mu(\epsilon^*p_B)\dfrac{A_2^V(q^2)}{m_B+m_V} \\
+ iq_\mu(\epsilon^*p_B)\dfrac{2m_V}{q^2}(A_3^V(q^2)-A_0^V(q^2))
+ \epsilon_{\mu\nu\rho\sigma}\epsilon^{*\nu}p_B^\rho p^\sigma \dfrac{2V^V(q^2)}{m_B+m_V},
\end{multline}
with the exact relations among the form factors
\begin{align}
&A_3^V(q^2)  =  \dfrac{m_B+m_V}{2m_V}A_1^V(q^2) - \dfrac{m_B-m_V}{2m_V}A_2^V(q^2), \\
&A_0^V(0)  =  A_3^V(0),\\
&\langle V|\partial_\mu A^\mu|B\rangle =  2m_V(\epsilon^*p_B)A_0^V(q^2),
\end{align}
where $V$ represents a $\rho$ or $\omega$ meson, $p_B$ is
the 4-momentum of the $B$ meson, and $p$, $\epsilon^*$ and $m_V$
are the 4-momentum, polarization 4-vector and mass of the vector
meson, respectively. Again, for light leptons ($e$ and $\mu$), the term
proportional to $q_\mu$ is negligible so that, in effect, the decay rate
depends only on the form factors $A_1(q^2)$, $A_2(q^2)$ and $V(q^2)$.

In the case \brholnu where the $\rho$-meson decays into two pions
$\rho\to \pi\pi$, the fully differential decay rate in the helicity
basis is:
\begin{multline}
\label{eq:rhodGdmV}
{{\d \Gamma} \over {\d q^2\ \d\!\cos\theta_{\ell}\ \d\!\cos\theta_{V}\ \d\chi\ \d m_V}} = \\
{{3} \over {8 (4 \pi)^4}} G_F^2 \vert V_{ub} \vert^2 {
{\vert {\vec{p}_{V}} \vert  q^2}
\over {m_B^2}} |\mathrm{BW}(m_V)|^2 
\times \biggl [
(1- \eta \cos \theta_{\ell})^2 \sin^2 \theta_V \vert H_+(q^2, m_V) \vert^2 
 \\
+(1+ \eta
\cos \theta_{\ell})^2 \sin^2 \theta_V \vert H_-(q^2, m_V) \vert^2   
+4 \sin^2 \theta_{\ell} \cos^2 \theta_{V} \vert H_0(q^2, m_V) \vert^2  \\
-4 \eta \sin \theta_{\ell} \sin \theta_{V}
\cos \theta_{V} \cos \chi  H_0(q^2, m_V) \times 
((1- \eta \cos \theta_{\ell})H_+(q^2, m_V) - \\
(1+\eta \cos \theta_{\ell})H_-(q^2, m_V)) 
-2 \sin^2 \theta_{\ell} \sin^2 \theta_V \cos {2 \chi} H_+(q^2, m_V) H_-(q^2, m_V)
\biggr ], 
\end{multline}
where $\mathrm{BW}(m_V)$ represents the resonance line shape, $m_V$ is
the invariant mass of the recoiling hadron, $|\vec{p}_V|$ is the magnitude
of the vector meson momentum in the $B$ meson rest frame,
$\theta_{\ell}$ is the polar angle of the lepton in the $W$ rest frame
with respect to the $W$ flight direction in the $B$ rest frame,
$\theta_{V}$ is the polar angle of one of the pseudoscalar daughters
in the rest frame of the vector meson with respect to the vector meson
flight direction in the $B$ rest frame, and $\chi$ is the angle
between the decay planes of the $W$ boson and the vector meson. The factor
$\eta$ is equal to $+1$ for semileptonic $B$-meson decays.

The differential decay rate for \bomegalnu, integrated over angular variables, is
\begin{multline}
{{\d \Gamma} \over {\d q^2 \d m_V}} = {{1} \over {96 \pi^3}} G_F^2 \vert V_{ub} \vert^2 {
{\vert {\vec{p}_{V}} \vert  q^2}
\over {m_B^2}} |\mathrm{BW}(m_V)|^2\times \\
 \biggl [
\vert H_0(q^2, m_V)\vert^2 + \vert H_+(q^2, m_V)\vert^2 + \vert H_-(q^2, m_V)\vert^2 \biggr ].
\end{multline}

In the above expressions, the helicity amplitudes are:
\begin{eqnarray}
H_{\pm}(q^2, m_V) &=& (m_B + m_V)A_1^V(q^2) \mp {{2m_B \vert \vec{p}_V \vert} \over 
{m_B+m_V}} V^V(q^2),  \nonumber\\
H_0(q^2, m_V) &=& {{m_B+m_V} \over {2 m_V \sqrt{q^2}}} \bigl [ (m_B^2 - m_V^2 - q^2)
A_1^V(q^2) - 4 {{m_B^2 \vert \vec{p}_V \vert^2} \over {(m_B+m_V)^2}} A_2^V(q^2) \bigr ]. 
\end{eqnarray}

For \brholnu decay, a relativistic Breit-Wigner function is used to
describe the $\rho$ line shape. The amplitude is
\begin{equation}
\mathrm{BW}(m_{\pi\pi}) = \dfrac{\pitot}{m_{\pi\pi}^2 - m_\rho^2 + im_{\pi\pi}\Gamma(m_{\pi\pi})}
\dfrac{B(\pitot)}{B(\pitotp)},
\end{equation}
where $m_{\rho}$ is the nominal $\rho$ mass,
\pitot is the pion momentum in the $\rho$ rest frame,  $\pitotp$
is the same but for fixed $m_{\pi\pi} = m_\rho$ and 
\begin{equation}
\Gamma(m_{\pi\pi}) =  \Gamma_0 \left(\dfrac{\pitot}{\pitotp}\right)^3 \dfrac{m_\rho}{m_{\pi\pi}}\left(\dfrac{B(\pitot)}{B(\pitotp)}\right)^2,
\end{equation}
where $\Gamma_0$ is the nominal $\rho$ width, and $B(x)$ is a Blatt-Weisskopf damping factor given by
\begin{equation}
B(x) = \dfrac{1}{\sqrt{1+R^2x^2}},
\end{equation}
with $R = 3$ (\GeVc)$^{-1}$. 

For the \bomegalnu decay, a simpler non-relativistic form of the Breit-Wigner function is
used for the $\omega$ line shape:
\begin{equation}
|\mathrm{BW}(m)|^2 = \dfrac{1}{2\pi}\dfrac{\Gamma}{(m-m_\omega)^2 + \left(\Gamma/2\right)^2},
\end{equation}
where $m_\omega$ and $\Gamma$ are the nominal mass and width of the $\omega$ meson.

\section{\label{datasamples}Data sample and simulation}
We use Belle data collected at the energy corresponding to the maximum of the
$\Upsilon(4S)$ resonance (10.58 GeV in centre-of-mass frame),
equivalent to an integrated luminosity of \IntLumShort.  Using
$\Br(\Upsilon(4S)\to\bzbz) = 0.486\pm0.006$ and
$\Br(\Upsilon(4S)\to\bpbm) = 0.514\pm0.006$~\cite{PDG}, we can estimate
the numbers of produced neutral and charged $B$-meson pairs,
$N_{\bzbz} = (373\pm7)\times10^{6}$ and $N_{\bpbm} =
(398\pm7)\times10^{6}$. We also utilise a sample of 79~fb$^{-1}$
collected below the \bb threshold to study the contribution of the
$e^+e^-\to q\bar{q}$ process, so-called continuum, where $q$ is
a $u$, $d$, $s$, or $c$ quark.

Monte Carlo~(MC) samples of $e^+e^-\to\Upsilon(4S)\to\bb$ and
continuum, equivalent to five times the integrated luminosity, are
used to study the major backgrounds. The simulation accounts for
changes in background conditions and beam collision parameters. Final
state radiation (FSR) from charged particles is modeled using the
PHOTOS package~\cite{photos}.

Dedicated MC samples are generated for charmless semileptonic decays,
which are not present in the samples mentioned above. The total number
of generated events for the signal MC is based on the number of \bb
pairs in data, scaled by a factor of 20 and assuming branching
fractions of $\Br(\bar{B}^0\to\xpulnu) = 1.709 \times 10^{-3}$ and
$\Br(B^-\to\xzulnu) = 1.835 \times 10^{-3}$.

Signal samples with \bpilnu, \brholnu and \bomegalnu decays are
modeled using Light Cone Sum Rule (LCSR) form factor
predictions~\cite{ball01,ball05}.  Other decays to exclusive meson
states are modeled using the updated quark model by
Isgur-Scora-Grinstein-Wise (ISGW2)~\cite{ISGW2}. The inclusive
component of charmless semileptonic decays is modeled to leading order
$\alpha_s$ based on a prediction in the Heavy-Quark Expansion
(HQE) framework~\cite{DeFazio}. The fragmentation process of the
resulting parton to the final hadron state is modeled using the PYTHIA
package~\cite{pythia62}.

In the analysis, for the \bpilnu decay modes, the signal MC events are
reweighted to reproduce the best parametrization by Bourrely, Caprini
and Lellouch (BCL), Eq.~39 in~\cite{bcl}, because the extrapolation of
the LCSR prediction from~\cite{ball01} is inadequate for the high
$q^2$ region.

\section{\label{selection}Event Selection}
Semileptonic $B$-meson decay candidates in events passing the full
reconstruction procedure are selected.  This procedure provides a
NeuroBayes output variable \NB that varies from zero to unity and tends
to have larger values in cases where the fully reconstructed tagging
candidate is a correctly reconstructed $B$ meson. To suppress
continuum events, the \NB variable, combined with 18 modified
Fox-Wolfram moment variables~\cite{SFW} in a neural net, is used to
form a variable, $\NBc$. Only high quality hadronic tag
candidates with $\lnNBc>-6$ are selected.  This corresponds to a
rejection of candidates with a probability to be a $B$ meson of less
than 0.25\%. An additional selection criterion is applied based on the
beam-constrained mass $\mbc = \sqrt{E_\mathrm{beam}^2/c^4 - (\sum_i
\vec{p}_i/c)^2}$, where $E_\mathrm{beam}$ is the beam energy in the 
centre of mass frame and the $\vec{p}_i$ are the 3-momenta of the
detected particles that form the $B$-meson candidate in the same
frame. The candidate must satisfy the condition $\mbc>5.27$~\GeVcc.

It is possible to have several $B$ candidates after full reconstruction.
In this case, depending on the
recoiling system of interest, we select the candidate with the highest
probability assigned by the full reconstruction algorithm.

All charged particles are required to originate from the region near
the interaction point (IP) of the electron and positron beams.
With respect to a cylindrical system with origin at the IP, axis of
symmetry $z$ aligned opposite the positron beam direction and radial
coordinate $r$, this region is defined as $|z_\mathrm{PCA}|<2$~cm and
$r_\mathrm{PCA}<0.4$~cm, where $z_\mathrm{PCA}$ and $r_\mathrm{PCA}$
are the coordinates of the point of closest approach (PCA) of the
reconstructed charged particle to the $z$ axis. All other charged
particles are ignored. After applying the above, we treat all
selected charged particles as originating from $B$-meson decays.

Electron candidates are identified using the ratio of the energy
detected in the ECL to the track momentum, the ECL shower shape,
position matching between the track and ECL cluster, the energy loss
in the CDC, and the response of the ACC~\cite{eid}. Muons are
identified based on their penetration range and transverse scattering
in the KLM detector~\cite{muid}. In the momentum region relevant to
this analysis, charged leptons are identified with an efficiency of
about 90\% while the probability to misidentify a pion as an electron
(muon) is 0.25\% (1.4\%). Charged pion candidates are selected with an
efficiency of 85\% and a kaon misidentification probability of 19\%,
based on the responses of the CDC, ACC and TOF subdetectors.

To reject leptons from $\gamma$-conversions in the detector material
and from $J/\psi$ and $\psi'$ decays, $M_{\ell\ell}$, the invariant mass of all
oppositely charged lepton ($\ell = e$ or $\mu$) pairs, is checked and
particles are vetoed if $M_{\ell\ell}<0.1$~\GeVcc,
$3.00\ \GeVcc <M_{\ell\ell}<3.12$~\GeVcc or $3.60\ \GeVcc <M_{\ell\ell}<3.75$~\GeVcc.

After a tag candidate has been selected, we look
for a lepton amongst the rest of the reconstructed particles not
already assigned to this tagging $B$ meson. For $B^{\pm}$, only
leptons with the correct charge are selected, whereas for $B^{0}$,
because of mixing, both lepton charges---in other words, all flavors
regardless of the reconstructed flavor of the neutral $B$-meson tag---are
accepted.  A chosen electron must have a momentum in the
laboratory frame $|\vec{p}_e^\mathrm{\ lab}|>300$~\MeVc, whilst a muon
must satisfy $|\vec{p}_\mu^\mathrm{\ lab}|>600$~\MeVc. These
thresholds are chosen based on the known performance properties of the
Belle particle identification algorithms. If several particles pass
these requirements, the particle with the highest probability to be a
lepton, as assigned by the reconstruction algorithm, is selected.

In the electron case, we consider all photons in the event that do not
belong to the tag side; if a photon is found whose direction lies
within a 3$^\circ$ cone around the electron direction, we add the
photon 4-momentum to that of the electron and exclude the
photon from further analysis. In cases where more than one photon is
found, only the nearest photon is merged with the electron.

From the tag side, we derive the signal $B$ meson momentum 4-vector in the
$\Upsilon(4S)$ rest frame using only the tag direction, by explicitly
requiring the invariant mass to be the $B$ meson mass:
\begin{equation}
 p_{B_\mathrm{sig}} \equiv
(E_{B_\mathrm{sig}}/c,\vec{p}_{B_\mathrm{sig}}) =
c\left(\frac{m_{\Upsilon(4S)}}{2},
-\frac{\vec{p}_{B_\mathrm{tag}}}{|\vec{p}_{B_\mathrm{tag}}|}
\sqrt{\left(\frac{m_{\Upsilon(4S)}}{2}\right)^2-m_B^2}\right).
\end{equation}

We select photons that are not assigned to the tag side with energy
in the laboratory frame of $E_{\gamma}>50$ MeV.  To form a $\pi^0$
candidate, we take all possible combinations of two photons; those
with invariant mass in the
range $|M_{\gamma\gamma}-m_{\pi^0}|<15$ MeV$/c^2$ are selected as
$\pi^0$ candidates. The selection is shown in Fig.~\ref{blabel}.

We require the number of signal-side charged particles, {\it i.e.,} charged
particles that have not been assigned to the tagging $B$ meson
candidate, to equal the number of charged particles expected for
the particular decay mode under study.

To calculate the residual energy, \eres, in the electromagnetic
calorimeter, we use photons that have not been assigned to either the
signal or tag sides. The photons are boosted from the laboratory frame
to the $\Upsilon(4S)$ rest frame and the energies are summed. For each
decay mode, the selection criterion on the residual energy is
optimised for maximal signal significance.

Combining the momentum 4-vectors of the selected hadron system and the
lepton on the signal side, and denoting the so-constructed pseudoparticle
 $Y$, we have $p_{Y} = p_\mathrm{hadrons}+ p_{\ell}$. We
can then calculate the cosine of the angle between the direction of
the signal side $B$ meson and the $Y$ in the $\Upsilon(4S)$ frame:
\begin{equation}
\cos{\theta_{BY}} =
\dfrac{2E_{B_\mathrm{sig}}E_{Y}/c^2-m_B^2c^2-m_{Y}^2c^2}{2|\vec{p}_{B_\mathrm{sig}}||\vec{p}_Y|},
\end{equation}
where $m_Y$ is the invariant mass of the pseudoparticle. For signal
decays, the condition $|\cos{\theta_{BY}}|<1$ must be
fullfilled. Allowing for resolution effects, we select events using
the loose selection $|\cos{\theta_{BY}}|\le 3$, keeping all correctly
reconstructed events and suppressing a majority of the
background. This choice provides us with enough background events to
fix background shapes in the fit procedure described below.

Having reconstructed the momentum 4-vectors of the candidate signal $B$
meson and pseudoparticle $Y$, we define the missing momentum 4-vector
as
\begin{equation}
 p_{\mathrm{miss}} \equiv
(E_{\mathrm{miss}}/c,\vec{p}_{\mathrm{miss}}) = 
p_{B_\mathrm{sig}} - p_{Y}.
\end{equation}
For a correctly reconstructed semileptonic decay,
$p_{\mathrm{miss}}$ represents the momentum 4-vector of a single
missing neutrino, with missing mass squared $\misq \equiv
 p_{\mathrm{miss}}^2/c^2$ consistent with zero.

The missing energy is required to satisfy $E_\mathrm{miss}>300$~\MeV
to avoid backgrounds where the $B$ meson decays hadronically but a
pion from this decay is misidentified as a muon.

For \bpizlnu candidates, we select the $\pi^0$ that gives the smallest
value of the magnitude of \misq, defined in this case as
$|p_{B_\mathrm{sig}}-p_\ell-p_{\pi^0}|^2$.  The angle
$\psi_{\gamma\gamma}^\mathrm{lab}$ between photons from the $\pi^0$
decay must satisfy $\cos{\psi_{\gamma\gamma}^\mathrm{lab}}>0.25$,
which mainly suppresses background where one photon candidate arises
from activity in the backward endcap and the other from the
forward endcap, forming a fake $\pi^0$. The residual
energy is required to satisfy the condition $\eres<0.6$~\GeV.

For \bpiplnu candidates, we require that both charged particles be
close to each other at the PCA:
$|z_\mathrm{PCA}^\ell-z_\mathrm{PCA}^\mathrm{\pi^+}|<1$~mm.  Because
the pion and lepton originate from the same vertex, they should have
matching $z$-coordinate values at the start of the track, in the
vicinity of the IP. We require the residual energy to satisfy
$\eres<1$~\GeV. In the case of the charged lepton being a muon, we test
the hypothesis that the selected pion is actually a muon from a
$J/\psi$ decay misidentified as a pion, and reject events where
$|M_{\mu\pi}-m_{J/\psi}|<20$ \MeVcc, assuming the muon hypothesis for
both particles.

For \brhozlnu candidates, we require all charged particles to originate
from the same vertex,
$|z_\mathrm{PCA}^\ell-z_\mathrm{PCA}^\mathrm{\pi^\pm}|<1$~mm and
$|z_\mathrm{PCA}^\mathrm{\pi^+}-z_\mathrm{PCA}^\mathrm{\pi^-}|<1$~mm.
The residual energy must satisfy $\eres<0.7$~\GeV. We select events
where the invariant mass of the two pions is around the nominal $\rho$
meson mass, requiring $|M_{\pi^+\pi^-}-m_{\rho}|<2\Gamma_\rho$ where
$m_\rho = 775.5$~\MeVcc and $\Gamma_\rho = 149.1$~\MeVcc are the
nominal $\rho$ mass and decay width, respectively.

For \brhoplnu candidates, we select the $\pi^+\pi^0$ pair with the
largest energy $E_{\pi^+\pi^0}$ in the $\Upsilon(4S)$ rest frame. The residual energy
must satisfy $\eres<0.7$~\GeV. The angle between the photons must
satisfy $\cos{\psi_{\gamma\gamma}^\mathrm{lab}}>0.4$. The two charged particles
must originate from the same vertex:
$|z_\mathrm{PCA}^\ell-z_\mathrm{PCA}^\mathrm{\pi^+}|<1$~mm. We
reject events where the invariant mass of the two particles, assuming the muon
hypothesis, lies close to the $J/\psi$ mass, {\it i.e.,} with
$|M_{\mu\pi}-m_{J/\psi}|<20$ \MeVcc. As for \brhozlnu, we require
$|M_{\pi^+\pi^0}-m_{\rho}|<2\Gamma_\rho$.

For \bomegalnu candidates where $\omega\to\pi^+\pi^-\pi^0$, we select
the $\pi^+\pi^-\pi^0$ combination containing the $\pi^0$ that has the
invariant mass closest to the nominal $\omega$ meson mass and require
$-40\ \MeVcc<M_{\pi^+\pi^-\pi^0}-m_{\omega}<30\ \MeVcc$, where
$m_\omega=782.65$ \MeVcc. The residual energy must satisfy
$\eres<0.5$~\GeV. The angle between the photons must satisfy
$\cos{\psi_{\gamma\gamma}^\mathrm{lab}}>0.25$. The charged particles
must originate from the same vertex:
$|z_\mathrm{PCA}^\ell-z_\mathrm{PCA}^\mathrm{\pi^\pm}|<1$~mm and
$|z_\mathrm{PCA}^\mathrm{\pi^+}-z_\mathrm{PCA}^\mathrm{\pi^-}|<1$~mm.

For $\bomegalnu$ candidates where $\omega\to\pi^0\gamma$, we consider
$\pi^0\gamma$ pairs for which $-40\
\MeVcc<M_{\pi^0\gamma}-m_{\omega}<30\ \MeVcc$ and then select the pair
that gives the smallest value of
$\misq=|{p}_{B_\mathrm{sig}}-{p}_{\ell}-{ p}_{\pi^0\gamma}|^2$. The
residual energy must satisfy $\eres<0.5$~\GeV. The angle between the
photons from the $\pi^0$ decay must satisfy
$\cos{\psi_{\gamma\gamma}^\mathrm{lab}}>0.4$.  The angle
$\theta_\gamma$ between the photon from the $\omega$ decay in the
$\omega$ rest frame and the $\omega$ direction in the laboratory frame
must satisfy $|\cos{\theta_\gamma}|<0.5$.

\section{Hadronic tag calibration}

In this analysis, we use charmed semileptonic $B$-meson decays to calibrate
the tagging efficiency, due to their large and well known branching
fractions. We can predict the number of events
$N(B\to\mathrm{hadrons},\,\bxclnu)$, where one $B$ meson is reconstructed
by the full reconstruction algorithm in a $B$-meson hadronic decay mode and
the other $B$ meson is reconstructed in an exclusive charmed semileptonic
mode.  We define
\begin{multline}
N(B\to\mathrm{hadrons},\bxclnu) = \\ N_{\bb}\times 
\Br(B\to\mathrm{hadrons}) \times \Br(\bxclnu) \times
\varepsilon^\mathrm{rec}(B\to\mathrm{hadrons},\bxclnu), 
\end{multline} where
$\varepsilon^\mathrm{rec}(B\to\mathrm{hadrons},\,\bxclnu)$ is the
reconstruction efficiency for the specific tag and signal modes.  To
correct for the specific hadronic tag modes, we factorise the
efficiency
\begin{equation} 
\varepsilon^\mathrm{rec}(B\to\mathrm{hadrons},\bxclnu) =
\varepsilon^\mathrm{rec}(B\to\mathrm{hadrons})\times
\varepsilon^\mathrm{rec}(\bxclnu)\times C,
\end{equation}
where $C$ is a correlation factor between the tag and signal sides
that accounts for the lower probability to reconstruct the tag
in the case where many particles are present on the signal side. We
assume that the MC accurately describes the product
$\varepsilon^\mathrm{rec}(\bxclnu)\times C$ because the dynamics of
exclusive \bxclnu decays are well known.

The decay rates in the MC used to calibrate the hadronic tag
efficiency are reweighted to the recent PDG values using the
correction factors given in Table~\ref{table:PDG2012}. We adjust the
number of produced \bzbz and
\bpbm pairs in the MC, which were produced with the assumption
of an equal production rate, using the $\Upsilon(4S)$ branching
fractions into $B$ mesons mentioned in Section~\ref{datasamples}.

\begin{table}
  \caption{\label{table:PDG2012}Branching fractions with uncertainties
      from PDG~\cite{PDG}, used for the hadronic tag calibration, as
        well as the branching fraction used in the Belle MC.}\vspace*{2mm}
  \begin{tabular}{l|c|c|c}
    \hline
    Decay process & $\Br^\mathrm{PDG}$, \%& $\Br^\mathrm{MC}$, \% & $\Br$ ratio \\ \hline\hline
    $B^- \to D^0 \ell^- \bar\nu_\ell$         & $ 2.23 \pm 0.11 $ &  2.31   & 0.965 \\
    $B^- \to D^{*0}\ell^- \bar\nu_\ell$       & $ 5.68 \pm 0.19 $ &  5.79   & 0.981 \\
    $\bar{B}^0 \to D^+\ell^- \bar\nu_\ell$    & $ 2.17 \pm 0.12 $ &  2.13   & 1.019 \\
    $\bar{B}^0 \to D^{*+}\ell^- \bar\nu_\ell$ & $ 5.05 \pm 0.12 $ &  5.33   & 0.947 \\
    $D^0 \to K^- \pi^+$                   & $ 3.87 \pm 0.05 $ &  3.82   & 1.013 \\
    $D^0 \to K^- \pi^+ \pi^0$             & $ 13.9 \pm 0.5  $ & 13.43   & 1.035 \\
    $D^0 \to K^- 2\pi^+ \pi^- $           & $ 8.07 \pm 0.20 $ &  7.155  & 1.128 \\
    $D^+ \to K^- 2\pi^+ $                 & $ 9.13 \pm 0.19 $ &  9.594  & 0.952 \\
    $D^+ \to K^- 2\pi^+ \pi^0$            & $ 5.99 \pm 0.18 $ &  6.03   & 0.993 \\
    $D^+ \to K^- 3\pi^+ \pi^-$            & $ 0.56 \pm 0.05 $ &  0.6252 & 0.896 \\
    $D^{*0} \to D^0 \pi^0$                & $ 61.9 \pm 2.9  $ &  61.9   & 1.000 \\
    $D^{*0} \to D^0 \gamma$               & $ 38.1 \pm 2.9  $ &  38.1   & 1.000 \\
    $D^{*+} \to D^0 \pi^+ $               & $ 67.7 \pm 0.5  $ &  67.7   & 1.000 \\
    $D^{*+} \to D^+ \pi^0 $               & $ 30.7 \pm 0.5  $ &  30.7   & 1.000 \\ \hline
  \end{tabular}
\end{table}

To evaluate the tag correction factor, we fit the \misq
distribution separately for each hadronic tag mode, split by charmed
semileptonic mode. For each tag mode, we calculate the average
correction factor over all charmed semileptonic modes and use it to
reweight events in the MC.

Overall, the tag efficiency correction is about
$\varepsilon^\mathrm{rec}_\mathrm{DATA}(B\to\mathrm{hadrons})/
\varepsilon^\mathrm{rec}_\mathrm{MC}(B\to\mathrm{hadrons}) \sim 0.75$ and
varies by several percent depending on the chosen semileptonic mode, due
to tag- and signal-side interference.  The statistical precision of the
calibration is 1.3\% for \bp and 1.8\% for \bz decay modes. We
estimate the systematic uncertainty due to the PDG branching fraction
uncertainties to be 3.0\% for \bp and 2.5\% for \bz decay modes.  To
select semileptonic decays with $D^{(*)}$ mesons, we use the particle
identification capabilities of the Belle detector. We estimate the
systematic uncertainty due to particle identification for the $B^+$
tag to be 2.3\% and for the $B^0$ tag 3.0\%.  The total uncertainty of
the tag correction, with correlations between modes included, is
estimated to be 4.2\% for $B^{+}$ and 4.5\% for $B^{0}$.
We do not count the lepton identification correction and its
uncertainty as part of the systematic uncertainty because 
it cancels in the ratio for the studied charmless
semileptonic decays.

\section{Signal extraction}
To obtain the number of signal events passing all selection criteria
for any given decay mode, we fit the \misq
distribution, for which signal events are expected to peak at $\misq =
0$.  We use a maximum likelihood technique~\cite{Barlow:1993dm} which
also takes into account finite MC statistics in the template
histograms that form the components of the fit.  The effect on the
fitting procedure of using MC \misq templates with finite statistics
is checked using a toy MC procedure.  We find that the fit
procedure itself does not introduce a bias for the decay modes studied, and
parameter uncertainties match expectations.

\subsection{\label{fitcomp}Components of the fit}
To describe the data \misq distributions, we divide the MC samples
into various components, each defining a template, depending on the
decay mode studied. To better describe the amount of \bxulnu cross-feed,
we adjust, where relevant, the MC branching fractions to
those obtained in this study.

For the \bpizlnu decay, we define the following components: \bpizlnu
signal, \bxulnu cross-feed, other $B$-meson decays and \qq continuum. The continuum
component is fixed to the MC prediction and the normalisations of
all other components are free parameters of the fit.

For the \bpiplnu decay, we define the following components: \bpiplnu
signal, \brhoplnu cross-feed, other \bxulnu cross-feed, other $B$-meson decays and \qq continuum.  The
continuum component is fixed to the MC prediction, the amount of
\brhoplnu cross-feed is fixed to the value obtained in the \brhoplnu
fit and all other components are free parameters of the fit.

For the \brhozlnu decay, we define the following components: \brhozlnu
signal, \bfzlnu, \bftwolnu, \bomegalnu, other \bxulnu cross-feed, \bpdslnu in which the $D^0$ decays to $K^-\pi^+$ or $\pi^+\pi^-$ final states, other $B$-meson decays
and \qq continuum.  The continuum, \bfzlnu and \bomegalnu components are small and fixed
to the MC prediction; the amounts of \bftwolnu, \bpdslnu in which the
$D^0$ decays to $K^-\pi^+$ or $\pi^+\pi^-$ final states and cross-feeds
are fixed to the values obtained from the invariant mass fit that is described later. All other components are free parameters of the
fit.

For the \brhoplnu decay, we define the following components: \brhoplnu
signal, \bpiplnu cross-feed, other \bxulnu cross-feed, \bzdslnu in which
the $D^+$ decays to $\pi^+\pi^0$, other $B$-meson decays and \qq continuum.  The continuum
component is fixed to the MC prediction, the amount of \bpiplnu cross-feed
is fixed to the values obtained in the \bpiplnu fit, and the
amount of \bzdslnu in which the $D^+$ decays to $\pi^+\pi^0$ is fixed to
the value obtained from the invariant mass fit. The normalisations of
all other components are free parameters of the fit.

For the \bomegalnu decay, we define the following components: \bomegalnu
signal ($\omega\!\to\!\pi^+\pi^-\pi^0$ or $\omega\!\to\!\pi^0\gamma$),
\bxulnu cross-feed, other $B$-meson decays and \qq continuum.
The continuum component is fixed to the MC prediction and all other
components are free parameters of the fit.

\subsection{Fit results}

\begin{figure}
  \includegraphics[width=0.595\textwidth]{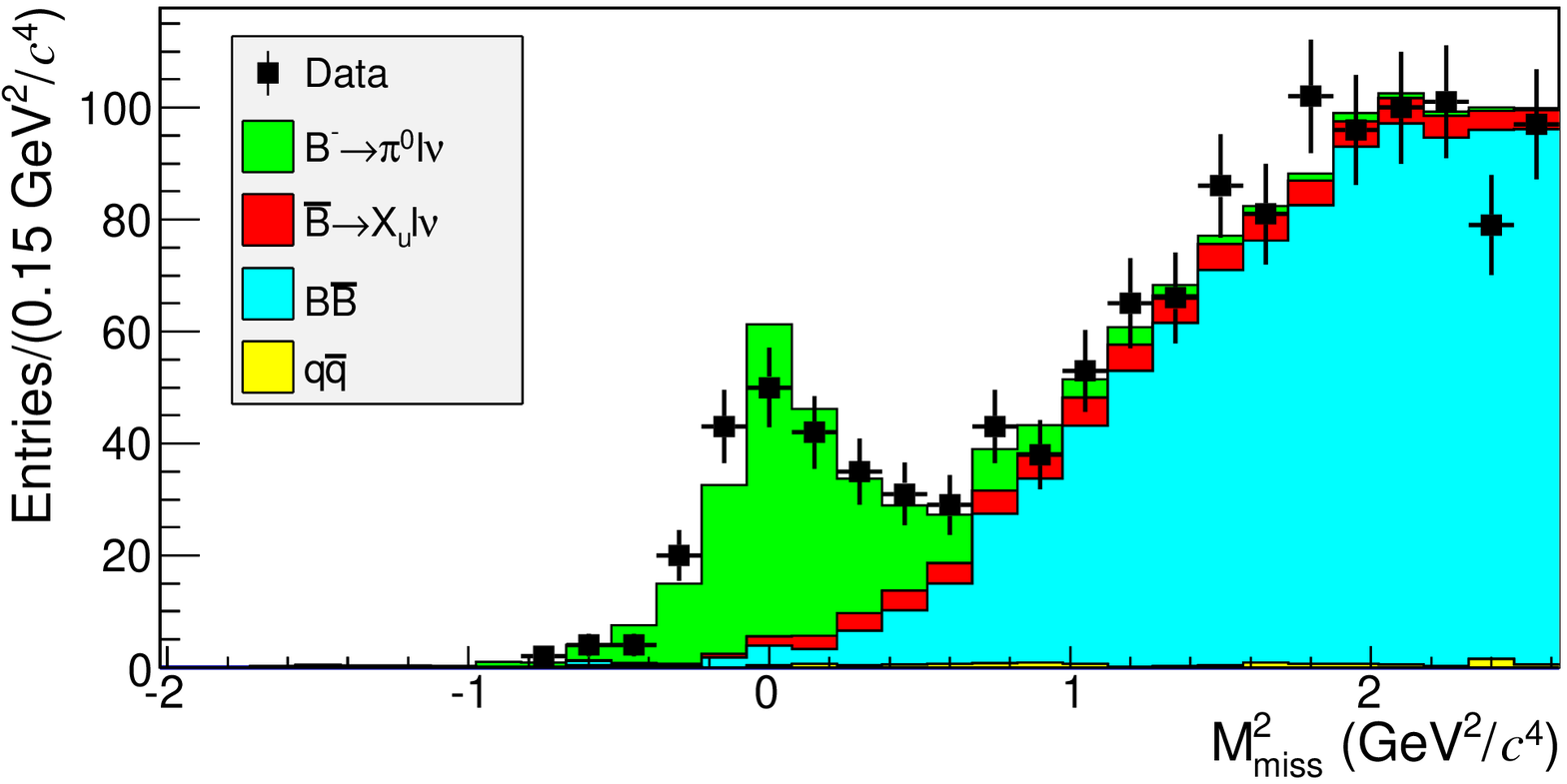}
  \includegraphics[width=0.595\textwidth]{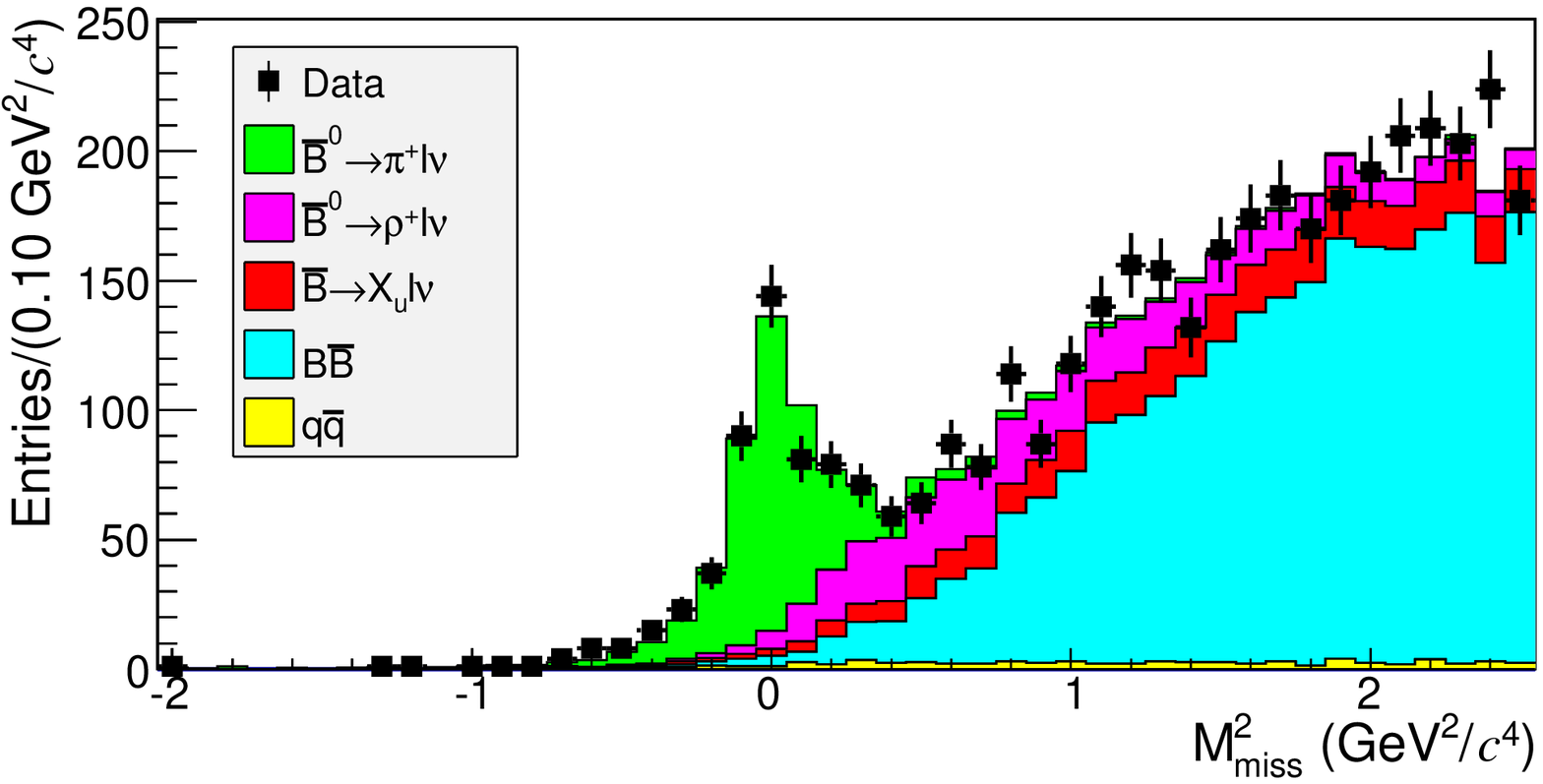}
  \caption{\label{fig:fitpilnu}(Color online) Fit to the \misq
  distributions in data for \bpizlnu decays (top) and \bpiplnu decays
  (bottom).  The fit components are described in the text.}
\end{figure}

\begin{figure}
  \includegraphics[width=0.595\textwidth]{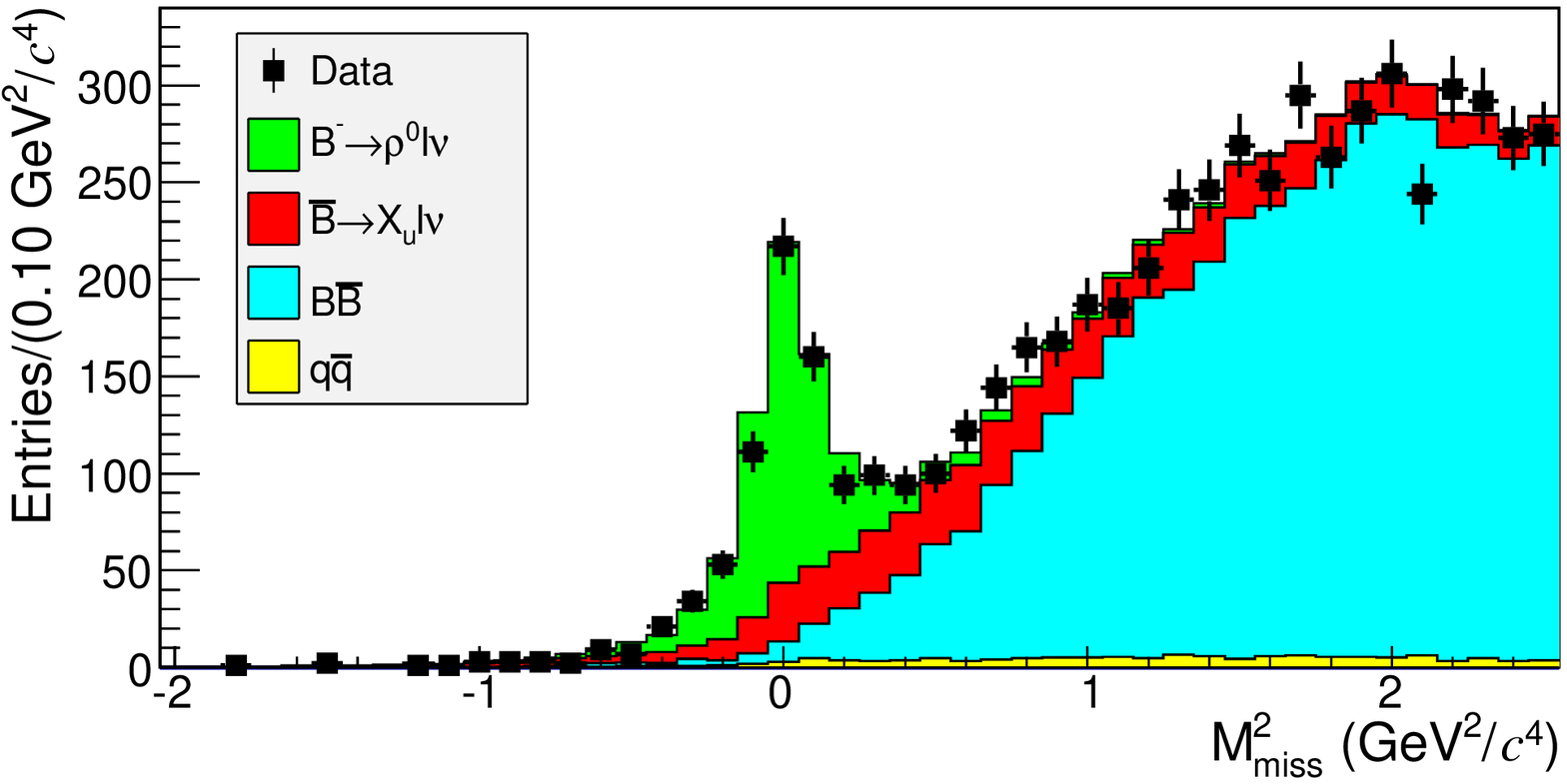}
  \includegraphics[width=0.595\textwidth]{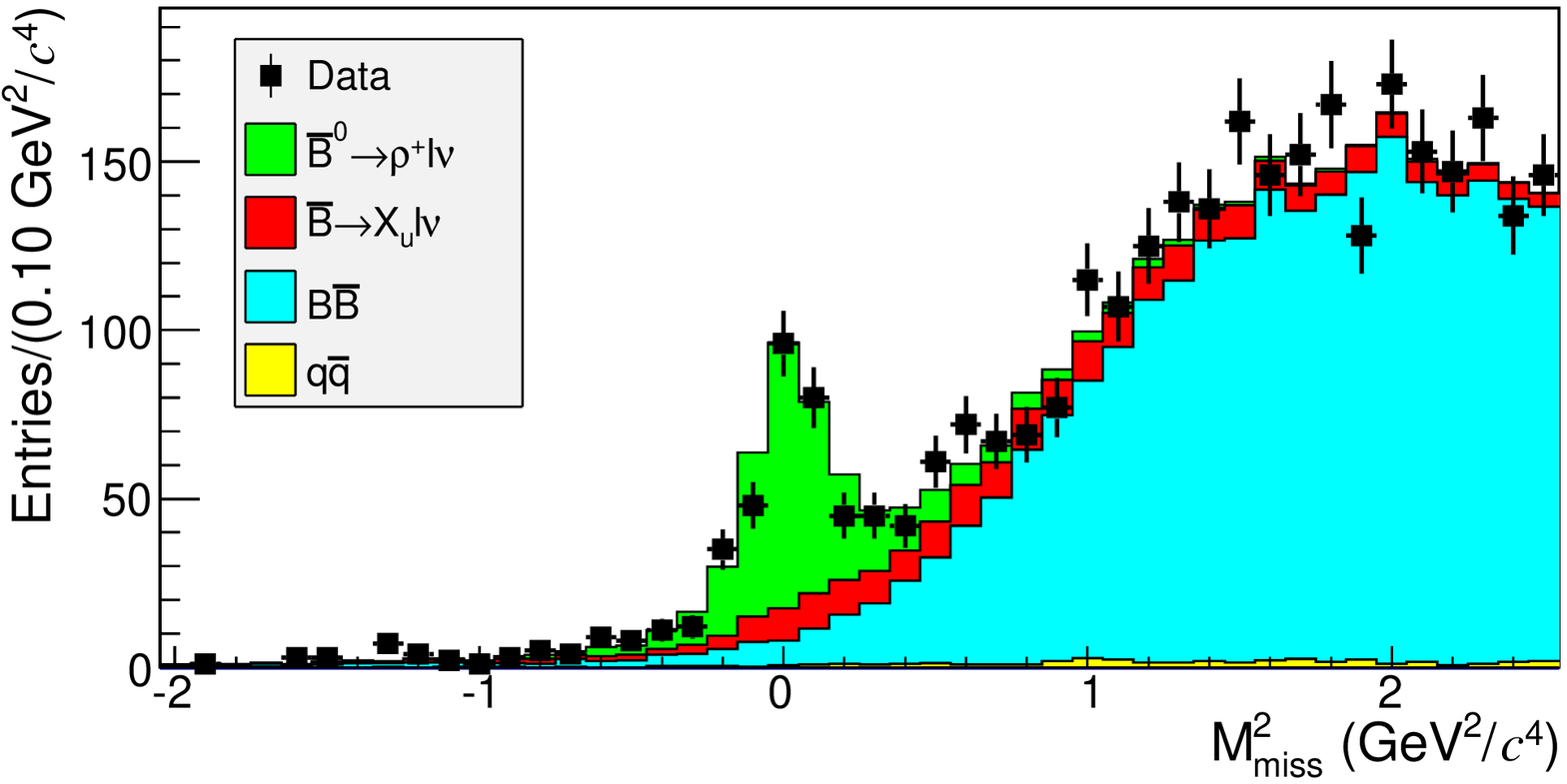}
  \caption{\label{fig:fitrholnu}(Color online) Fit to the \misq
  distribution in data for \brhozlnu decays (top) and \brhoplnu
  (bottom).  The fit components are described in the text.}
\end{figure}

\begin{figure}
  \includegraphics[width=0.595\textwidth]{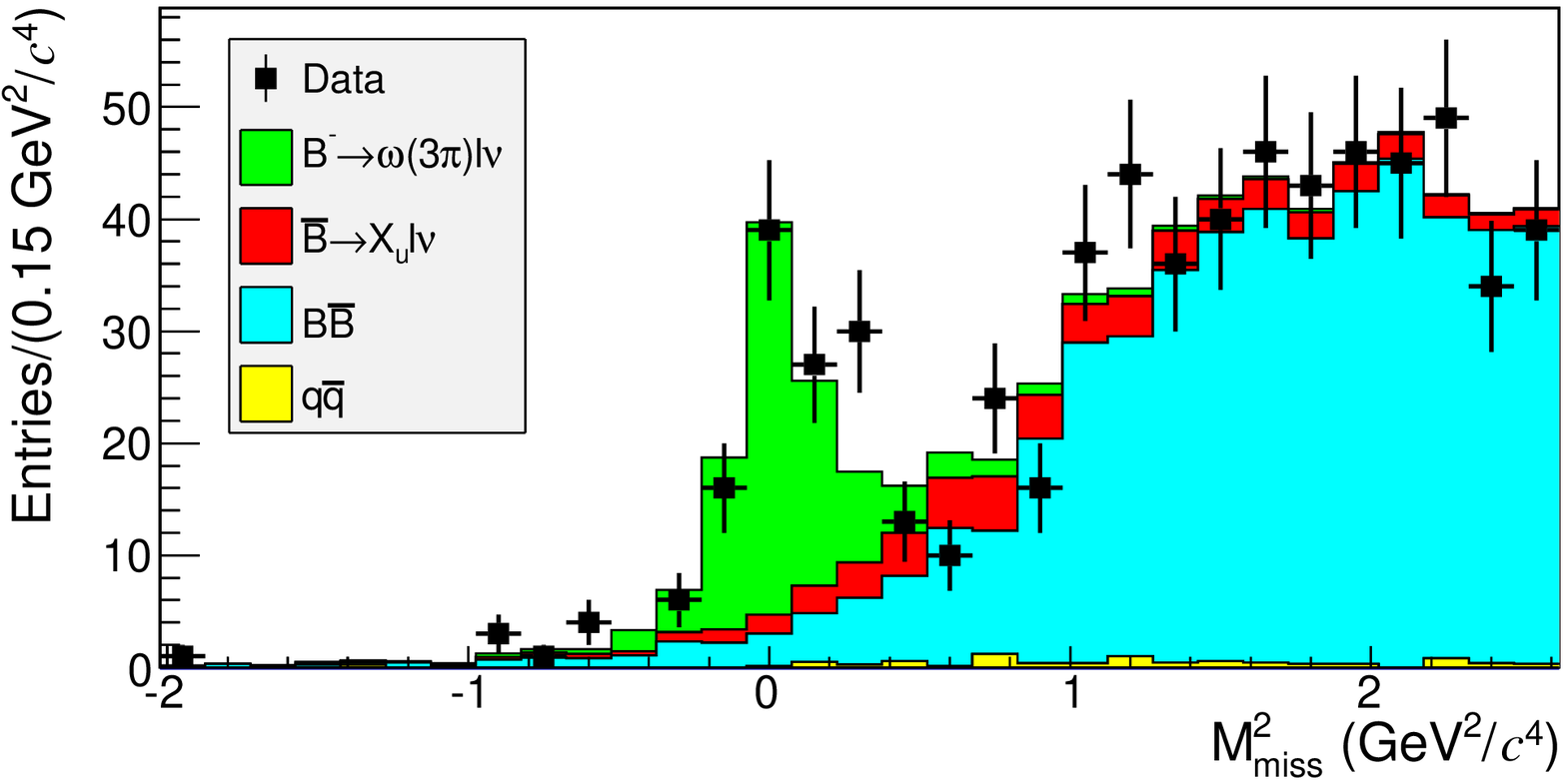}
  \includegraphics[width=0.595\textwidth]{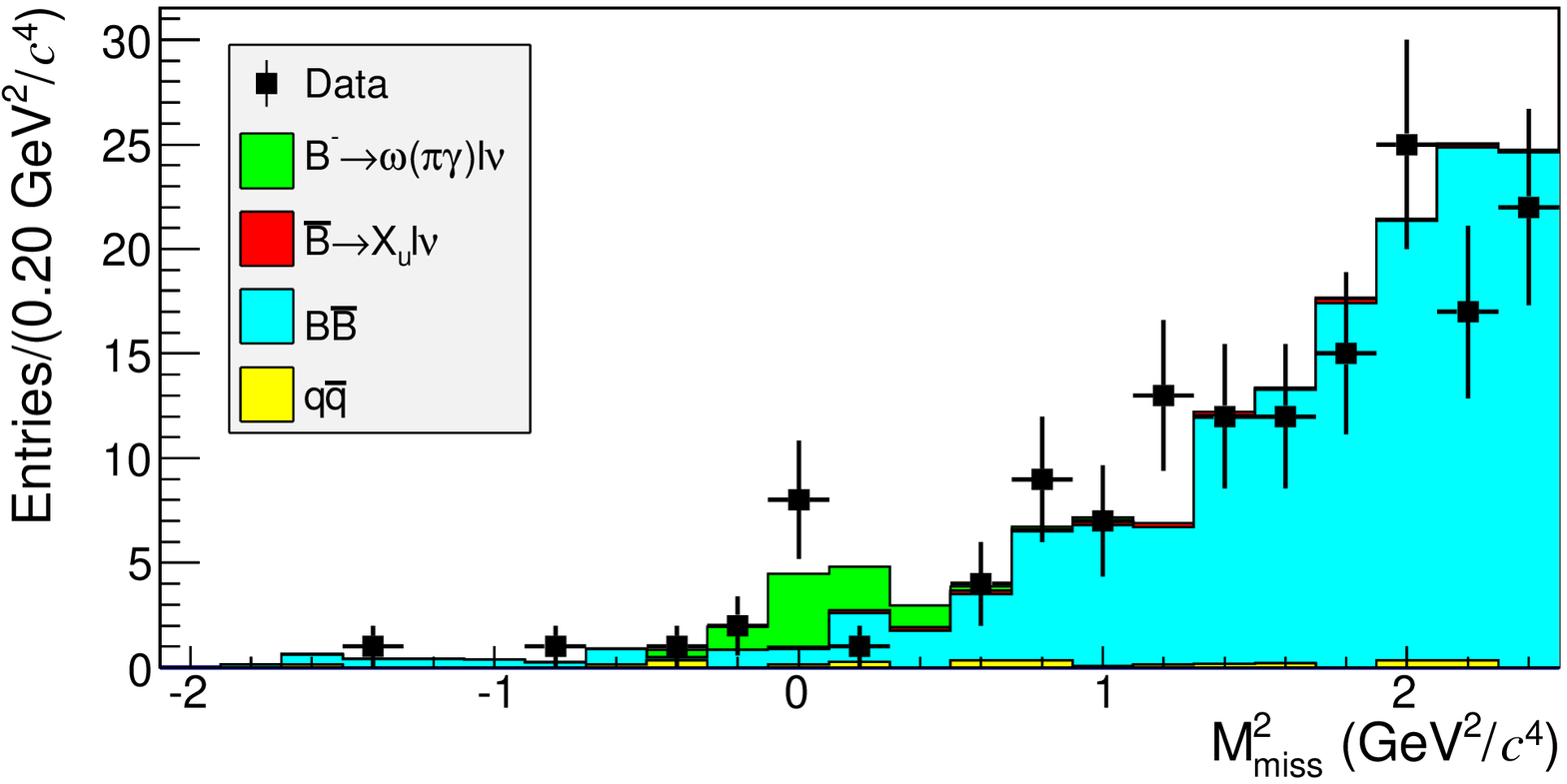}
  \caption{\label{fig:fitomegalnu}(Color online) Fit to the \misq
  distribution in data for \bomegalnu decays where
  $\omega\to\pi^+\pi^-\pi^0$ (top) and $\omega\to\pi^0\gamma$
  (bottom). The fit components are described in the text.}
\end{figure}

The fitted \misq distributions are shown in
Fig.~\ref{fig:fitpilnu} for \bpilnu decays, in
Fig.~\ref{fig:fitrholnu}  for \brholnu decays and in
Fig.~\ref{fig:fitomegalnu} for \bomegalnu decays.

The parameter values obtained from the fit, as well as the values of
the fixed parameters, are presented in Tables \ref{table:pi0yield}--\ref{table:omegayield}.

\begin{table}
  \caption{\label{table:pi0yield}Fitted yields for \bpizlnu decays.}
  \centering\vspace*{2mm}
  \begin{tabular}{ccc}
    \hline
    Component & Yield \\\hline\hline
    
$\bpizlnu$ & $\hphantom{1}232.2 \pm 22.6$ \\
$\bxulnu$ & $\hphantom{1}100.0 \pm 86.7$ \\
$\bb$ & $1993.4 \pm 90.7$ \\
$\qq$ & \hphantom{19}18.5(fixed) \\

    \hline
    $\chi^2/\textrm{ndf}$ & 56.3/50 \\
    \hline
  \end{tabular}
  \caption{\label{table:picyield}Fitted yields for \bpiplnu decays.}
  \centering\vspace*{2mm}
  \begin{tabular}{ccc}
    \hline
    Component & Yield \\\hline\hline
    
$\bpiplnu$ & $\hphantom{1}462.6 \pm \hphantom{1}27.7$ \\
$\brhoplnu$ &\hphantom{1} 514.5(fixed) \\
$\bxulnu$ & $\hphantom{1}599.5 \pm 198.4$ \\
$\bb$ & $5511.6 \pm 200.7$ \\
$\qq$ & \hphantom{1}111.8(fixed) \\

    \hline
    $\chi^2/\textrm{ndf}$ & 76.0/76 \\
    \hline
  \end{tabular}
\end{table}

\begin{table}
  \caption{\label{table:rho0yield}Fitted yields for \brhozlnu decays.}
  \centering\vspace*{2mm}
  \begin{tabular}{ccc}
    \hline
    Component & Yield \\\hline\hline
    
$\brhozlnu$ & $\hphantom{1}621.7 \pm \hphantom{1}35.0$  \\
$\bxulnu$ & $\hphantom{1}757.3 \pm 109.0$ \\
$\bb$ & $6901.6 \pm 128.9$  \\
$\bftwolnu$   & 13.3(fixed) \\
$\bdkplnu$   & 25.1(fixed) \\
$\bpdpplnu$   & 1.2(fixed) \\
$\bomegapplnu$   & 6.1(fixed) \\
$\bfzlnu$   & 9.5(fixed) \\
$\qq$   & 169.9(fixed) \\

    \hline
    $\chi^2/\textrm{ndf}$ & 59.5/52 \\
    \hline
  \end{tabular}
  \caption{\label{table:rhocyield}Fitted yields for \brhoplnu decays.}
  \centering\vspace*{2mm}
  \begin{tabular}{ccc}
    \hline
    Component & Yield \\\hline\hline
    
$\brhoplnu$ & $343.3 \pm 28.3$  \\
$\bxulnu$ & $243.4 \pm 91.6$  \\
$\bb$ & $4039.7 \pm 105.1$  \\
$\qq$   & 59.2(fixed)  \\
$\bpiplnu$   & 10.5(fixed)  \\
$\bzdpplnu$   & 1.3(fixed)  \\

    \hline
    $\chi^2/\textrm{ndf}$ & 84.4/65 \\
    \hline
  \end{tabular}
\end{table}

\begin{table}
  \caption{\label{table:omegayield}Fitted yields for \bomegalnu decays.}
  \centering\vspace*{2mm}
  \begin{tabular}{ccc}
    \hline
    &    \multicolumn{2}{c}{Yield} \\
    Component &  $\omega\!\to\!\pi^+\pi^-\pi^0$ & $\omega\!\to\!\pi^0\gamma$ \\\hline\hline
    
    $\bomegalnu$ & $\hphantom{1}96.7 \pm 14.5$ & $\hphantom{28}9.0 \pm \hphantom{1}4.0$\\
    $\bxulnu$ & $\hphantom{1}62.3 \pm 38.0$  & $\hphantom{28}2.2 \pm 11.4$\\
    $\bb$ & $763.6 \pm 43.2$  & $287.4 \pm 19.0$\\
    $\qq$ &  10.8(fixed) & 4.4(fixed) \\

    \hline
    $\chi^2/\textrm{ndf}$ & 55.8/43 & 41.4/32\\
    \hline
  \end{tabular}
\end{table}

\begin{figure*}
  \subfigure[Various kinematic variables for the \brhozlnu decay in the region $|\misq|<0.25$~\GeVcc.]{
  \begin{tabular}{cc}
      \includegraphics[width=0.41\textwidth]{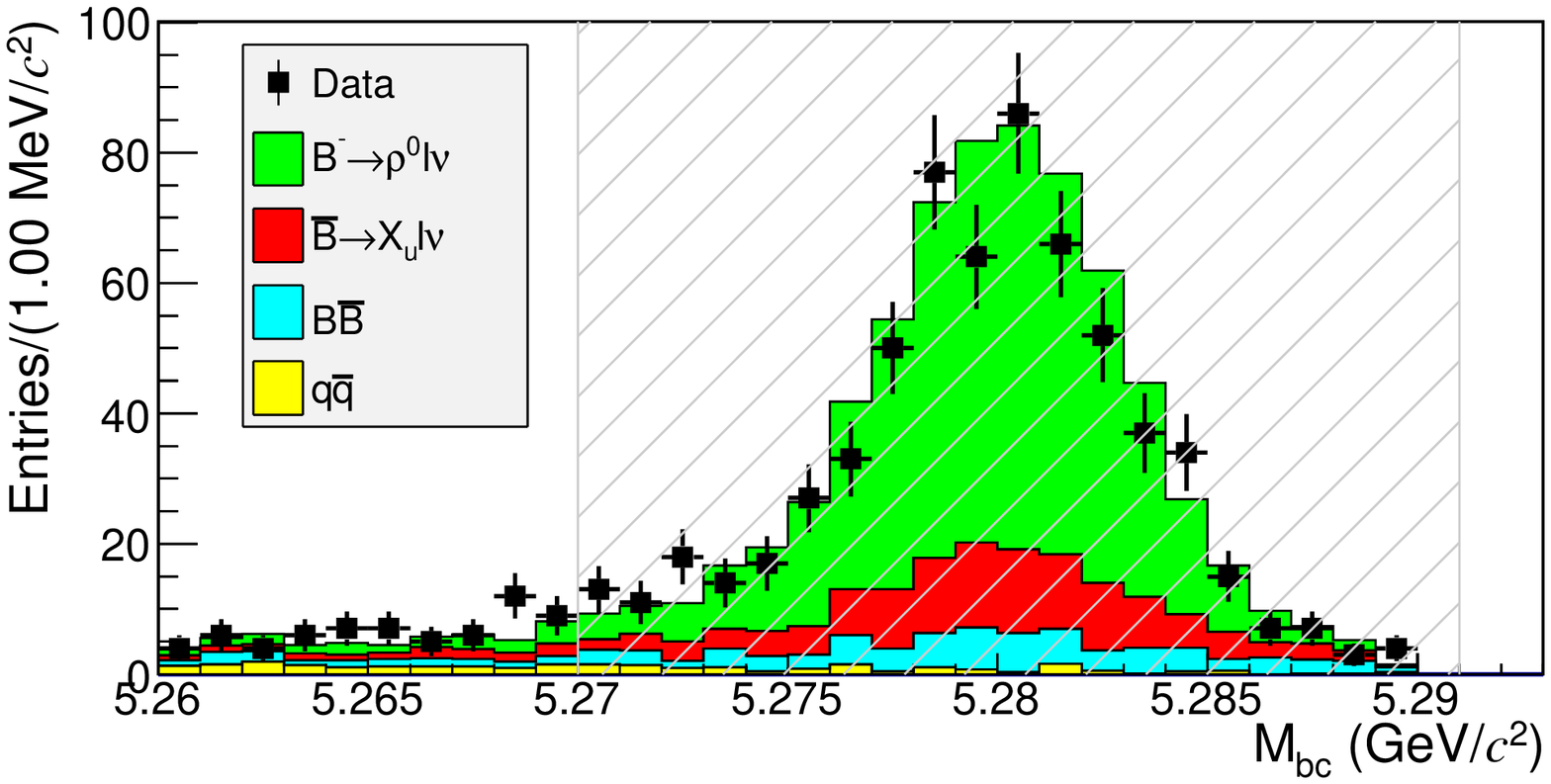} &
      \includegraphics[width=0.41\textwidth]{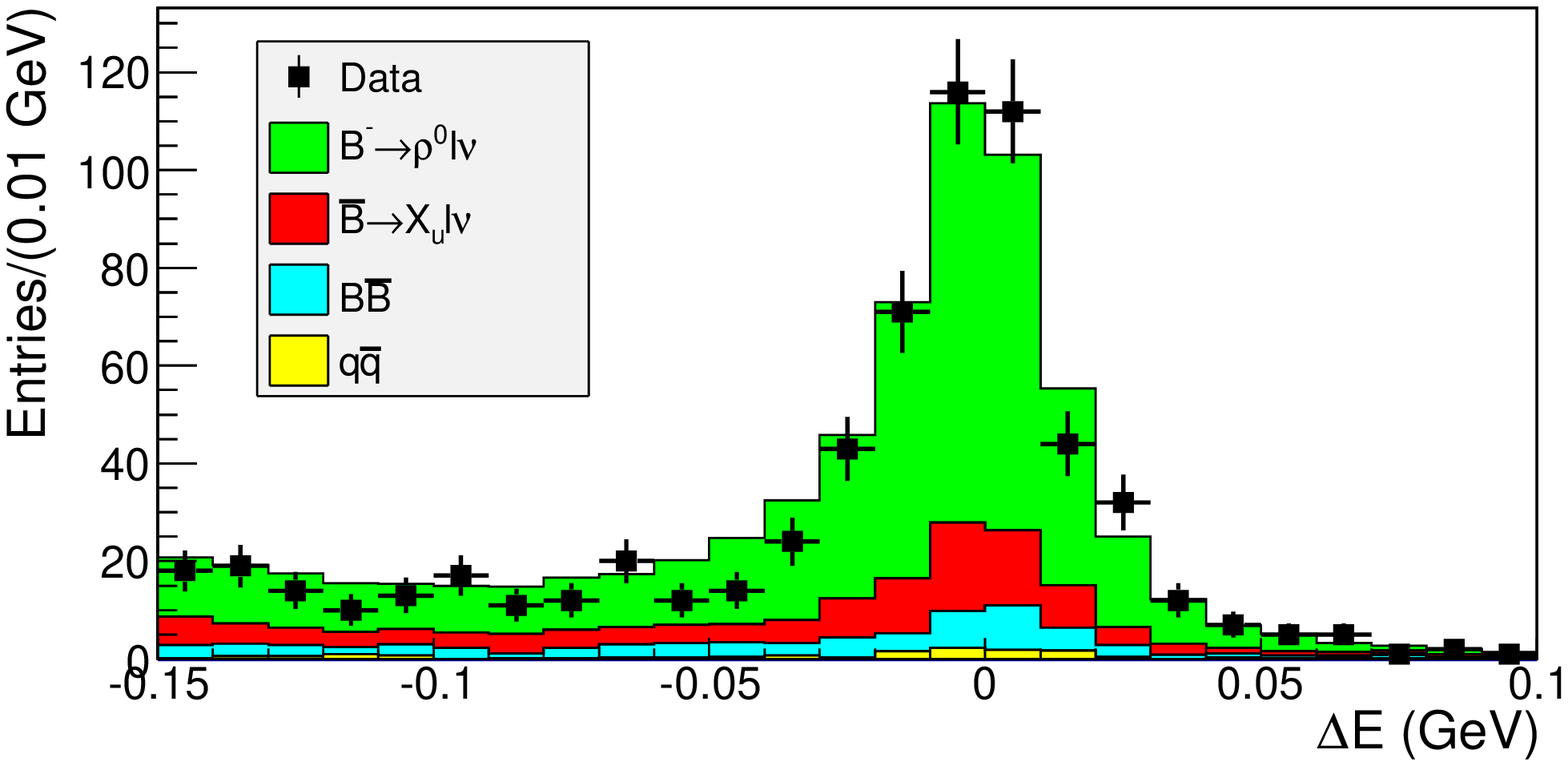} \\

      \includegraphics[width=0.41\textwidth]{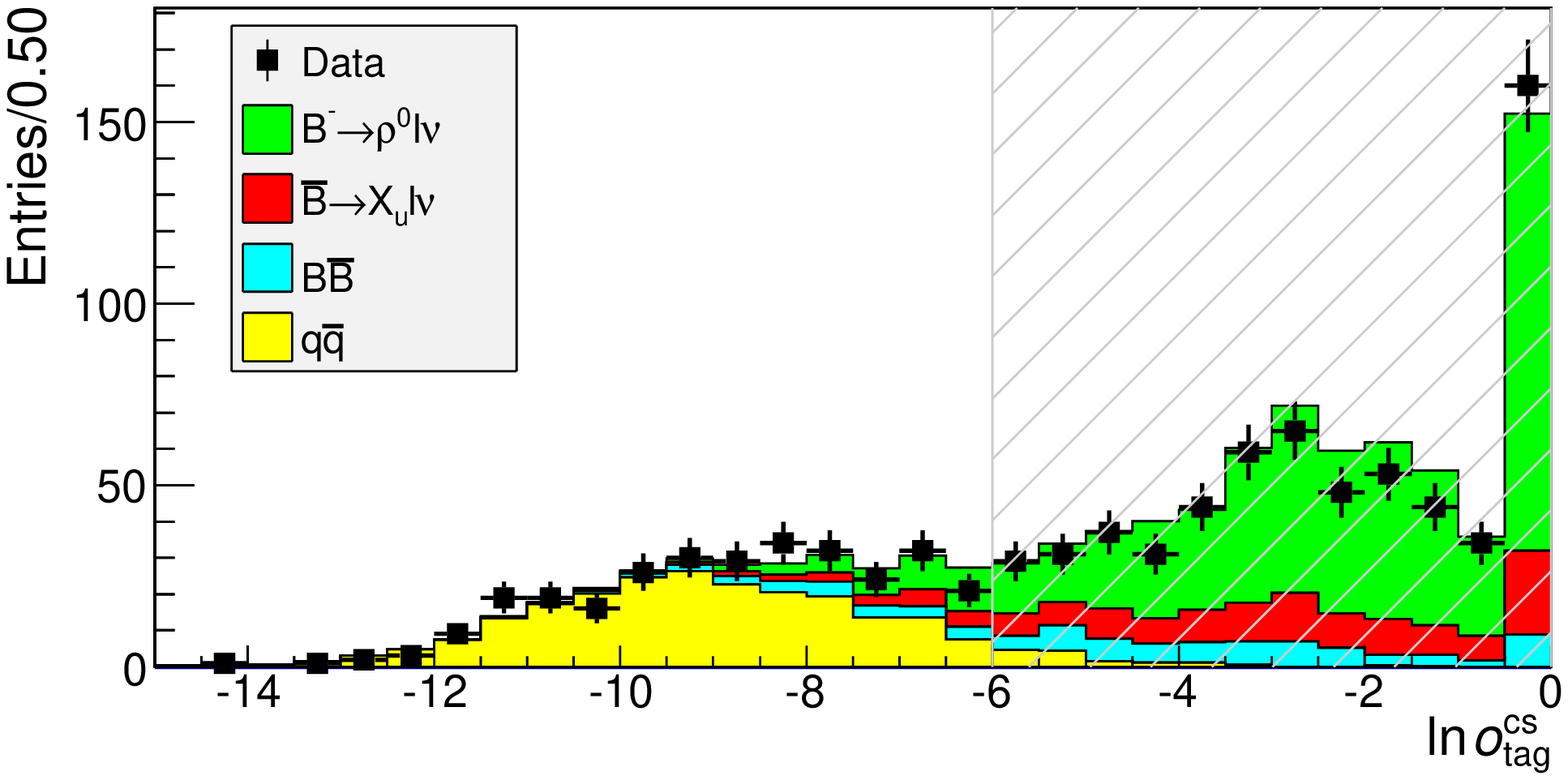} &
      \includegraphics[width=0.41\textwidth]{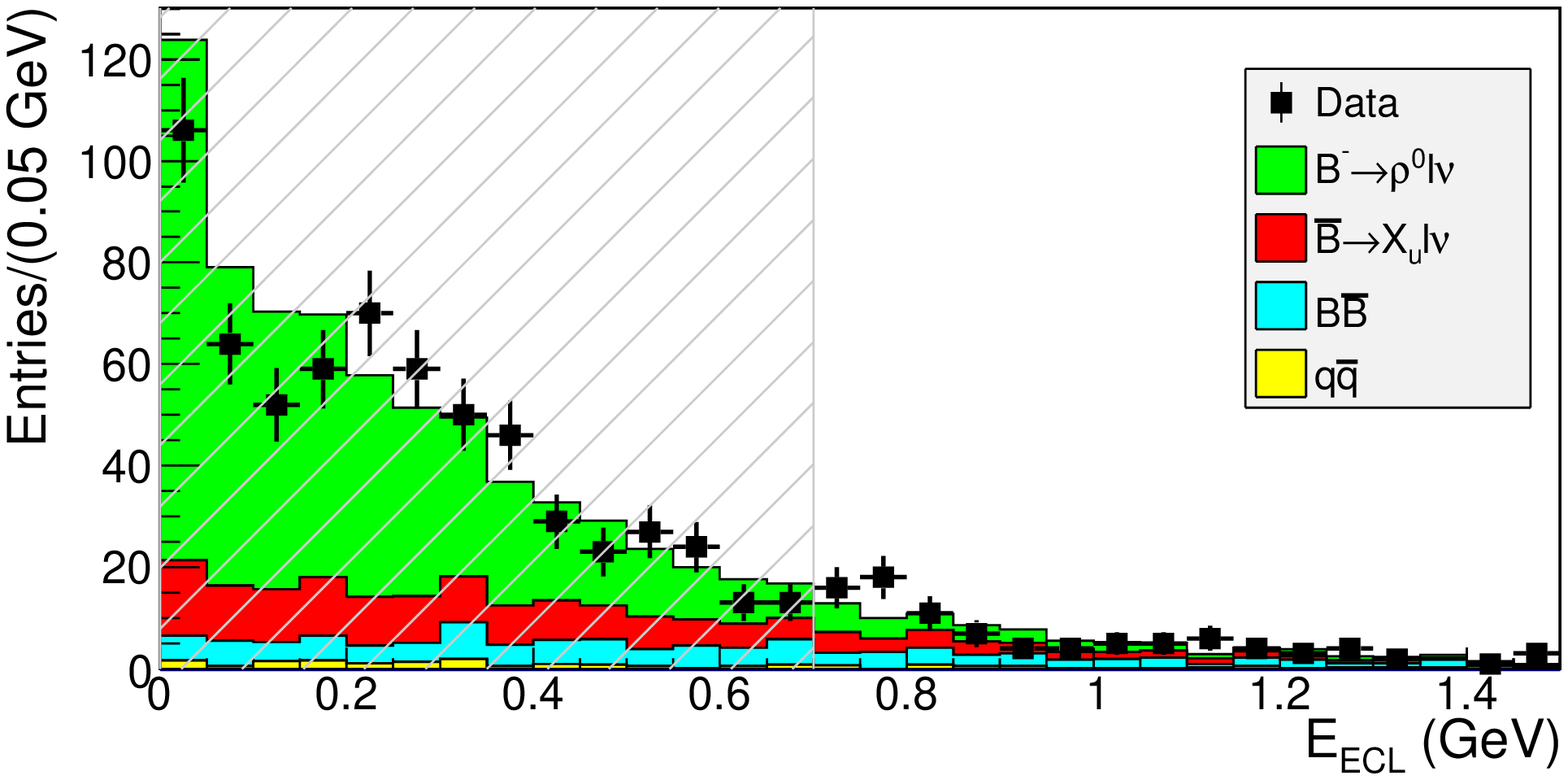} \\

      \includegraphics[width=0.41\textwidth]{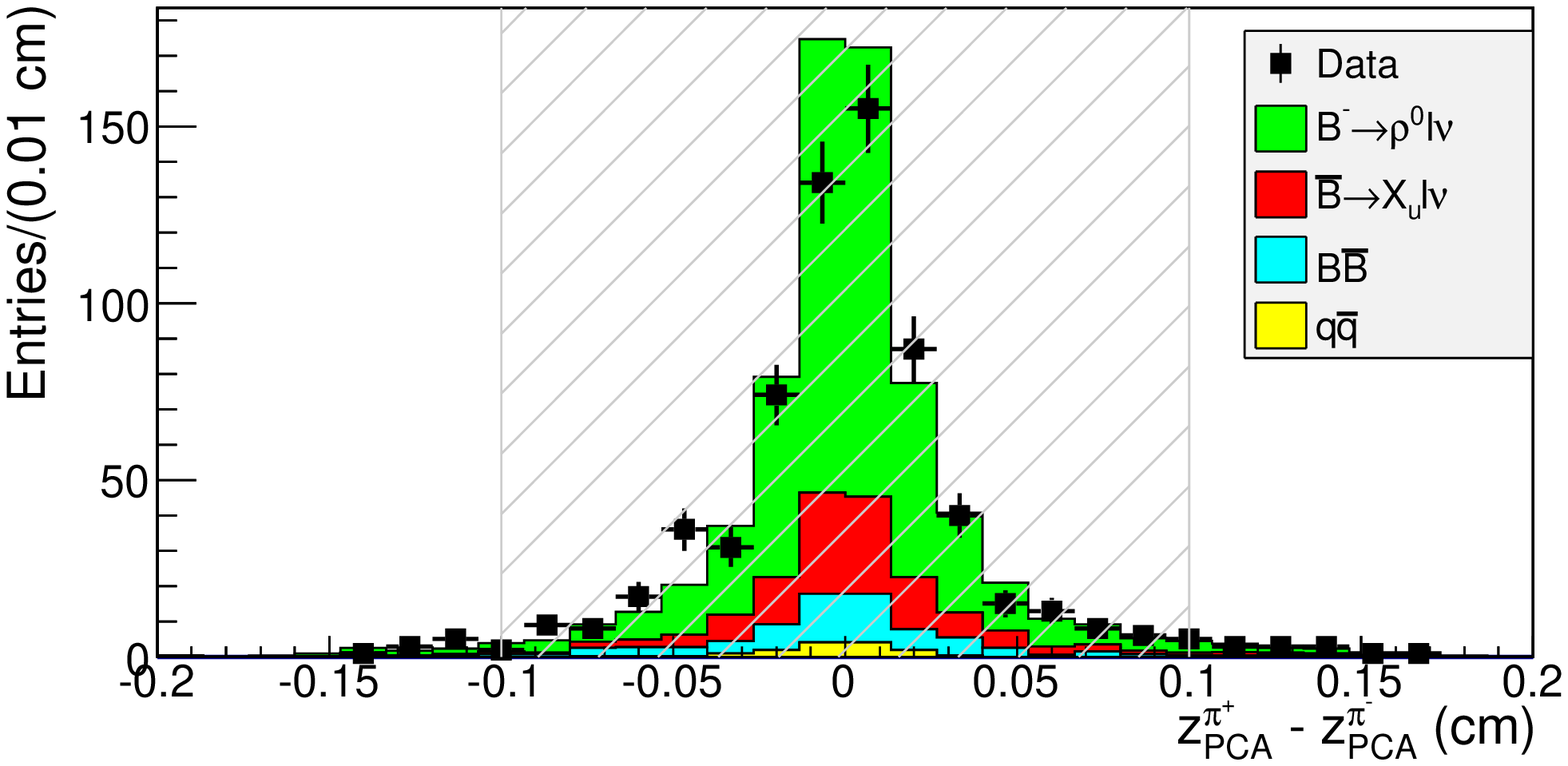} &
      \includegraphics[width=0.41\textwidth]{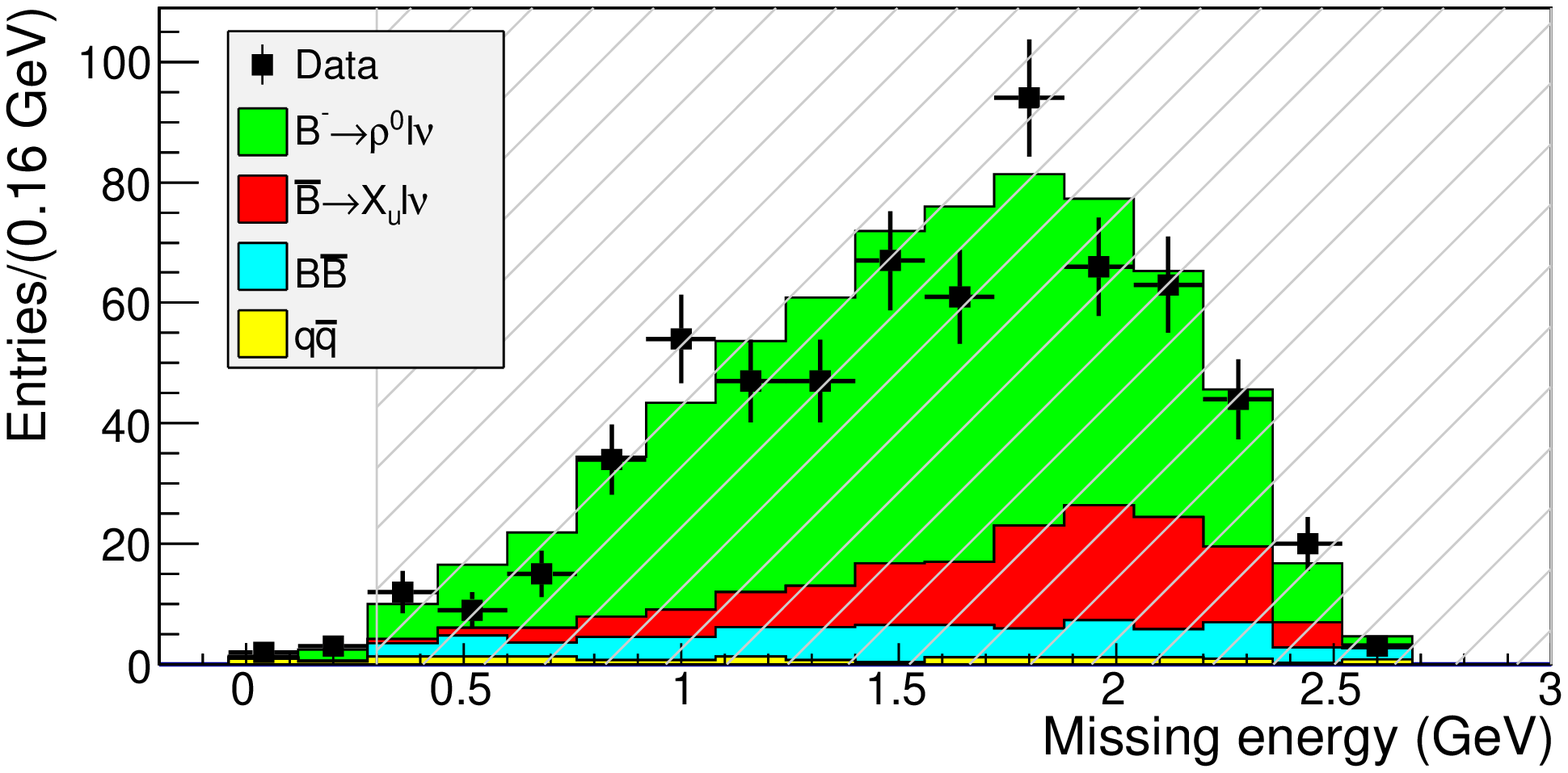} \\
      \end{tabular}
    \label{alabel}
  }
  \subfigure[Invariant mass of two photons and angles between two photons
     for the \bpizlnu decay in the region $|\misq|<0.5$~\GeVcc.]{
    \includegraphics[width=0.41\textwidth]{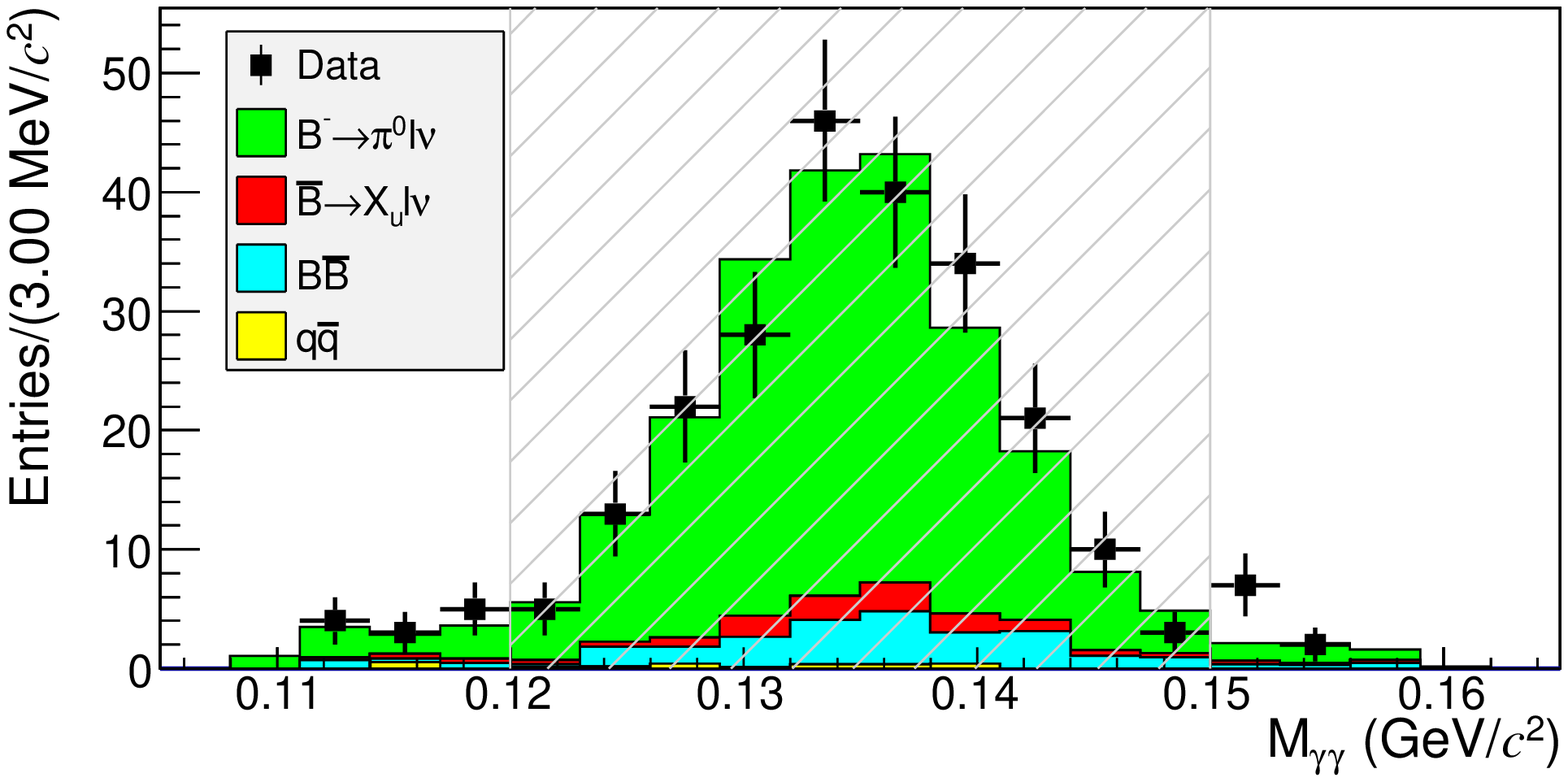}
    \includegraphics[width=0.41\textwidth]{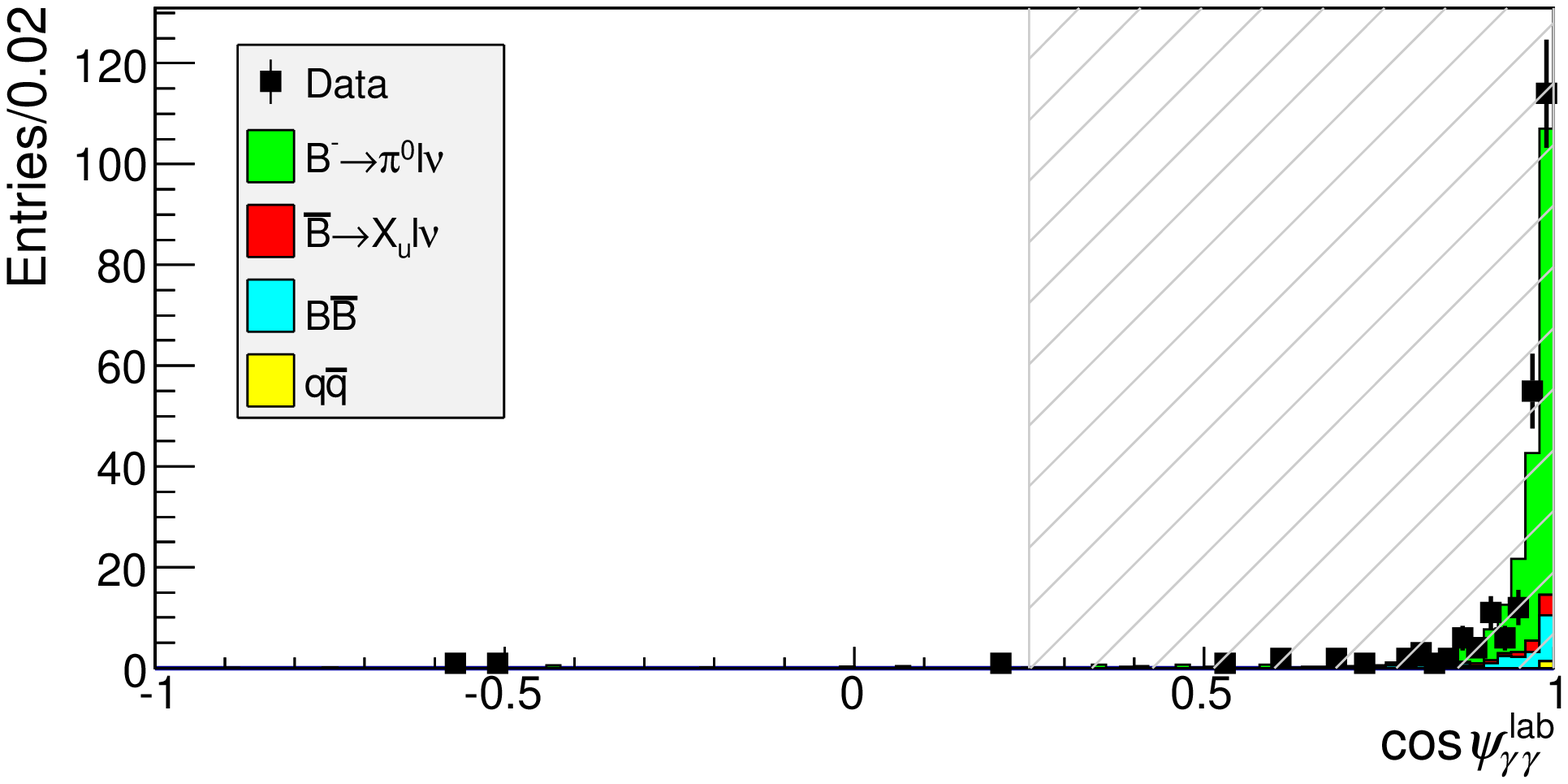}
    \label{blabel}
  } 
  \subfigure[Invariant mass of three pions and angle between photons
    from the $\pi^0$ decay for the \bomegalnu decay in the region $|\misq|<0.5$~\GeVcc.]{
    \includegraphics[width=0.41\textwidth]{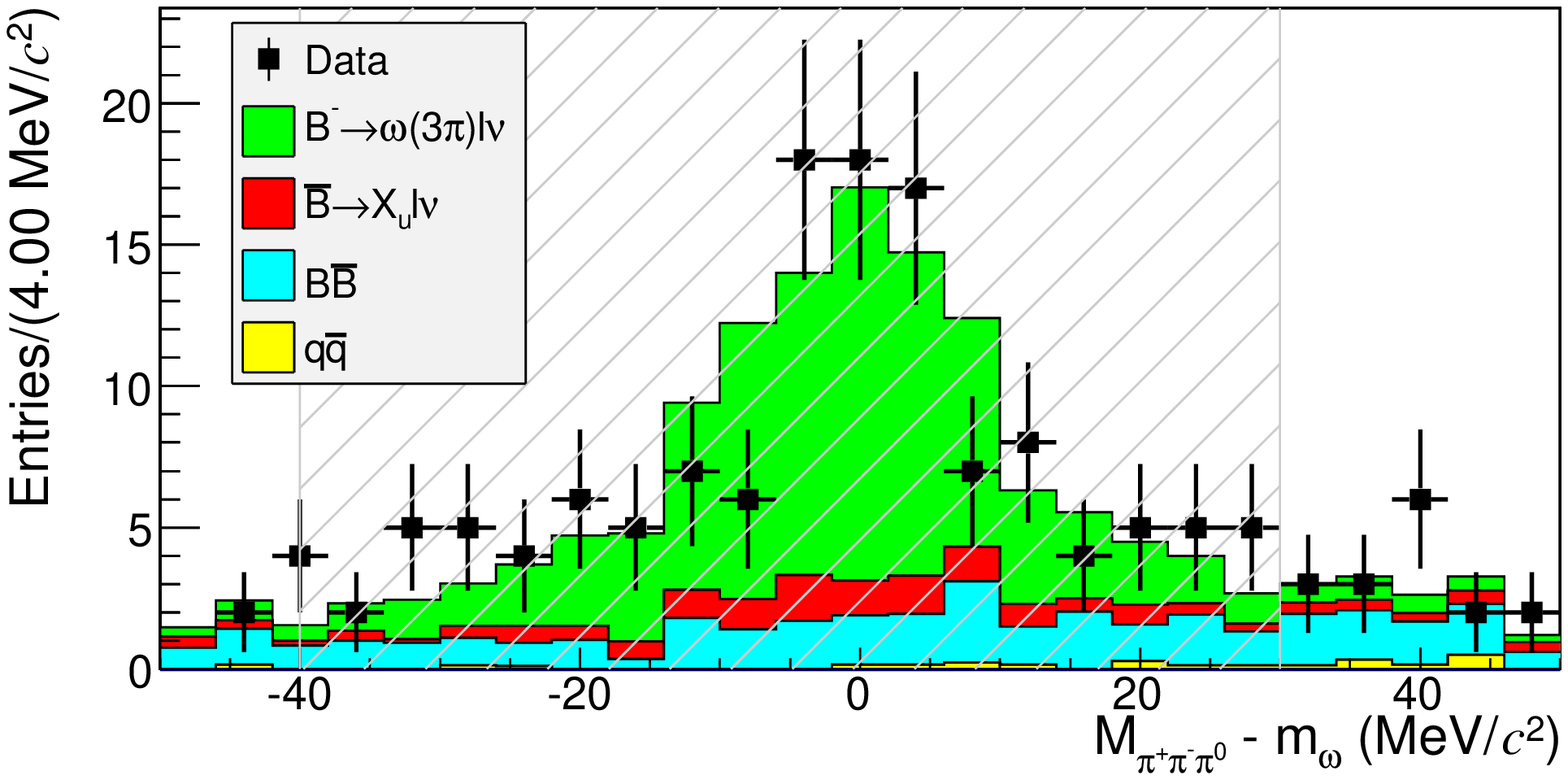}
    \includegraphics[width=0.41\textwidth]{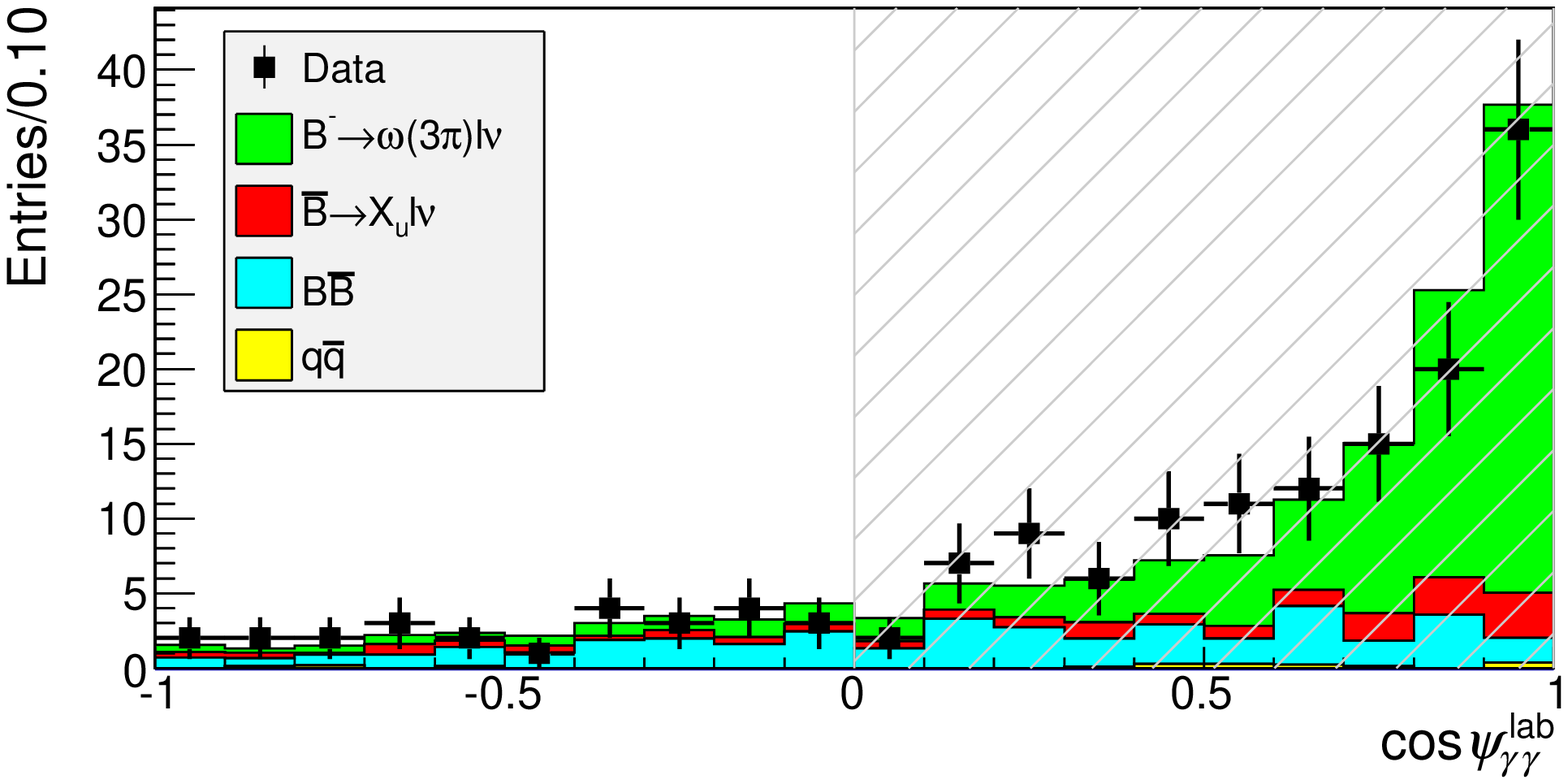}
    \label{clabel}
  }
  \caption{\label{fig:dtmcomp} (Color online) Data/MC comparison for
  \subref{alabel} \brhozlnu, \subref{blabel} \bpizlnu and
  \subref{clabel} \bomegalnu decays, where corresponding components
  are scaled according to the result of the fit to the \misq
  distribution. Where present, the hatched region shows the selection
  criterion on the corresponding variable. In each case \misq is
  required to be close to zero and all other selection criteria are
  applied except the one on the variable plotted.}
\end{figure*}

Figure~\ref{fig:dtmcomp} shows various kinematic variables as well as
the selection criteria for several decay modes. In these figures, the
MC components have been scaled according to the fit result.  The same
distributions for other decay modes also show similar level of data/MC
agreement and are not shown here.

Since the signal MC has been generated in proportion to the number of
\bb pairs, the assumption of efficiency cancellation lets us evaluate the
branching fraction of specific semileptonic decay modes over the entire
$B$ meson decay phase space as follows:
\begin{equation}
 \Br^\mathrm{DATA}(B\to X_u\ell\nu) = \Br^\mathrm{MC}(B\to X_u\ell\nu)
\dfrac{\nfit}{\nmc},
\end{equation}
where \nfit is the fitted signal yield, and
\nmc is the number of signal events in MC with
efficiency corrections applied.  The fitted signal yields and
corresponding MC predictions, along with the resulting branching
fractions, are summarised in Table~\ref{table:br}. Because of the
marginal contribution of the \bomegapglnu decay, as seen in this
table, we do not consider this mode further. The systematic
uncertainties are described below.
\begin{table}
\caption{\label{table:br}The fitted yields and MC predictions, MC and
resulting branching fractions in units of $10^{-4}$. The experimental
branching fractions are shown with both statistical (first) and
experimental systematic uncertainties (second).}
\centering\vspace*{2mm}
\begin{tabular}{cccccc}
  \hline
  $X_u$ & \nfit & \nmc & $\Br^\text{MC}$ & $\Br^\text{DATA}$ \\\hline\hline
  
$\pi^0$ & $ 232.2 \pm 22.6 $ & 211.1 & 0.73 & $ 0.80 \pm 0.08 \pm 0.04$ \\ 
$\pi^+$ & $ 462.6 \pm 27.7 $ & 421.3 & 1.36 & $ 1.49 \pm 0.09 \pm 0.07$ \\ 
$\rho^0$ & $ 621.7 \pm 35.0 $ & 505.1 & 1.49 & $ 1.83 \pm 0.10 \pm 0.10$ \\ 
$\rho^+$ & $ 343.3 \pm 28.3 $ & 295.1 & 2.77 & $ 3.22 \pm 0.27 \pm 0.24$ \\ 
$\omega(3\pi)$ & $ 96.7 \pm 14.5 $ & 104.1 & 1.15 & $ 1.07 \pm 0.16 \pm 0.07$ \\ 
$\omega(\pi^0\gamma)$ & $ 9.0 \pm 4.0 $ & 9.8 & 1.15 & $ 1.06 \pm 0.47 \pm 0.07$ \\ 
$\omega$(average) &   &   & 1.15 & $ 1.07 \pm 0.15 \pm 0.07$ \\ 

  \hline
\end{tabular}
\end{table}

\subsection{Signal extraction in bins of $q^2$}
\begin{figure*}
  \includegraphics[width=\textwidth]{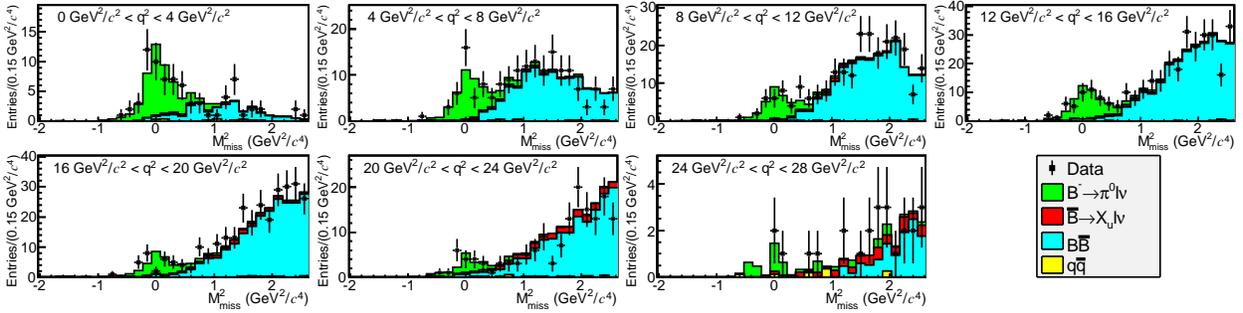}
  \caption{\label{fig:q2mmfitpi0}(Color online) Projection onto the \misq axis of the fitted distribution to
    data for \bpizlnu decay, for 7 bins in $q^2$.}
\end{figure*}

\begin{figure*}
  \includegraphics[width=\textwidth]{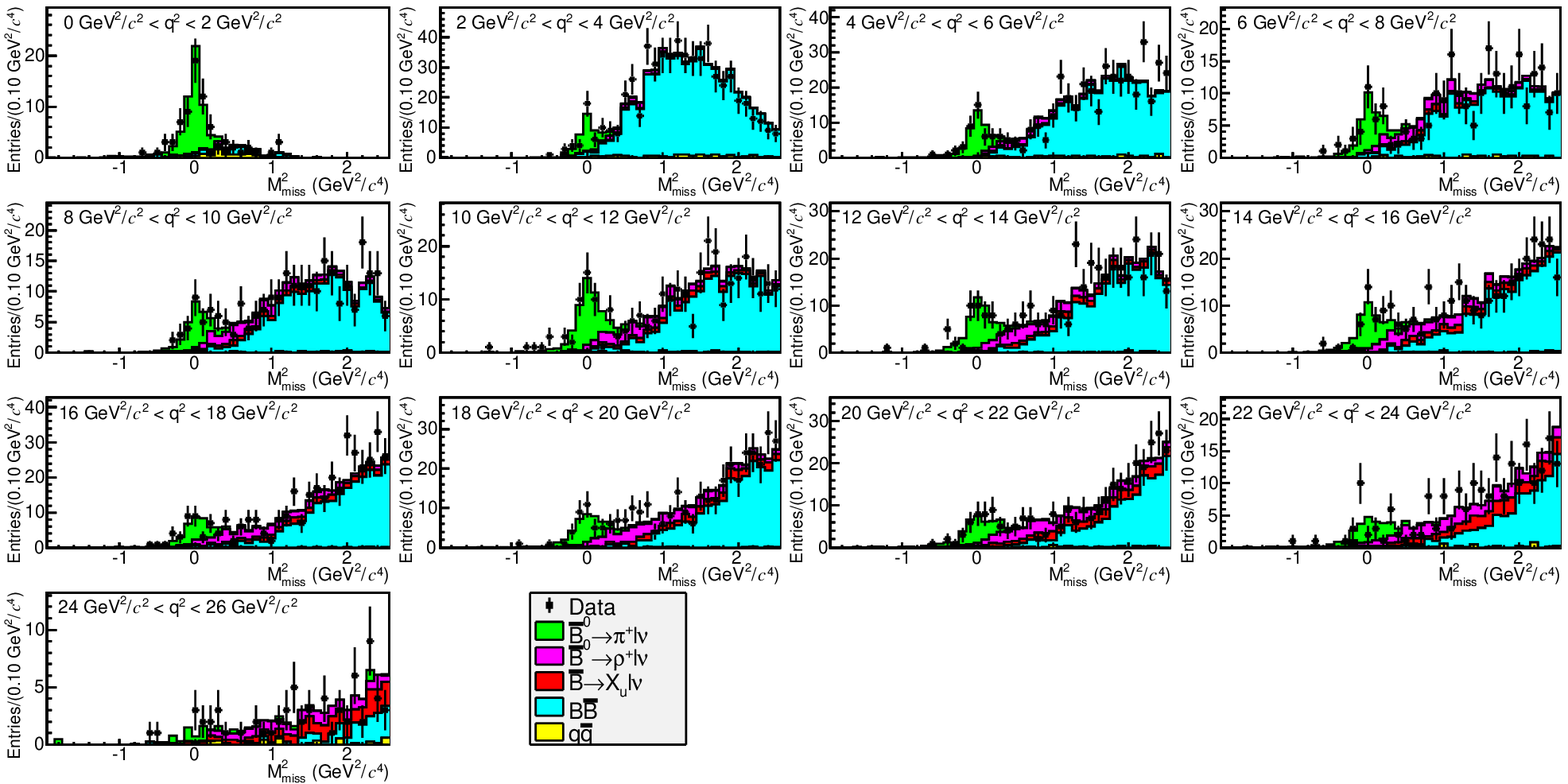}
  \caption{\label{fig:q2mmfitpic}(Color online) Projection onto the \misq axis of the fitted distribution to
    data for \bpiplnu decay, for 13 bins in $q^2$.}
\end{figure*}

\begin{figure*}
  \includegraphics[width=\textwidth]{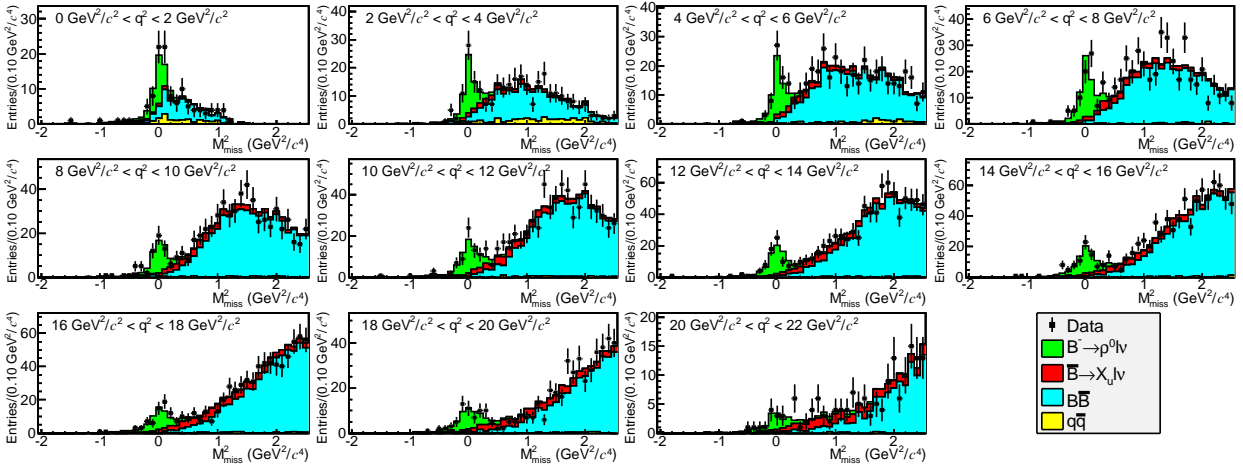}
  \caption{\label{fig:q2mmfitrho0}(Color online) Projection onto the \misq axis of the fitted distribution to
    data for \brhozlnu decay, for 11 bins in $q^2$.}
\end{figure*}

\begin{figure*}
  \includegraphics[width=\textwidth]{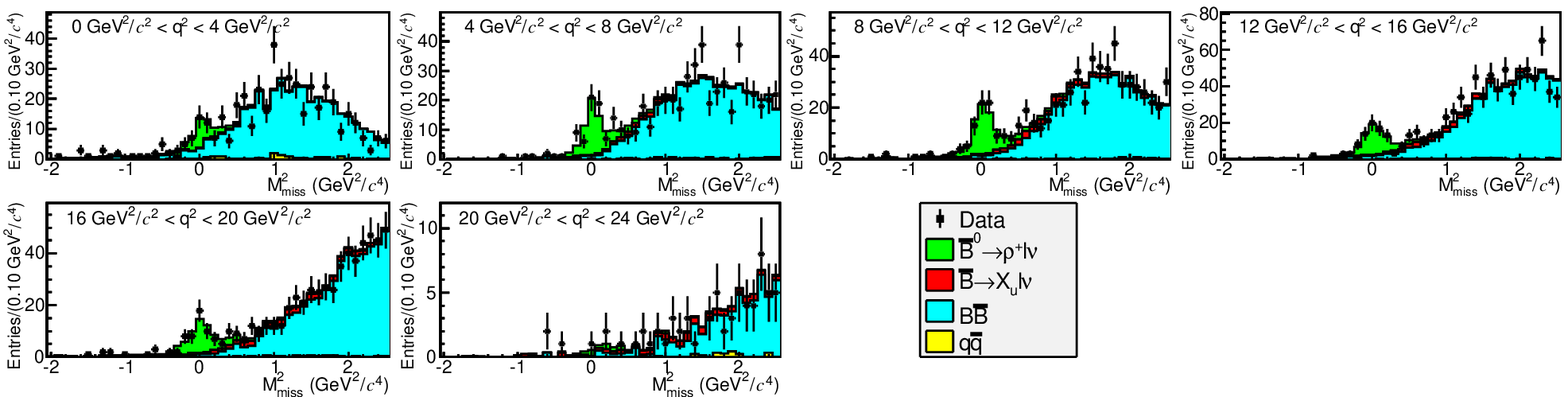}
  \caption{\label{fig:q2mmfitrhoc}(Color online) Projection onto the \misq axis of the fitted distribution to
    data for \brhoplnu decay, for 6 bins in $q^2$.}
\end{figure*}

\begin{figure*}
  \includegraphics[width=\textwidth]{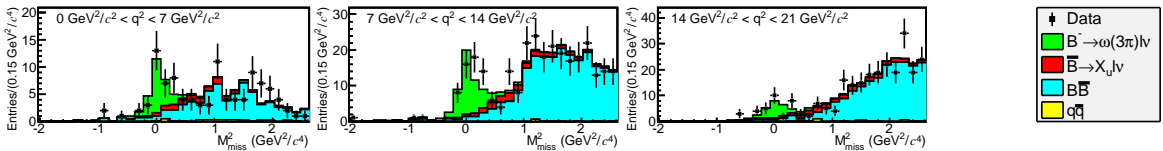}
  \caption{\label{fig:q2mmfitom3p}(Color online) Projection onto the \misq axis of the fitted distribution to
    data for \bomegalnu decay, for 3 bins in $q^2$.}
\end{figure*}

We also perform signal extractions in bins of $q^2$. To obtain the
number of signal events, we perform a two-dimensional binned maximum
likelihood fit in the $q^2$-\misq plane. The shapes of the fit
components are taken from MC and they are organised as described
above. The signal component, as well as the \bb component, is allowed
to float in each $q^2$ bin independently. The rest of the components
are varied or fixed in the same manner as in the $q^2$-independent
\misq fit and the parameters of the fit are the yields of each
component in the entire $q^2$-\misq plane. The \misq distributions are
shown in Fig.~\ref{fig:q2mmfitpi0} for \bpizlnu decays,
Fig.~\ref{fig:q2mmfitpic} for \bpiplnu decays,
Fig.~\ref{fig:q2mmfitrho0} for \brhozlnu decays,
Fig.~\ref{fig:q2mmfitrhoc} for \brhoplnu decays and
Fig.~\ref{fig:q2mmfitom3p} for \bomegappplnu decays.
We correct for the effects of finite detector resolution and
bremsstrahlung on the $q^2$ distributions using a simple unfolding
procedure described in our previous untagged
measurement~\cite{BelleUntaggedPi}, involving the product of the
inverse response matrix that is built using the true and reconstructed $q^2$
variables, with the vector of yields as a function of $q^2$.  The
extracted yields and partial branching fractions, as well as the full
statistical correlation matrices, are given in the Appendix. A
comparison of the yields and resulting branching fractions obtained
from the fits in bins of $q^2$ with those obtained from fits to the
entire $q^2$ range is shown in Table~\ref{table:brcomp}. As can be
seen, they are in excellent agreement.

\begin{table*}
  \caption{\label{table:brcomp}Comparison of signal yields and resulting
    branching fractions for the full $q^2$ range and obtained by summing over $q^2$ bins.
     Here $\varepsilon$ is the total efficiency for the entire $q^2$ range.}\vspace*{2mm}
  \begin{tabular}{
  @{\hspace{0.25cm}}l@{\hspace{0.4cm}}
  @{\hspace{0.25cm}}c@{\hspace{0.4cm}}
  @{\hspace{0.25cm}}c@{\hspace{0.4cm}}
  @{\hspace{0.25cm}}c@{\hspace{0.4cm}}
  @{\hspace{0.25cm}}c@{\hspace{0.4cm}}
  @{\hspace{0.25cm}}c@{\hspace{0.4cm}}
  }
    \hline
    $X_u$ & $N$ & $\sum\Delta N$ & $\varepsilon$ & $\Br$     & $\sum\Delta \Br$ \\
          &     &                & $10^{-3}$     & $10^{-4}$ &  $10^{-4}$       \\
    \hline
    $\pi^0$  & $ 232.2 \pm 22.6 $ & $233.3\pm20.6$ & $1.83\pm0.03$ & $ 0.80 \pm 0.08 \pm 0.04$ & $0.81 \pm 0.07 \pm 0.04$ \\ 
    $\pi^+$  & $ 462.6 \pm 27.7 $ & $461.1\pm27.4$ & $2.07\pm0.02$ & $ 1.49 \pm 0.09 \pm 0.07$ & $1.49 \pm 0.09 \pm 0.07$ \\ 
    $\rho^0$ & $ 621.7 \pm 35.0 $ & $621.9\pm34.8$ & $2.13\pm0.02$ & $ 1.83 \pm 0.10 \pm 0.10$ & $1.84 \pm 0.10 \pm 0.10$ \\ 
    $\rho^+$ & $ 343.3 \pm 28.3 $ & $350.2\pm27.3$ & $0.72\pm0.01$ & $ 3.22 \pm 0.27 \pm 0.24$ & $3.26 \pm 0.26 \pm 0.24$ \\ 
    $\omega$ & $ \hphantom{0}96.7 \pm 14.5 $ & $ \hphantom{0}99.0\pm15.0$ & $0.64\pm0.01$ & $ 1.07 \pm 0.16 \pm 0.07$ & $1.13 \pm 0.18 \pm 0.07$ \\ 
    \hline
  \end{tabular}
\end{table*}

\subsection{Discussion}
In the case of exact isospin symmetry, there are known relations between
hadronic states with different isospin projections.
For \bpilnu decays,
\begin{equation}
2\times\dfrac{\Br(\bpizlnu)}{\Br(\bpiplnu)}\dfrac{\tau_{B^0}}{\tau_{B^+}} =
1 
\end{equation}
and similarly, for \brholnu decays,
\begin{equation}
2\times\dfrac{\Br(\brhozlnu)}{\Br(\brhoplnu)}\dfrac{\tau_{B^0}}{\tau_{B^+}}
= 1.
\end{equation}
Using the lifetime ratio of neutral to charged $B$ mesons from
the PDG~\cite{PDG}, $\tau_{B^+}/\tau_{B^0} = 1.079\pm0.007$, we can
test the isospin relations using the obtained branching fractions and
taking into account correlations between measurements:
\begin{eqnarray}
2\times\dfrac{\Br(\bpizlnu)}{\Br(\bpiplnu)}\dfrac{\tau_{B^0}}{\tau_{B^+}} & = &
1.00 \pm 0.13_\mathrm{tot} 
\end{eqnarray}
and
\begin{eqnarray}
2\times\dfrac{\Br(\brhozlnu)}{\Br(\brhoplnu)}\dfrac{\tau_{B^0}}{\tau_{B^+}} & = & 1.06\times(1\pm 0.13_\mathrm{tot}),\ \ 
\end{eqnarray}
which are in a good agreement with the predictions. Using isospin
relations, we can present results for the combined branching fractions,
taking into account correlations in the systematic uncertainties:
\begin{eqnarray}
  \Br(\bpiplnu) & = & (1.49 \pm 0.08_\mathrm{stat} \pm 0.07_\mathrm{syst})\times10^{-4}\nonumber \\
  & = & (1.49 \pm 0.10_\mathrm{tot})\times10^{-4} 
\end{eqnarray}
and 
\begin{eqnarray}
  \Br(\brhoplnu) & = & (3.34 \pm 0.16_\mathrm{stat} \pm 0.17_\mathrm{syst})\times10^{-4}\nonumber \\
  & = & (3.34 \pm 0.23_\mathrm{tot})\times10^{-4},
\end{eqnarray}
where the total error is obtained by adding the statistical and
systematic uncertainties in quadrature.
For \bpilnu decay, this measurement agrees with recent untagged results from
Belle~\cite{BelleUntaggedPi} and \babar~\cite{BaBarUntaggedPiOmegaEta} at a
similar level of precision.  For \brholnu decay,
the branching fraction is 43\% ($2.7\sigma$) higher than the current PDG
value $\Br^\mathrm{PDG}(\brholnu) = (2.34\pm0.15\pm0.24)\times10^{-4}$
and the precision is almost twice that of the PDG value.

The branching fraction of \bomegalnu decay is in good agreement with the 
PDG value $\Br^\mathrm{PDG}(\bomegalnu) = (1.15\pm0.17)\times10^{-4}$
and has the same precision.

\begin{figure}
  \includegraphics[width=0.595\textwidth]{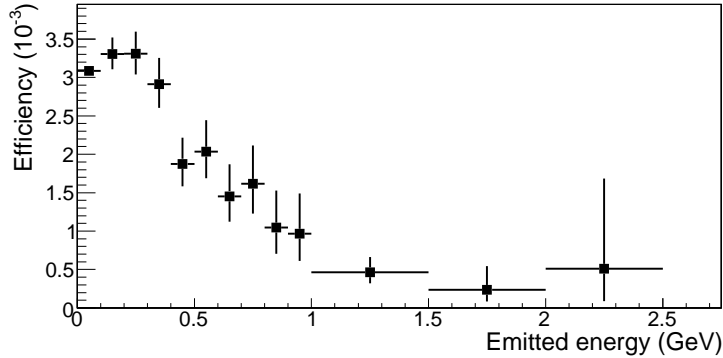}
  \caption{\label{fig:effvsenergy}Detection efficiency for \bpiplnu
  decay as a function of the total emitted energy carried by photons.}
\end{figure}

We note that the obtained branching fractions are fully inclusive of
soft photon emission. As an example, the dependence of the
reconstruction efficiency of the \bpiplnu decay on the energy carried
away by photons is shown in Fig.~\ref{fig:effvsenergy}. It is seen
that the detection efficiency is constant for total emitted energy below
300 MeV, where the internal bremsstrahlung process should dominate; MC
describes this process using the PHOTOS package. For higher emitted
energies, the efficiency drops and should naturally suppress possible
direct (or structure-dependent) emission, which is not included in MC.

\section{Systematic uncertainties}
The systematic uncertainties are organized into two
categories: those related to detector simulation, such as the charged
track reconstruction efficiency, particle identification and neutral
cluster reconstruction; and effects of the form factor model used in
the MC.

The difference between the track reconstruction efficiency for data
and MC is estimated using partially reconstructed $D^*$ mesons and
found to be 0.35\% per charged particle track. This difference is
assigned as the track reconstruction systematic error.

We estimate the $\pi^0$ reconstruction efficiency uncertainty to be
2\%, from a dedicated study of $\eta$ decays in the modes
$\eta\to\pi^+\pi^-\pi^0$ and $\eta\to\gamma\gamma$.

By evaluating the full reconstruction tag efficiency using charmed
semileptonic decays of $B$ mesons, we effectively measure the ratio of
the branching fractions between charmed and charmless semileptonic
decays. In this case, the systematic uncertainty due to the lepton
identification mostly cancels. However, the momentum spectra of
charmed and charmless semileptonic decays are not the same and this
leads to a small difference in the lepton identification efficiency,
which we assign as a systematic uncertainty. We conservatively
estimate this uncertainty to be the difference between the
efficiencies for charmed and charmless semileptonic decay modes and
the value is 1\%.

The $K/\pi$ separation uncertainty appears in the analysis when we
apply a kaon track veto. To obtain the effect of the kaon veto for
each decay, the pion angular and momentum distributions are convoluted
with the efficiency obtained from a dedicated study of particle
identification efficiency using $D^*$ decays. We estimate the
uncertainty from the kaon veto to be $\sim 1$\% per pion track.

We estimate the uncertainty from a possible shape variation in the
\misq template histogram for the \bb component, due to inaccuracies in
the charmed semileptonic decay branching fractions used in the MC, by
varying these fractions according to the PDG values~\cite{PDG}.
The variation in the extracted yields is $\leq 0.1$\%. Due to the
smallness of this variation, this uncertainty is not listed in the
summary systematic table.

We also establish that our branching fraction results are not sensitive
to the particular values chosen for variables used in our selection
criteria by varying these within reasonable limits and comparing the results
obtained. We find that the variations in the extracted branching fractions 
stay within statistical fluctuations.

We assign a systematic uncertainty for the modeling of charmless
semileptonic cross-feed for the \bpiplnu decay and \brholnu decays
where it was fixed in the fit procedure.

\subsection{Background to the \brholnu decay}
In the MC simulation, the inclusive component of charmless semileptonic $B$ meson
decays is generated using a HQE model, producing partons that are
subsequently hadronized into various hadronic final states using the
PYTHIA6.2 package~\cite{pythia62}.  The most difficult background for
\brholnu is \bpipilnu with a non-resonant pion pair, because
it is completely indistinguishable from signal when looking at the
\misq distribution alone.

\begin{figure}
  \includegraphics[width=0.695\textwidth]{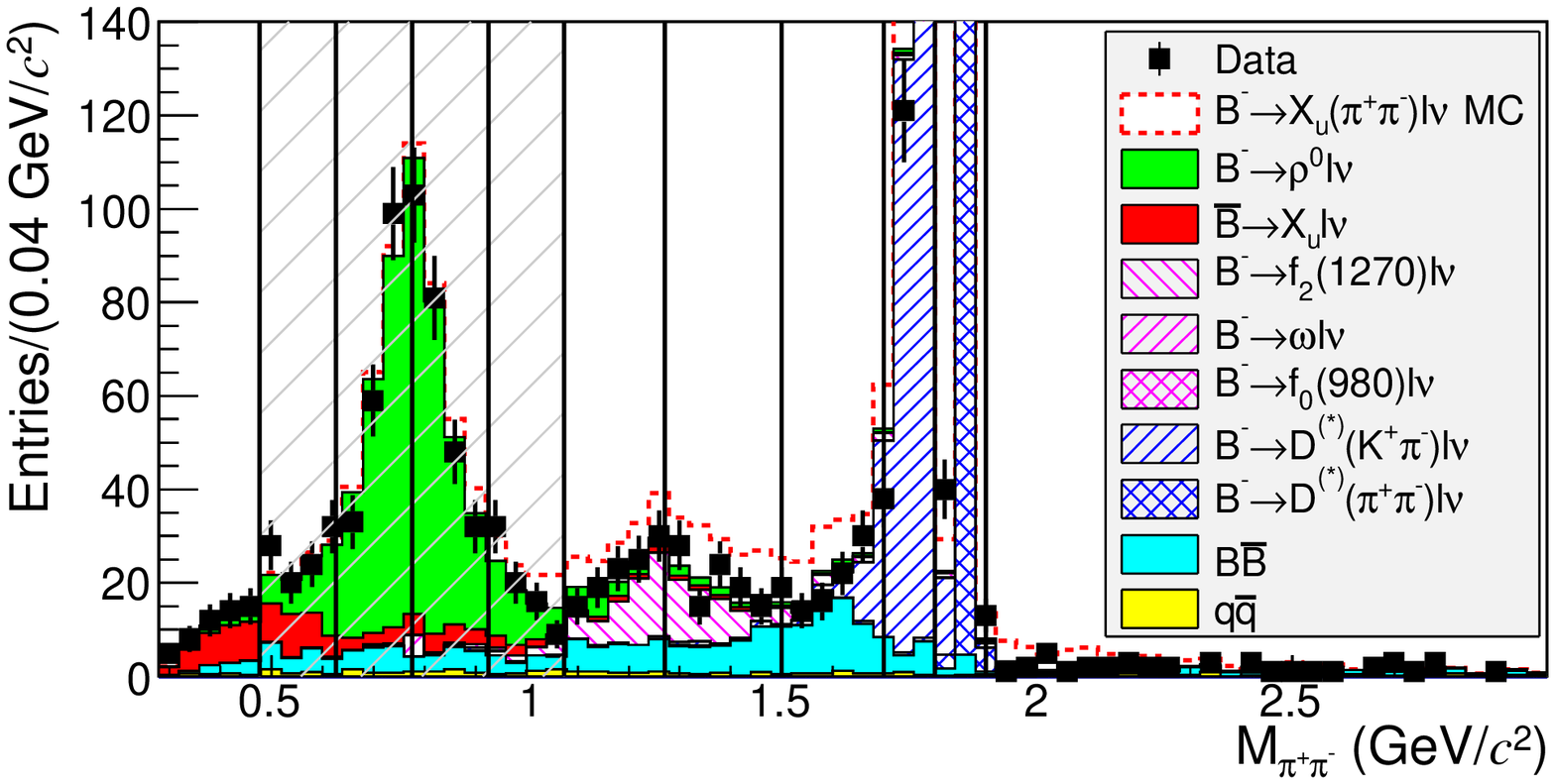}
  \includegraphics[width=0.695\textwidth]{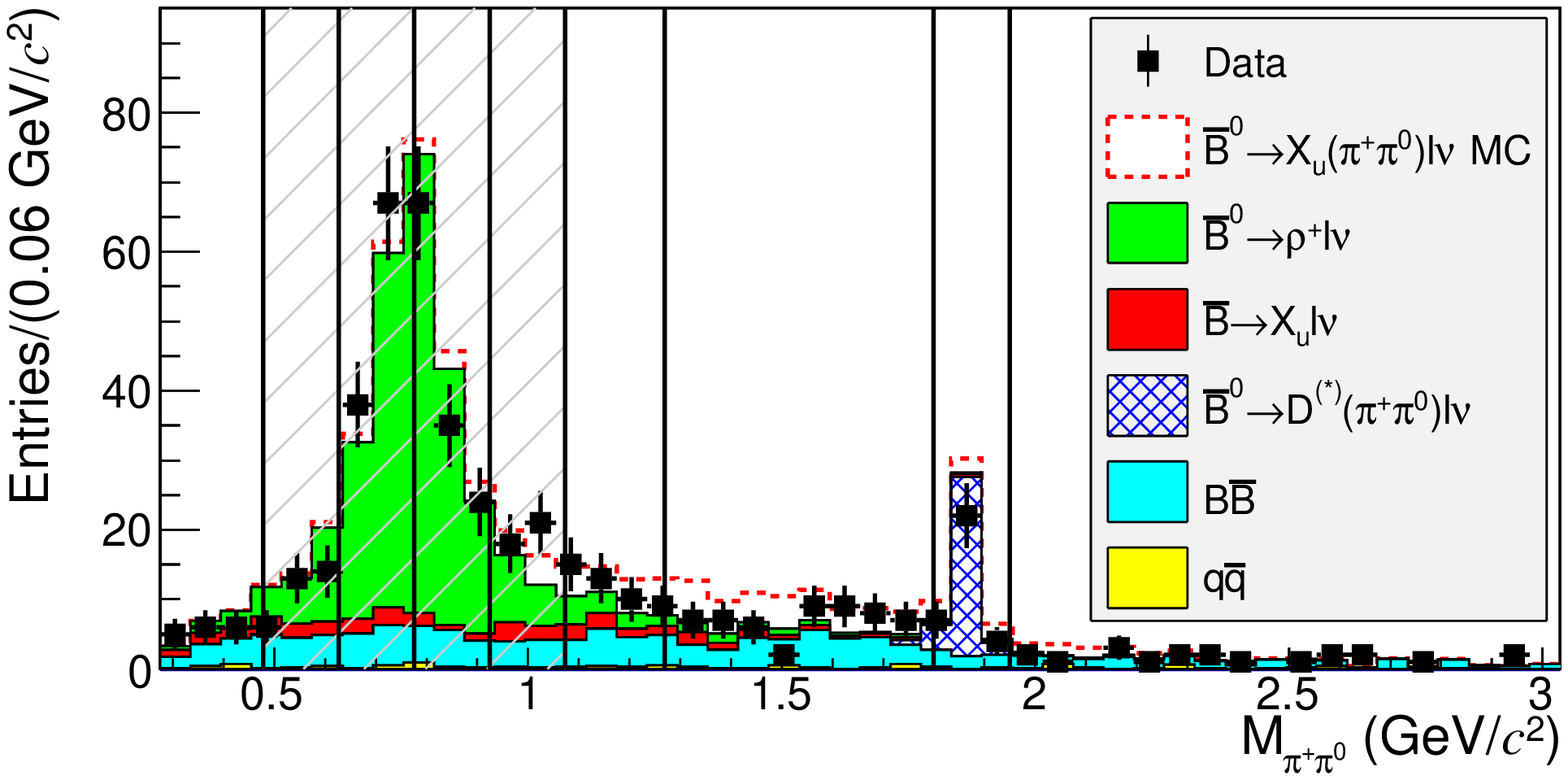}
  \caption{\label{fig:invmassproj}(Color online) Projection of the fitted
    distribution to data for the \brholnu decay onto the $M_{\pi\pi}$
    axis.  Top plot -- \brhozlnu, bottom plot -- \brhoplnu. The
    inclusive component \bpipilnu predicted by MC is
    shown on the top of the stack. Vertical lines show the bins in
    invariant mass used during the fit procedure. The hatched region
    shows the actual selection criterion on the invariant mass.}
\end{figure}

To estimate the possible \bpipilnu non-resonant
component, we perform a binned two-dimensional maximum likelihood fit
to the $M_{\pi\pi}$-\misq distribution both for \brhozlnu and
\brhoplnu decays, where the bin size in \misq is the same as
shown in Fig.~\ref{fig:fitrholnu} and the bins in invariant
mass are shown as vertical lines in
Fig.~\ref{fig:invmassproj}. Additionally, this fit allows us to fix
the yield of the
\bftwolnu decay where $f_2\to\pi^+\pi^-$, and \bpdslnu where $D^0\to
K^+\pi^-$ and the kaon was misidentified as a pion, as well as
$D^0\to\pi^+\pi^-$. The $M_{\pi\pi}$ projections of the fitted
distributions in the region $|\misq|<0.25$~\GeVcc,
with the \bpipilnu component fixed to zero, are shown
in Fig.~\ref{fig:invmassproj}. For illustration purposes, the yield
of \bpipilnu predicted by MC is shown at the top of
the stack. Relevant numbers extracted from the fit are shown in
Table~\ref{table:twodfitresults}. It can be seen that the numbers of
\brhozlnu and \brhoplnu decays are in excellent agreement with those obtained from
the \misq distribution fit. The results show that for
the \brhozlnu decay, the inclusive component decaying into two pions
is overestimated in the current MC scheme; at our present sensitivity,
the yield is consistent with zero. The extracted number of \bftwolnu
decays is more than 5$\sigma$ away from zero and almost 3 times larger than the
ISWG2 model prediction, but we cannot claim that the peak in data
around 1.3~\GeVcc is completely saturated by \bftwolnu decays, and to do
this an additional dedicated study is needed. From the above, we
estimate the uncertainty from the \bpipilnu non-resonant
 cross-feed to be 1\%.  For the \brhoplnu decay, the fit
cannot completely rule out a two-pion inclusive component, but it
shows that it is overestimated in MC by at least a factor of two. As a
result of the fit, we estimate the uncertainty from the \bpipilnu cross-feed to be 5\%.
In light of the above, in this analysis we excluded the inclusive
component that decays to two pions from the generated MC event
samples.

\begin{table}
\caption{\label{table:twodfitresults}
Results of the two-dimensional fit to the $M_{\pi\pi}$-\misq
distribution, for the scenarios where the $B\to X_u(\pi\pi)\ell\nu$
non-resonant component is either floated or fixed to zero.
$N_{\pi\pi}^\mathrm{fit}$ and $N_{\pi\pi}^\mathrm{MC}$ are the numbers
of \bpipilnu decays from the fit and MC prediction,
respectively, $N_{f_2}$ and $N_{f_2}^\mathrm{MC}$ are the numbers of
\bftwolnu decays from the fit and MC prediction, respectively, and
$N_\rho^\mathrm{sel}$ is the number of \brholnu decays from the fit
within the invariant mass selection criterion.}
\vspace*{2mm}
\begin{tabular}{cccccc}
  \hline
  $X_u$  & $N_{\pi\pi}^\mathrm{fit}$ & $N_{\pi\pi}^\mathrm{MC}$ & $N_{f_2}$ & $N_{f_2}^\mathrm{MC}$ & $N_\rho^\mathrm{sel}$ \\
  \hline
  \multirow{2}{*}{$\rho^0$} & $45.8\pm45.4$ & \multirow{2}{*}{334.9} & $128.0\pm34.0$ & \multirow{2}{*}{58.4} &$620.8\pm34.4$ \\
                             & 0             &                        & $154.4\pm22.2$ & & $627.0\pm33.8$ \\
  \hline
  \multirow{2}{*}{$\rho^+$} & $51.4\pm23.0$  & \multirow{2}{*}{125.6} & \multicolumn{2}{c}{\multirow{2}{*}{N.A.}} & $327.4\pm27.8$ \\
                             & 0              &                        & & & $344.0\pm27.8$ \\
  \hline
\end{tabular}
\end{table}

The other backgrounds to the \brhozlnu decay, \bomegalnu where
$\omega\to\pi^+\pi^-$ and \bfzlnu where $f_0\to\pi^+\pi^-$, are
expected to be small. Thus, we assign a conservative 100\% uncertainty
for the \bomegalnu and \bfzlnu components, predicted by the
MC simulation. We estimate the total uncertainty from $X_u$ cross-feed to the
\brhozlnu decay branching fraction measurement to be 2.4\%.

\subsection{\label{sysrhocveto}Cross-feed from \brhoplnu to \bpiplnu decays}
There is a large cross-feed from \brhoplnu to \bpiplnu decays because
those decays have an identical track topology, with one lepton and one
charged pion on the signal side.

For the \bpiplnu decay mode, the uncertainty in the \brhoplnu cross-feed
of 0.9\% is estimated from the difference in the yields obtained
from the \misq fit where the \brhoplnu component was
first fixed using the PDG branching fraction value and then to
the value obtained in this analysis.

In order to estimate how well the MC describes the shape of this
background, we study the effect of vetoing events in the \bpiplnu
\misq distribution that also pass the selection
criteria for \brhoplnu events. To suppress as many cross-feed events
as possible, we use looser selection criteria to identify \brhoplnu
decays and compare this to the default selection. About 43\% of MC
generated \brhoplnu decays are removed from the \bpiplnu sample in
this case. The ratio $N^\mathrm{DATA}/N^\mathrm{MC}$ changes by $\sim
2$\% with an expected uncertainty of about 1.4\% due to signal
counting. We also observe that the veto changes the background shapes
under the signal peak considerably, without introducing a large effect
on the data/MC ratio. We can therefore say that the MC reproduces the
\brhoplnu shapes sufficiently well to justify not assigning an additional
systematic uncertainty due to this shape variation.

\subsection{\label{systcont}Continuum description}
\begin{figure*}
  \includegraphics[width=0.495\textwidth]{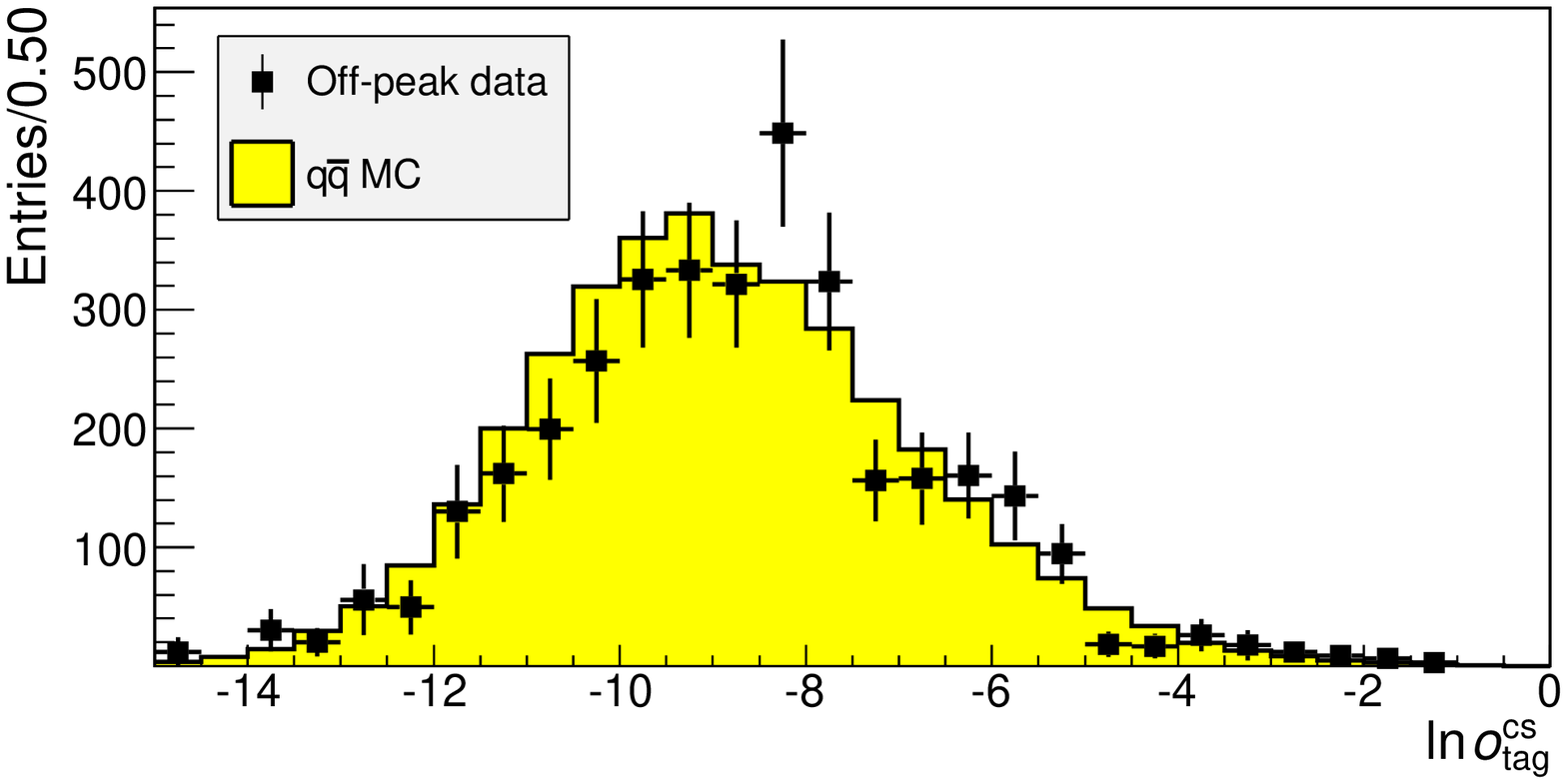}
  \includegraphics[width=0.495\textwidth]{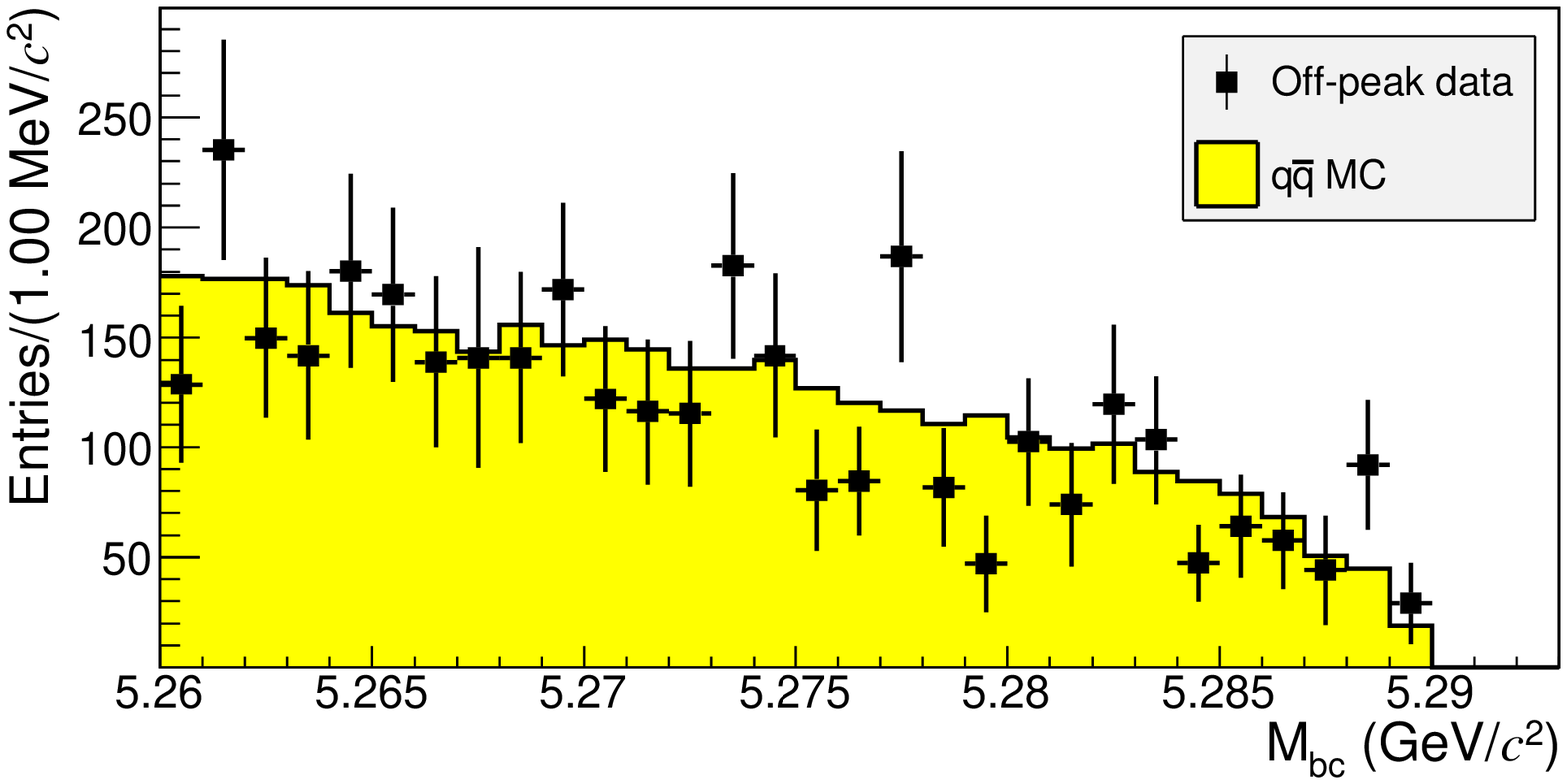}
  \includegraphics[width=0.495\textwidth]{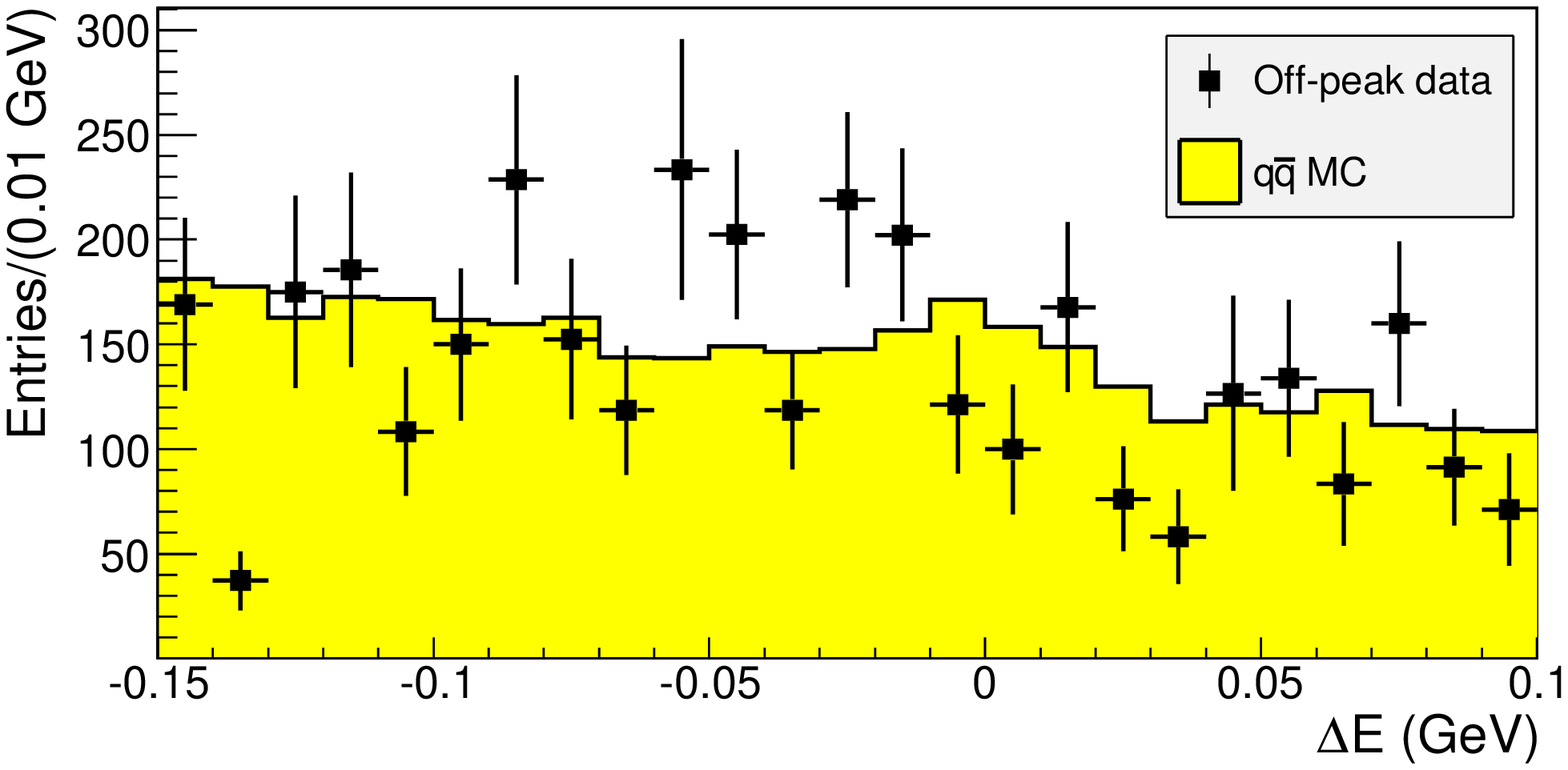}
  \includegraphics[width=0.495\textwidth]{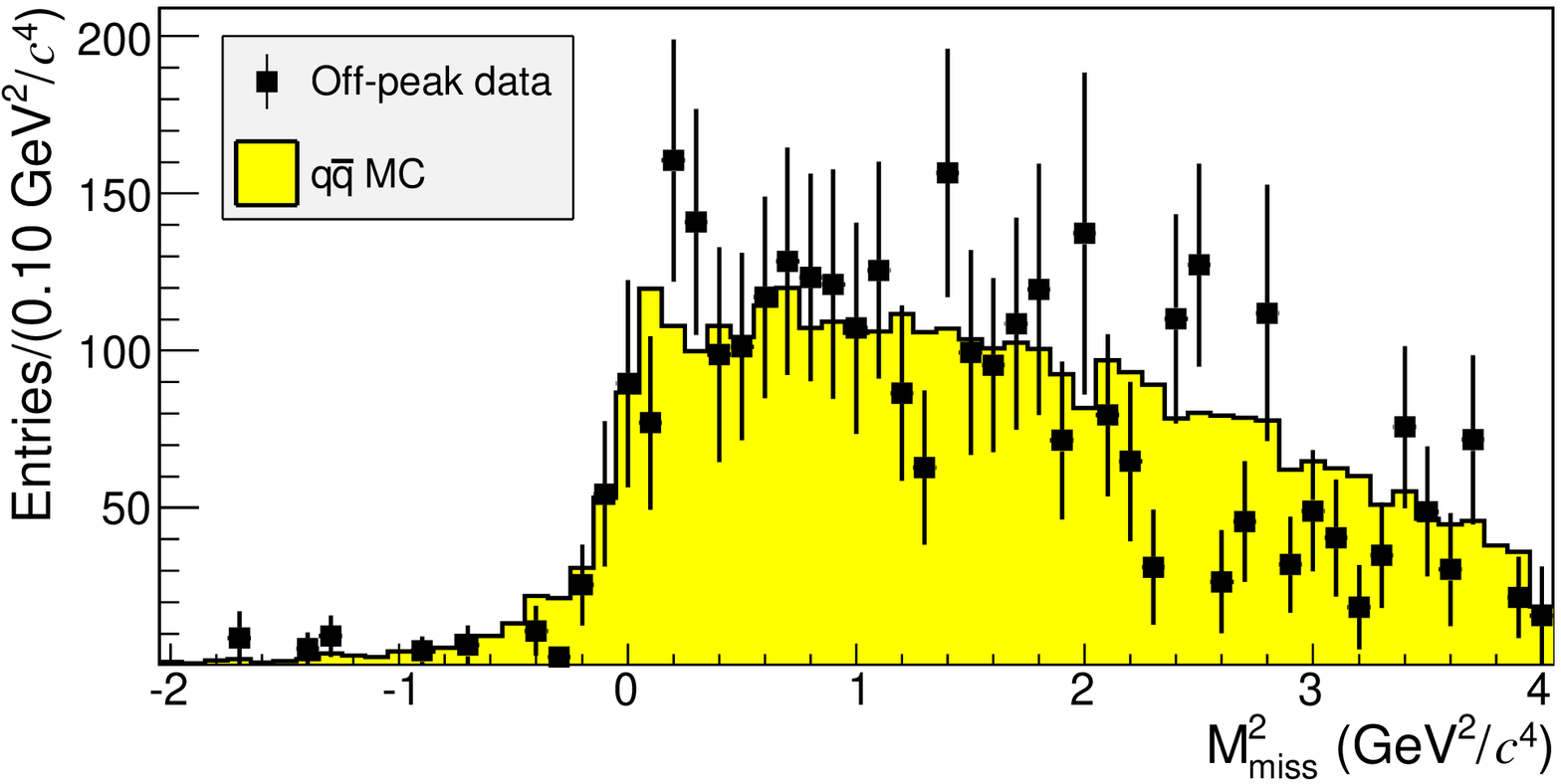}
  \caption{\label{fig:continuumrho0}(Color online) Comparison of \qq MC and off-peak
    data, using loose \brhozlnu decay selection criteria.}
\end{figure*}

A check of the continuum description uncertainty is made using 79
fb$^{-1}$ of off-peak data, collected at a collision energy 60 MeV
below the $\Upsilon(4S)$ peak. It is difficult
to compare data and MC directly within the selection criteria used in
the analysis because of the low off-peak sample size. We loosen the
selection criteria on the \NBc and \mbc variables and compare yields
and distributions between \qq continuum MC and off-peak data. The
total number of selected events for each studied decay mode is given
in Table~\ref{table:continuum}. A comparison of several distributions
for the \brhozlnu decay is shown in Fig.~\ref{fig:continuumrho0}. Decay
distributions for other decay modes convey a similar picture. As can
be seen, the data/MC agreement in most cases is at the level of
10\%. Also, the \mbc and \misq distributions are found to be in good
agreement.

\begin{table}
\caption{\label{table:continuum}Comparison of the number of selected
continuum events using loose criteria for the studied processes. $N_\qq^\mathrm{MC}$
is the MC prediction, and $N^\mathrm{data}$ is the number of off-peak data scaled
according to the luminosity.}
\vspace*{2mm}
\begin{tabular}{
  @{\hspace{0.25cm}}c@{\hspace{0.4cm}}
  @{\hspace{0.25cm}}c@{\hspace{0.4cm}}
  @{\hspace{0.25cm}}c@{\hspace{0.4cm}}
  @{\hspace{0.25cm}}c@{\hspace{0.4cm}}
}
\hline
$X_u$ & $N_\qq^\mathrm{MC}$ & $N^\mathrm{DATA}$ & Ratio \\
\hline
$\pi^0$ & 1196$\pm$15 & 1371$\pm$110 & 0.87$\pm$0.07 \\
$\pi^+$ & 2982$\pm$24 & 3045$\pm$164 & 0.98$\pm$0.05 \\
$\rho^0$ & 3655$\pm$27 & 3490$\pm$176 & 1.05$\pm$0.05 \\
$\rho^+$ & 2999$\pm$24 & 2561$\pm$151 & 1.17$\pm$0.07 \\
$\omega(3\pi)$ &  401$\pm$9 &  361$\pm$57 & 1.11$\pm$0.18 \\
$\omega(\pi\gamma)$ &  234$\pm$7 &  232$\pm$45 & 1.01$\pm$0.20 \\
\hline
\end{tabular}
\end{table}

It is difficult to claim that within the tighter, default selection
criteria used in this analysis, the MC describes continuum events with
the same level of agreement.  Because of this, the uncertainty due to the
continuum description is checked by scaling the continuum template
histogram by a factor of 50\% and examining the effect on the
fits. Based on this, the systematic uncertainty due to the continuum
description in MC is found to be less than 1\% for all decay modes.

\subsection{\label{ffuncertainty}Model uncertainty}
We estimate the uncertainty related to the form factor shape of
\bpilnu as the difference in efficiency when comparing the BCL~\cite{bcl}
and KMOW~\cite{kmow} form factor parametrizations.  For \brholnu and
\bomegalnu decays, we estimate the uncertainty as the difference in the
total efficiencies obtained using the LCSR calculation~\cite{ball05} and
the ISGW2 model result~\cite{ISGW2}.  The \bomegalnu decay has a slightly larger
uncertainty than the other decays due to the steeper efficiency
dependence on $q^2$.

\subsection{Summary of systematic uncertainties}
The summary of the systematic uncertainties is given in
Table~\ref{table:systsum}.  The total uncertainty is obtained by
summing the individual uncertainties in quadrature.

For branching fractions evaluated in bins of $q^2$, we assume for each
source of systematic uncertainty (with model uncertainty excluded)
that the size of the uncertainty is the same for all $q^2$ bins. As a
consequence, the total systematic uncertainty is also assumed to be
fully correlated between different $q^2$ bins, {\it i.e.}, it does not
affect the shape of the measured $q^2$ spectrum. As there is one
dominating source of systematic error---the tag calibration---this
assumption should be valid to a good approximation.

To average between different isospin modes, we assume partial
correlation in the tag efficiency calibration uncertainty (100\%
correlation in particle ID and 30\% correlation in branching fractions
uncertainties used for tag calibration), and 100\% correlation in
track reconstruction, lepton ID and kaon veto. The other components of
the systematic uncertainty listed in Table~\ref{table:systsum} are
assumed to be uncorrelated. 
Under this assumption, the systematic correlation
between the \bpiplnu and \bpizlnu modes is 49\% and
between the \brhoplnu and \brhozlnu modes 35\%.
\begin{table*}[htb]
  \caption{\label{table:systsum}Systematic uncertainties for
    the branching fraction results in percent.}
    \vspace*{2mm}
    \begin{tabular}{
	l@{\hspace{0.8cm}}
	*{4}{@{\hspace{0.7cm}}c@{\hspace{0.7cm}}}
	@{\hspace{0.7cm}}c@{\hspace{0.cm}}}
      \hline \hline
      Source of uncertainty & \multicolumn{5}{c}{Assigned systematic uncertainty for \bxulnu decays} \\
      $X_u$ & $\pi^+$ & $\pi^0$ & $\rho^+$ & $\rho^0$ & $\omega(3\pi)$ \\
      \hline
      {\bf Detector Simulation:} & & & & & \\
      Track reconstruction             & 0.35 & -     & 0.35 & 0.7 & 0.7 \\
      $\pi^0$ reconstruction           &     -  & 2.0 &  2.0 & -      &  2.0 \\
      Lepton identification            &  1.0 & 1.0 &  1.0 & 1.0  &  1.0 \\
      Kaon veto                        &  0.9 &  -    &  1.0 & 2.0  &  2.0 \\
      Continuum description            &  1.0 & 0.5 &  0.5 & 0.7  &  0.0 \\
      $X_u$ cross-feed                 &  0.9 &  -    &  5.0 & 2.4  &   -    \\
      Tag calibration                  &  4.5 & 4.2 &  4.5 & 4.2  &  4.2 \\\hline
      Combined                         &  4.9 & 4.8 &  7.2 & 5.4  &  5.2 \\
      \hline
      {\bf Form Factor Shapes:}        &  1.1 & 1.9 &  1.7 & 1.3  &  3.8 \\
      \hline 
      {\bf Total systematic error}     &  5.0 & 5.1 &  7.4 & 5.6  &  6.4 \\
      \hline \hline
    \end{tabular}
\end{table*}

\section{\label{vub}Extraction of $|V_{ub}|$}

To extract a value for $|V_{ub}|$, we use the expression
\begin{equation}
|V_{ub}| = \sqrt{\dfrac{C_v\Delta\Br}{\tau_B \Delta\zeta}},
\end{equation}
where $C_v = 2$ for $B^+$ decay modes and $C_v = 1$ for $B^0$ decay
modes, $\tau_B$ is the lifetime of the corresponding $B$ meson,
$\Delta\Br$ is the measured partial branching fraction within a given
region of $q^2$ and $\Delta\zeta =\int\d\Gamma/|V_{ub}|^2$ is the
normalised partial decay width calculated within that $q^2$ region for
each model.  Values of $|V_{ub}|$ extracted from charmless
semileptonic partial branching fractions within the $q^2$ region valid
for each model are given in Table~\ref{table:ulnuvub}. For low
$q^2$, the form factor predictions are made using LCSR
calculations while, for high $q^2$, the predictions are
calculated using Lattice QCD (LQCD). Some calculations give form
factor predictions corresponding to the entire $q^2$ region. It is
evident that the main contribution to the $|V_{ub}|$ uncertainty
comes from theory. The experimental uncertainty is typically
2-3 times less than the corresponding theoretical one. The lifetime
values $\tau_{B^0} = 1.519\pm0.007$~ps and $\tau_{B^+} =
1.641\pm0.008$~ps are taken from the PDG~\cite{PDG}.

\begin{table*}
  \caption{\label{table:ulnuvub}Values of the CKM matrix element
    $|V_{ub}|$ based on rates of exclusive \bxulnu decays and
    theoretical predictions of form factors within various $q^2$
    ranges. The first uncertainty is statistical, the second is experimental
    systematic and the third is theoretical. The theoretical uncertainty
    for the ISGW2 model is not available.}  \centering
    \vspace*{2mm}
  \begin{tabular}{cccccccc}
    \hline
    $X_u$  & Theory & $q^2$ & \nfit & \nmc & $\Delta\Br$ & $\Delta\zeta$ & $|V_{ub}|$ \\
    & & \GeVcc &&& $10^{-4}$ & ps$^{-1}$ & $10^{-3}$\\
    \hline \hline
\multirow{4}{*}{ $ \pi^0 $ }& LCSR~\cite{kmow} & $ < 12 $ &  $ 119.6 \pm 16.2 $ & 116.5 &  $ 0.423 \pm 0.057 $ & $4.59^{+1.00}_{-0.85}$  &  $ 3.35 \pm 0.23 \pm 0.09^{+0.36}_{-0.31} $ \\
& LCSR~\cite{Ball:2004ye} & $ < 16 $ &  $ 168.2 \pm 18.9 $ & 153.5 &  $ 0.588 \pm 0.066 $ & $5.44^{+1.43}_{-1.43}$ &  $ 3.63 \pm 0.20 \pm 0.10^{+0.60}_{-0.40} $ \\
& HPQCD~\cite{hpqcd} &\multirow{2}{*}{ $ > 16 $ }&\multirow{2}{*}{  $ 58.6 \pm 10.5 $ }&\multirow{2}{*}{ 57.6 }&\multirow{2}{*}{  $ 0.196 \pm 0.035 $ }& $2.02^{+0.55}_{-0.55}$ &  $ 3.44 \pm 0.31 \pm 0.09^{+0.59}_{-0.39} $ \\
& FNAL~\cite{Bailey:2008wp} &&&&& $2.21^{+0.47}_{-0.42}$ &  $ 3.29 \pm 0.30 \pm 0.09^{+0.37}_{-0.30} $ \\
\hline
\multirow{4}{*}{ $ \pi^+ $ }& LCSR~\cite{kmow} & $ < 12 $ &  $ 247.2 \pm 18.9 $ & 233.1 &  $ 0.808 \pm 0.062 $ & $4.59^{+1.00}_{-0.85}$  &  $ 3.40 \pm 0.13 \pm 0.09^{+0.37}_{-0.32} $ \\
& LCSR~\cite{Ball:2004ye} & $ < 16 $ &  $ 324.2 \pm 22.6 $ & 305.1 &  $ 1.057 \pm 0.074 $ & $5.44^{+1.43}_{-1.43}$ &  $ 3.58 \pm 0.12 \pm 0.09^{+0.59}_{-0.39} $ \\
& HPQCD~\cite{hpqcd} &\multirow{2}{*}{ $ > 16 $ }&\multirow{2}{*}{  $ 141.3 \pm 16.0 $ }&\multirow{2}{*}{ 116.1 }&\multirow{2}{*}{  $ 0.445 \pm 0.050 $ }& $2.02^{+0.55}_{-0.55}$ &  $ 3.81 \pm 0.22 \pm 0.10^{+0.66}_{-0.43} $ \\
& FNAL~\cite{Bailey:2008wp} &&&&& $2.21^{+0.47}_{-0.42}$ &  $ 3.64 \pm 0.21 \pm 0.09^{+0.40}_{-0.33} $ \\
\hline
\multirow{4}{*}{ $\rho^0$ }& LCSR~\cite{ball05} & $ < 16 $ &  $ 476.5 \pm 30.5 $ & 420.0 &  $ 1.431 \pm 0.091 $ & $13.7^{+3.4}_{-3.4}$ &  $ 3.56 \pm 0.11 \pm 0.09^{+0.54}_{-0.37} $ \\
& BM~\cite{Beyer:1998ka} &\multirow{3}{*}{ \frange }&\multirow{3}{*}{  $ 621.7 \pm 35.0 $ }&\multirow{3}{*}{ 505.1 }&\multirow{3}{*}{  $ 1.834 \pm 0.103 $ }& $15.8^{+2.3}_{-2.3}$ &  $ 3.76 \pm 0.11 \pm 0.10^{+0.31}_{-0.25} $ \\
& UKQCD~\cite{ukqcd} &&&&& $16.5^{+3.5}_{-2.3}$ &  $ 3.68 \pm 0.10 \pm 0.10^{+0.29}_{-0.34} $ \\
& ISGW2~\cite{ISGW2} &&&&& 14.1 &  $ 3.98 \pm 0.11 \pm 0.10\hphantom{^{+?.??}_{-?.??}} $ \\
\hline
\multirow{4}{*}{ $\rho^+$ }& LCSR~\cite{ball05} & $ < 16 $ &  $ 268.8 \pm 25.0 $ & 245.7 &  $ 2.574 \pm 0.239 $ & $13.7^{+3.4}_{-3.4}$ &  $ 3.51 \pm 0.16 \pm 0.13^{+0.53}_{-0.36} $ \\
& BM~\cite{Beyer:1998ka} &\multirow{3}{*}{ \frange }&\multirow{3}{*}{  $ 343.3 \pm 28.3 $ }&\multirow{3}{*}{ 295.1 }&\multirow{3}{*}{  $ 3.222 \pm 0.266 $ }& $15.8^{+2.3}_{-2.3}$ &  $ 3.66 \pm 0.15 \pm 0.14^{+0.30}_{-0.24} $ \\
& UKQCD~\cite{ukqcd} &&&&& $16.5^{+3.5}_{-2.3}$ &  $ 3.59 \pm 0.15 \pm 0.13^{+0.28}_{-0.33} $ \\
& ISGW2~\cite{ISGW2} &&&&& 14.1 &  $ 3.87 \pm 0.16 \pm 0.15\hphantom{^{+?.??}_{-?.??}} $ \\
\hline
\multirow{2}{*}{ $ \omega $ }& LCSR~\cite{ball05} & $ < 12 $ &  $ 61.3 \pm 11.4 $ & 71.6 &  $ 0.611 \pm 0.113 $ & $7.88^{+1.86}_{-1.86}$ &  $ 3.08 \pm 0.29 \pm 0.11^{+0.44}_{-0.31} $ \\
& ISGW2~\cite{ISGW2} & \frange &  $ 96.7 \pm 14.5 $ & 104.1 &  $ 1.069 \pm 0.160 $ & 14.1 &  $ 3.03 \pm 0.23 \pm 0.11\hphantom{^{+?.??}_{-?.??}}$ \\
\hline
    \end{tabular}
\end{table*}

We also perform a $|V_{ub}|$ determination with a model-independent
description of \bpilnu decays assuming isospin symmetry and the BCL form
factor parametrization~\cite{bcl}, the most recent LQCD
calculation by the FNAL/MILC collaboration~\cite{Bailey:2008wp}, and the vector
form factor value for $f_+(q^2)$ at $q^2 = 0$ calculated in the LCSR framework
from~\cite{bharucha}. We define a goodnes of fit as follows:
\begin{equation}
  \chi^2 = \chi^2_{\bpilnu} + \chi^2_\mathrm{LCSR} + \chi^2_\mathrm{LQCD}.
\end{equation}
Here, $\chi^2$ for the \bpiplnu and \bpizlnu decay modes is given by:
\begin{equation}
\chi^2_{\bpilnu} = \sum_{i,j} \delta \Br_i (C^\mathrm{EXP})^{-1}_{ij} \delta \Br_j,
\end{equation}
where $C^\mathrm{EXP}=C^\mathrm{EXP}_\mathrm{stat}+C^\mathrm{EXP}_\mathrm{syst}$
is the full experimental variance matrix.  The statistical variance
matrix $C^\mathrm{EXP}_\mathrm{stat}$ is presented in the Appendix.
The systematic variance matrix $C^\mathrm{EXP}_\mathrm{syst}$ is
obtained from the uncertainties given in Table~\ref{table:systsum},
excluding the form factor shape uncertainty.

The difference between the measured and predicted partial
branching fractions $\delta \Br_i$ in the $q^2$ range $\Delta q^2_i$ is given by:
\begin{equation}
\delta \Br_i= \Br^\mathrm{exp}_i - \dfrac{\tau_{B}}{C_v}\int_{\Delta q^2_i}\dfrac{G_F^2}{24\pi^3}|V_{ub}|^2\left|f_+(q^2,\vec{b})\right|^2\pitot^3\d q^2,
\end{equation}
where $C_v = 2$ for \bpizlnu and $C_v = 1$ for \bpiplnu, and
$f_+(q^2,\vec{b})$ is expressed using the BCL form factor
parametrization~\cite{bcl}:
\begin{equation}
f_+(q^2,\vec{b})= \frac{1}{1-q^2/ m_{B^*}^2}\,\sum\limits_{k= 0}^{K}
b_k(t_0)\, z(q^2)^k\,.
\end{equation}
The factor in front of the sum describes the pole due to the
presence of the $B^*$ vector resonance with mass $m_{B^*} =
5.325$~\GeVcc; the vector $\vec{b}=(b_0,b_1,b_2,\ldots,b_K)$ for a chosen
value of $K$ represents the set of parameters to be determined by the
fit. The function $z(q^2)\equiv z(q^2,t_0)$ is given by
\begin{equation}
z(q^2, t_0)=\frac{\sqrt{t_+-q^2}-\sqrt{t_+-t_0}}{\sqrt{t_+-q^2}+ \sqrt{t_+-t_0}}\ ,
\end{equation}
where $t_+= (m_B + m_\pi)^2$ and the optimal choice for $t_0$ is 
$t_0 = t_\mathrm{opt} \equiv (m_B + m_\pi)(\sqrt{m_B} - \sqrt{m_\pi})^2$, 
which provides a mapping of the
physical region $0 < q^2 < (m_B-m_\pi)^2$ onto the symmetric interval
$|z|<0.279$ in the complex $z$-plane.  The last parameter $b_K$ in the vector
$\vec{b}$ is constrained by angular momentum conservation at the $B\pi$
threshold:
\begin{equation}
b_{K}=-\frac{(-1)^{K}}{K}\sum\limits_{k=0}^{K-1} (-1)^k k b_k,
\end{equation}
leaving only $(b_0,b_1,b_2,\ldots,b_{K-1})$ free. Unitarity and
crossing symmetry properties of the form factor constrain the $\vec{b}$
parameters:
\begin{equation}
\sum\limits_{j,k=0}^{K} B_{jk}b_j b_k \leq 1,
\end{equation}
where the coefficients $B_{0k}$, $0\leq k \leq 14$ are given in Table~\ref{table:bjk}.
\begin{table*}
  \caption{\label{table:bjk}The elements $B_{0k}$ calculated with $t_0
  = t_\mathrm{opt}$ in units of $10^{-4}$. The other elements can be
  obtained by the relation $B_{j(j+k)} = B_{0k}$ and the symmetry
  property $B_{jk} = B_{kj}$.}
    \vspace*{2mm}
  \begin{tabular}{cccccccccccccccc}
    \hline
    $k$ &0 &1 &2 &3 &4 &5 &6 &7 &8 &9 &10 &11 &12 &13 &14 \\
    \hline
    $B_{0k}$& 197.08 &41.93 &$-$109.16 &$-$58.89 &$-$2.24 &12.18 &11.09 & 5.02 & 2.13 & 0.26 &$-$0.07 &$-$0.27 &$-$0.16 &$-$0.14 &$-$0.07\\
    \hline
  \end{tabular}
\end{table*}

The contribution to the $\chi^2$ function from the LQCD points is:
\begin{equation}
\chi^2_\mathrm{LQCD} = \sum_{i,j}\delta f_+^\mathrm{LQCD}(q^2_i)\,(C^\mathrm{LQCD})^{-1}_{ij}\,\delta f_+^\mathrm{LQCD}(q^2_j),
\end{equation}
 where $C^\mathrm{LQCD}$ is the full variance matrix provided with the LQCD
points $f_+^\mathrm{LQCD}(q^2_i)$, and 
\begin{equation}
\delta f_+^\mathrm{LQCD}(q^2_i) = f_+^\mathrm{LQCD}(q^2_i) - f_+(q^2_i,\vec{b}).
\end{equation}
The LQCD points are highly correlated and more than half of the
eigenvalues of the LQCD covariance matrix are extremely small (of the
order of $10^{-6}$ of the largest eigenvalue) or even negative. To
treat this situation, we omit some of the LQCD points as suggested in
Ref.~\cite{BaBarUntaggedPiRho}, leaving only 4 points out of 12. This
allows us to build the contribution to the $\chi^2$ from LQCD data.

The LCSR contribution to the total $\chi^2$ is:
\begin{equation}
\chi^2_\mathrm{LCSR} = \left(\dfrac{f_+^\mathrm{LCSR}(0)- f_+(0,\vec{b})}{\delta f_+^\mathrm{LCSR}(0)}\right)^2,
\end{equation}
where $f_+^\mathrm{LCSR}(0) = 0.261^{+0.020}_{-0.023}$
from~\cite{bharucha}. In this formalism, the free parameters are $|V_{ub}|$ and 
the real coefficients $b_k$, $0\leq k<K$; thus, the total number of free
parameters is $N=K+1$.

\begin{figure}
  \includegraphics[width=0.595\textwidth]{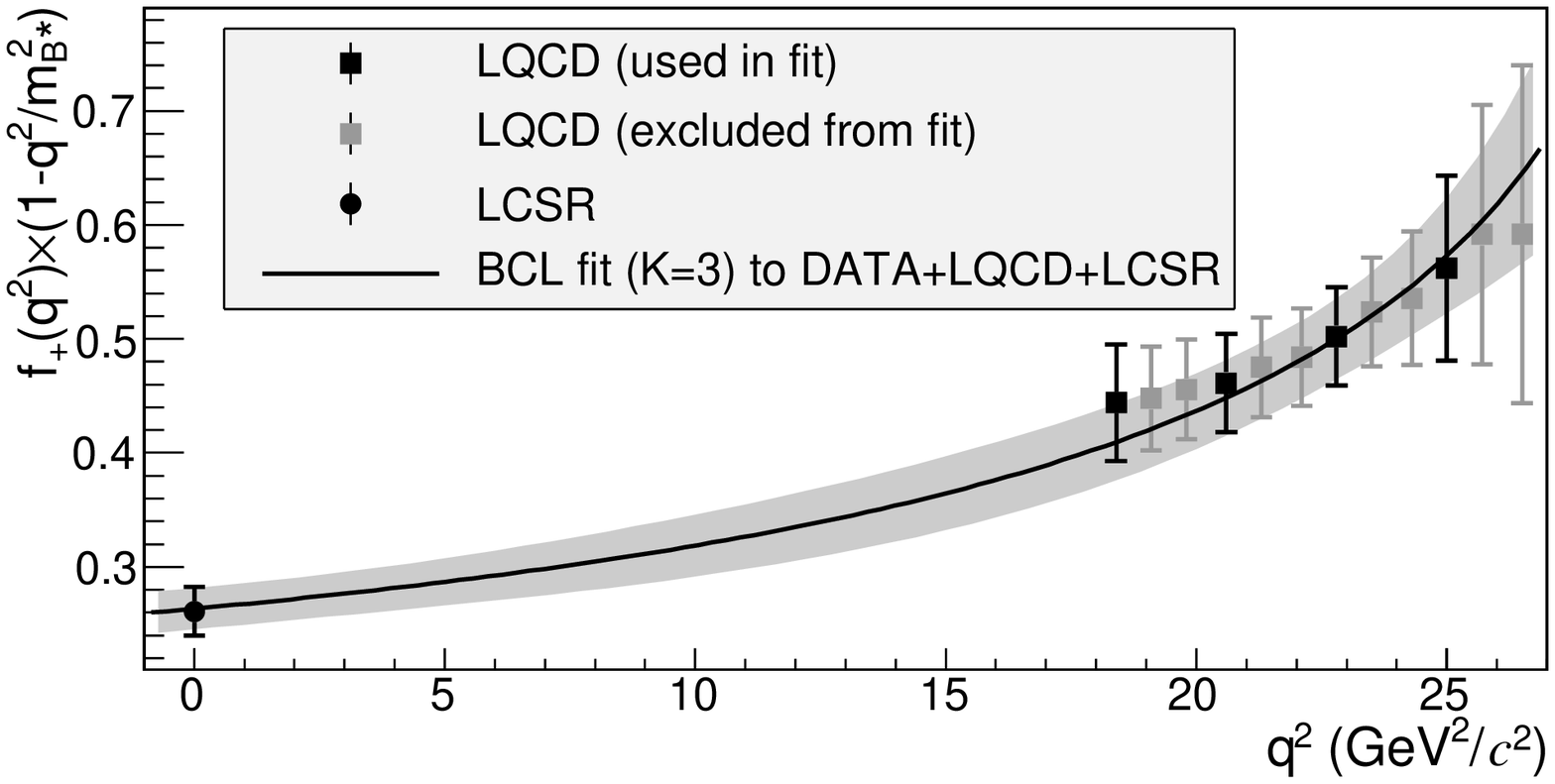}
  \includegraphics[width=0.595\textwidth]{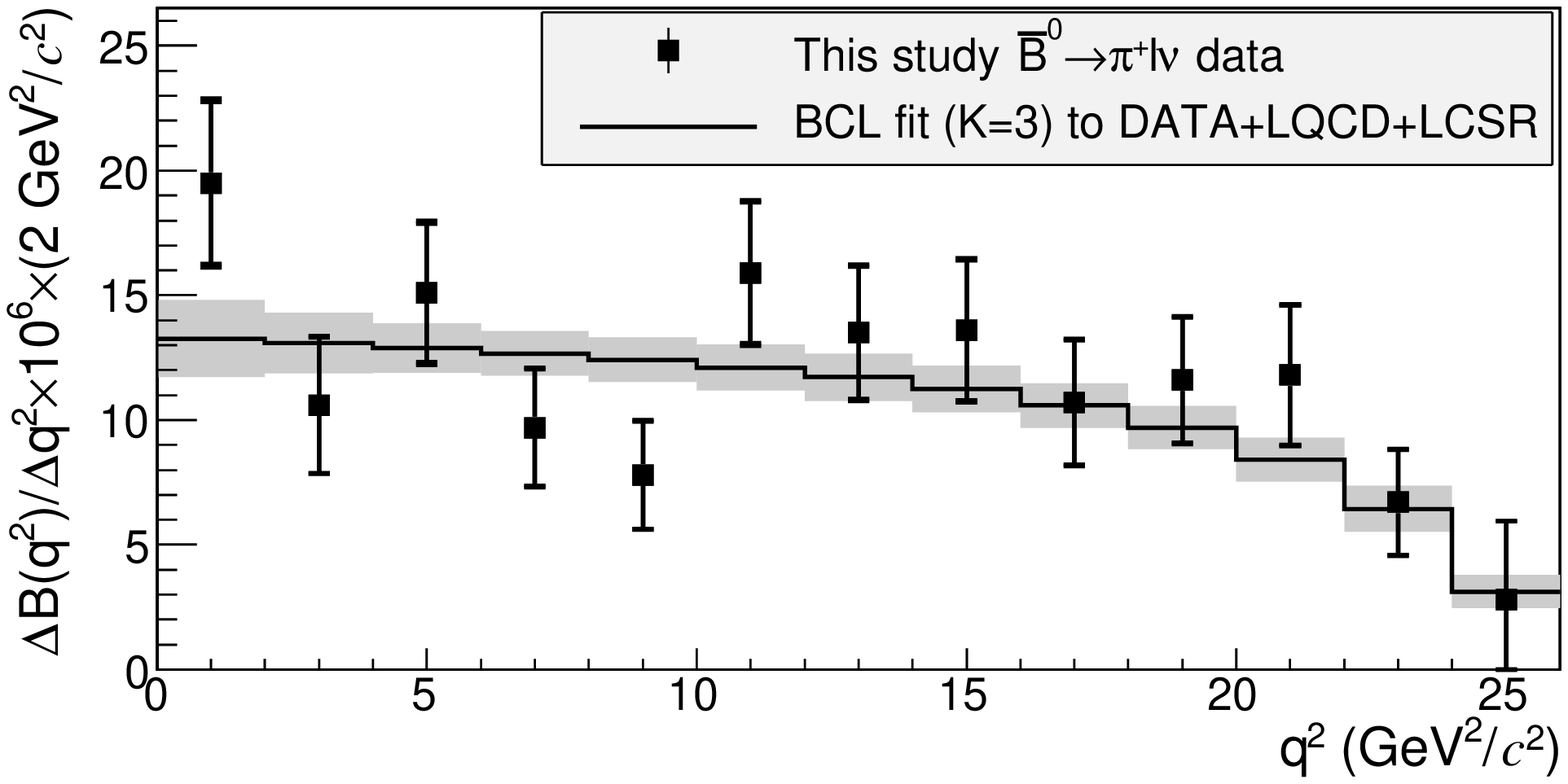}
  \includegraphics[width=0.595\textwidth]{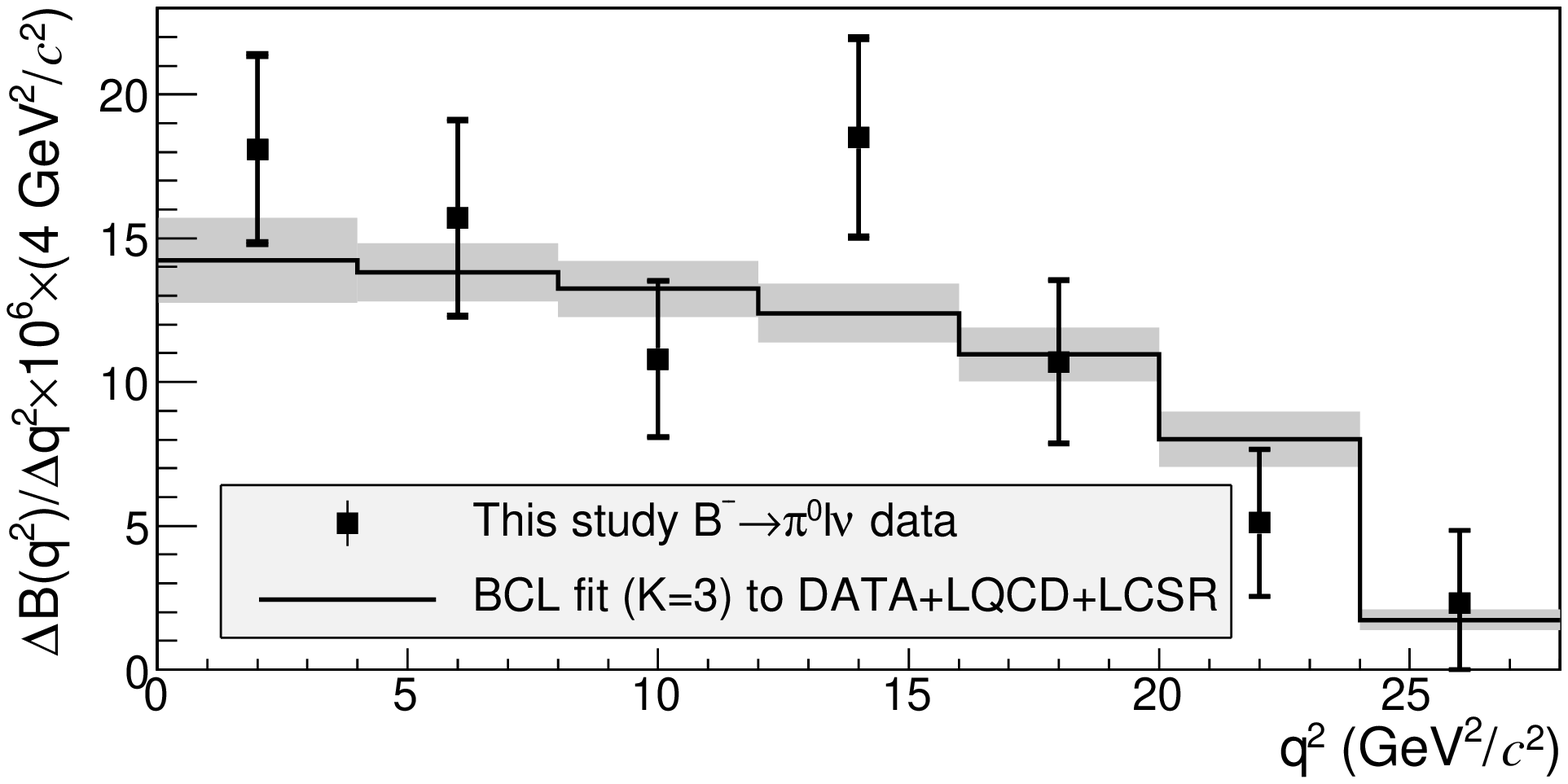}
  \caption{\label{fig:vubq2fit}Fit to data obtained in this analysis,
  LQCD points and the LCSR prediction at $q^2=0$ using the BCL
  parametrization, with the number of free parameters $N=4$. The
  shaded regions represent the uncertainties of the fit.}
\end{figure}

A typical fit is shown in Fig.~\ref{fig:vubq2fit}, using \bpiplnu
and \bpizlnu data, LQCD points and the LCSR form factor prediction
at $q^2=0$ with $N=4$.

The fit results obtained using different numbers of parameters in the
BCL parametrization, along with all possible combinations of \bpiplnu
and \bpizlnu differential branching fractions obtained in this
analysis, as well as the theoretical predictions, are shown in
Table~\ref{table:vubq2fit}. It can be seen that the values of
$|V_{ub}|$ are in agreement for the different data sets,
indicating that the input data are self consistent at the current level
of precision. For $N>4$, the form factor parametrization starts to
oscillate, reflecting the statistical fluctuations in the data, and
does not satisfy the unitarity condition on the parameters $\sum
B_{jk}b_j b_k \leq 1.$
\begin{table*}
  \caption{\label{table:vubq2fit}The fit results for different numbers
  of parameters $N$ and various sets of data with full reconstruction
  tagging and theoretical predictions. The ``+'' sign indicates that the
  corresponding data set was included in the fit and the ``$-$'' sign
  otherwise.}
    \vspace*{2mm}
  \begin{tabular}{ccccccccccc}
    \hline
    N & LCSR & LQCD & $\pi^+$ & $\pi^0$ & $b_0$ & $b_1$ & $b_2$ & $|V_{ub}|\times10^3$ & $B_{jk}b_jb_k$ & $\chi^2/\mbox{ndf}$ \\
    \hline
3&$+$&$+$&$+$&$+$&$\hphantom{-}0.450\pm0.030$&$-0.588\pm0.081$&&$3.40\pm0.23$&$0.0146$&22.82/22\\ 
3&$-$&$+$&$+$&$+$&$\hphantom{-}0.455\pm0.039$&$-0.586\pm0.083$&&$3.35\pm0.35$&$0.0146$&22.79/21\\ 
3&$+$&$-$&$+$&$+$&$\hphantom{-}0.444\pm0.054$&$-0.575\pm0.131$&&$3.44\pm0.35$&$0.0141$&22.53/18\\ 
3&$+$&$+$&$-$&$+$&$\hphantom{-}0.439\pm0.032$&$-0.530\pm0.098$&&$3.50\pm0.27$&$0.0125$&\hphantom{1}6.36/9\hphantom{1}\\ 
3&$+$&$+$&$+$&$-$&$\hphantom{-}0.459\pm0.031$&$-0.635\pm0.090$&&$3.39\pm0.24$&$0.0165$&16.11/15\\ 
3&$-$&$+$&$-$&$+$&$\hphantom{-}0.462\pm0.040$&$-0.467\pm0.126$&&$3.17\pm0.41$&$0.0110$&\hphantom{1}5.40/8\hphantom{1}\\ 
3&$-$&$+$&$+$&$-$&$\hphantom{-}0.451\pm0.039$&$-0.645\pm0.092$&&$3.50\pm0.39$&$0.0167$&15.97/14\\ 
3&$+$&$-$&$-$&$+$&$\hphantom{-}0.379\pm0.063$&$-0.369\pm0.173$&&$3.83\pm0.45$&$0.0071$&\hphantom{1}5.07/5\hphantom{1}\\ 
3&$+$&$-$&$+$&$-$&$\hphantom{-}0.485\pm0.067$&$-0.702\pm0.176$&&$3.27\pm0.36$&$0.0197$&15.60/11\\ 
4&$+$&$+$&$+$&$+$&$\hphantom{-}0.438\pm0.033$&$-0.701\pm0.162$&$\hphantom{-}0.171\pm0.577$&$3.52\pm0.29$&$0.0152$&22.20/21\\ 
4&$-$&$+$&$+$&$+$&$\hphantom{-}0.443\pm0.041$&$-0.700\pm0.164$&$\hphantom{-}0.177\pm0.583$&$3.47\pm0.40$&$0.0153$&22.16/20\\ 
4&$+$&$-$&$+$&$+$&$\hphantom{-}0.427\pm0.051$&$-0.802\pm0.225$&$\hphantom{-}0.566\pm0.685$&$3.65\pm0.40$&$0.0299$&21.18/17\\ 
4&$+$&$+$&$-$&$+$&$\hphantom{-}0.437\pm0.038$&$-0.545\pm0.197$&$-0.198\pm0.745$&$3.52\pm0.33$&$0.0115$&\hphantom{1}6.35/8\hphantom{1}\\ 
4&$+$&$+$&$+$&$-$&$\hphantom{-}0.446\pm0.035$&$-0.756\pm0.177$&$\hphantom{-}0.190\pm0.643$&$3.53\pm0.31$&$0.0177$&15.51/14\\ 
4&$-$&$+$&$-$&$+$&$\hphantom{-}0.457\pm0.042$&$-0.517\pm0.207$&$\hphantom{-}0.005\pm0.777$&$3.19\pm0.43$&$0.0090$&\hphantom{1}5.31/7\hphantom{1}\\ 
4&$-$&$+$&$+$&$-$&$\hphantom{-}0.438\pm0.042$&$-0.760\pm0.174$&$\hphantom{-}0.171\pm0.640$&$3.63\pm0.45$&$0.0176$&15.40/13\\
4&$+$&$-$&$-$&$+$&$\hphantom{-}0.378\pm0.064$&$-0.380\pm0.414$&$-0.145\pm1.286$&$3.84\pm0.55$&$0.0066$&\hphantom{1}5.07/4\hphantom{1}\\ 
4&$+$&$-$&$+$&$-$&$\hphantom{-}0.454\pm0.061$&$-1.008\pm0.263$&$\hphantom{-}0.875\pm0.798$&$3.58\pm0.43$&$0.0580$&13.69/10\\ 

    \hline
  \end{tabular}
\end{table*}

\begin{figure*}
  \includegraphics[width=0.495\textwidth]{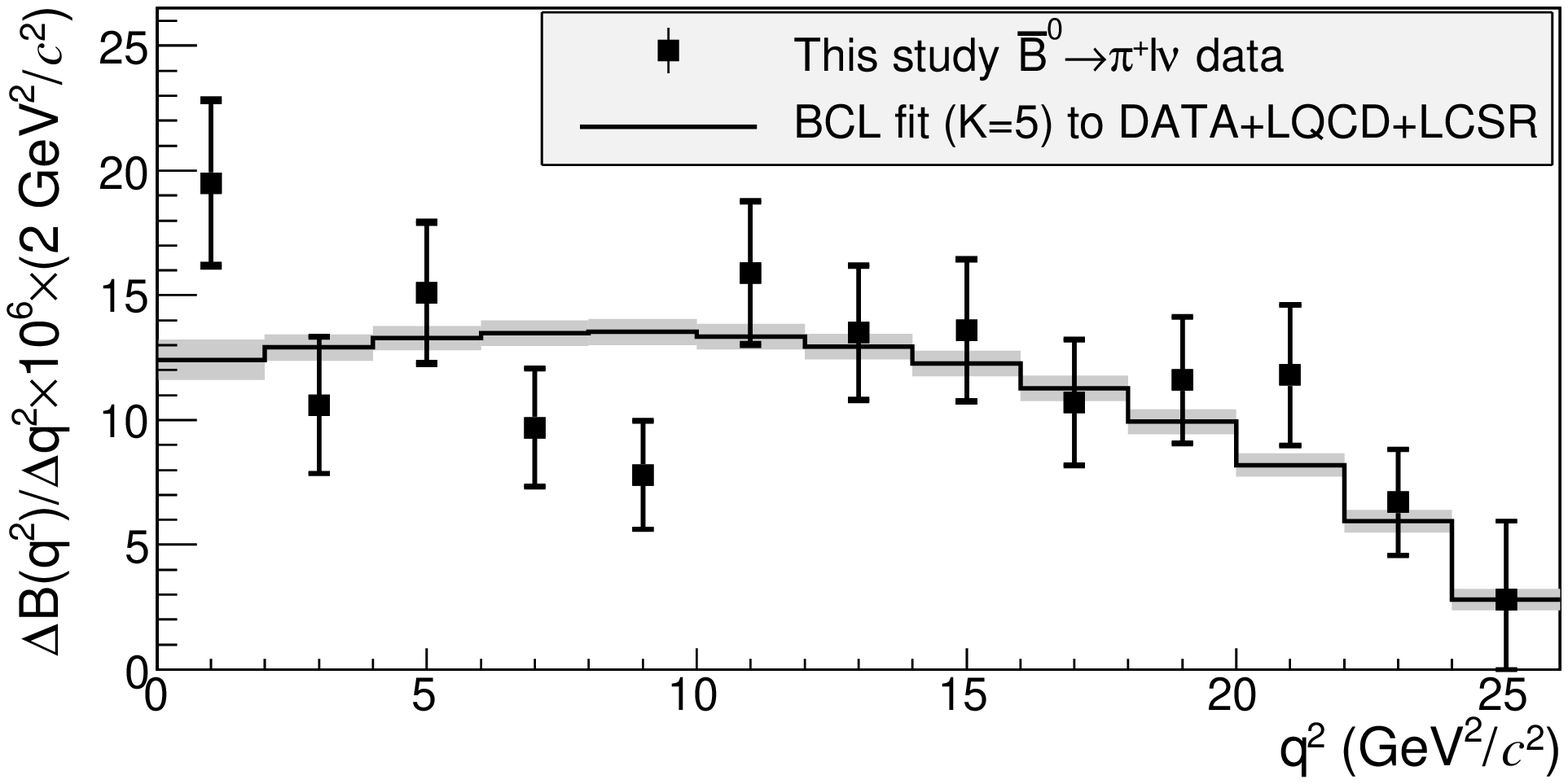}
  \includegraphics[width=0.495\textwidth]{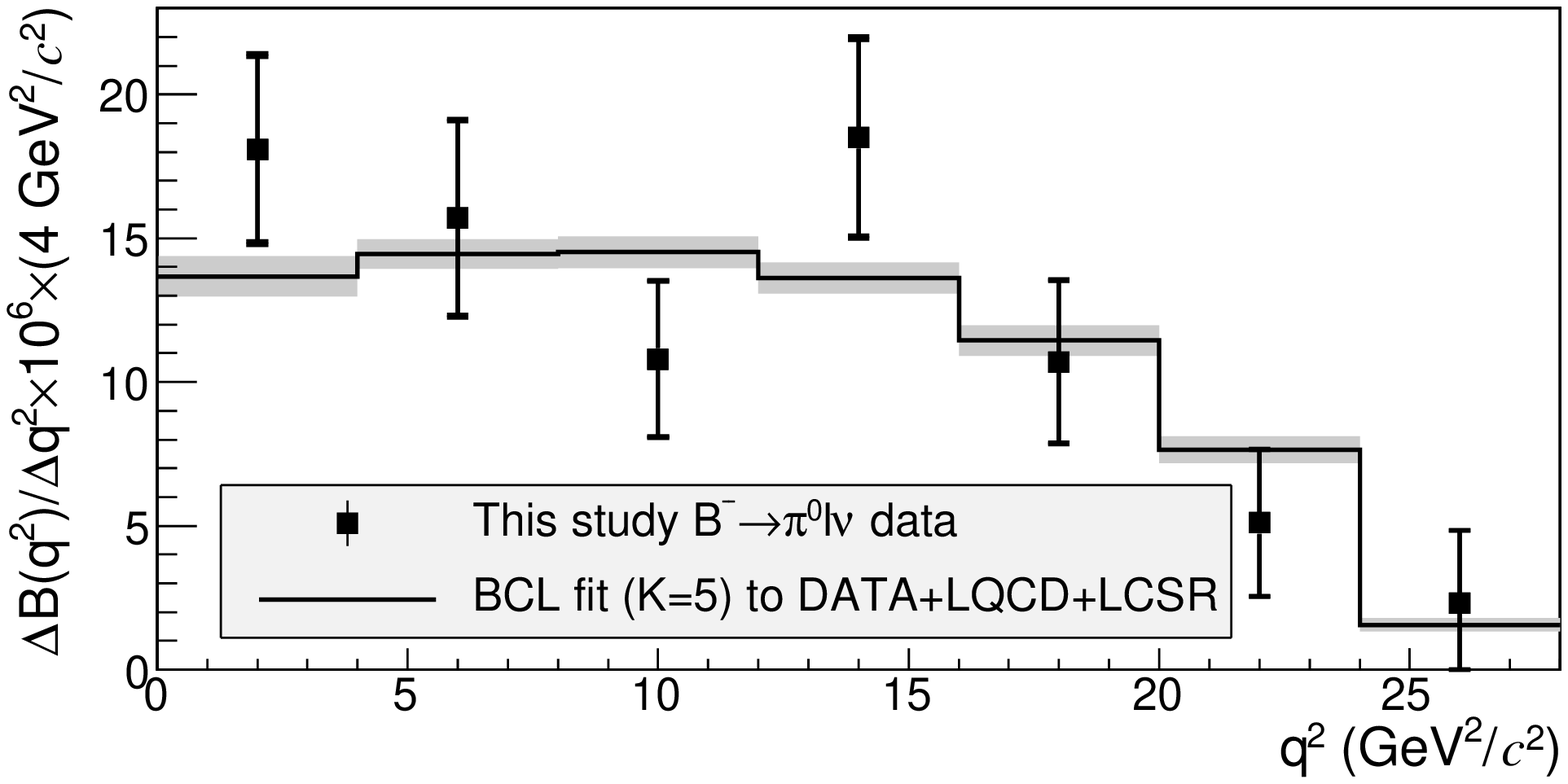}
  \includegraphics[width=0.495\textwidth]{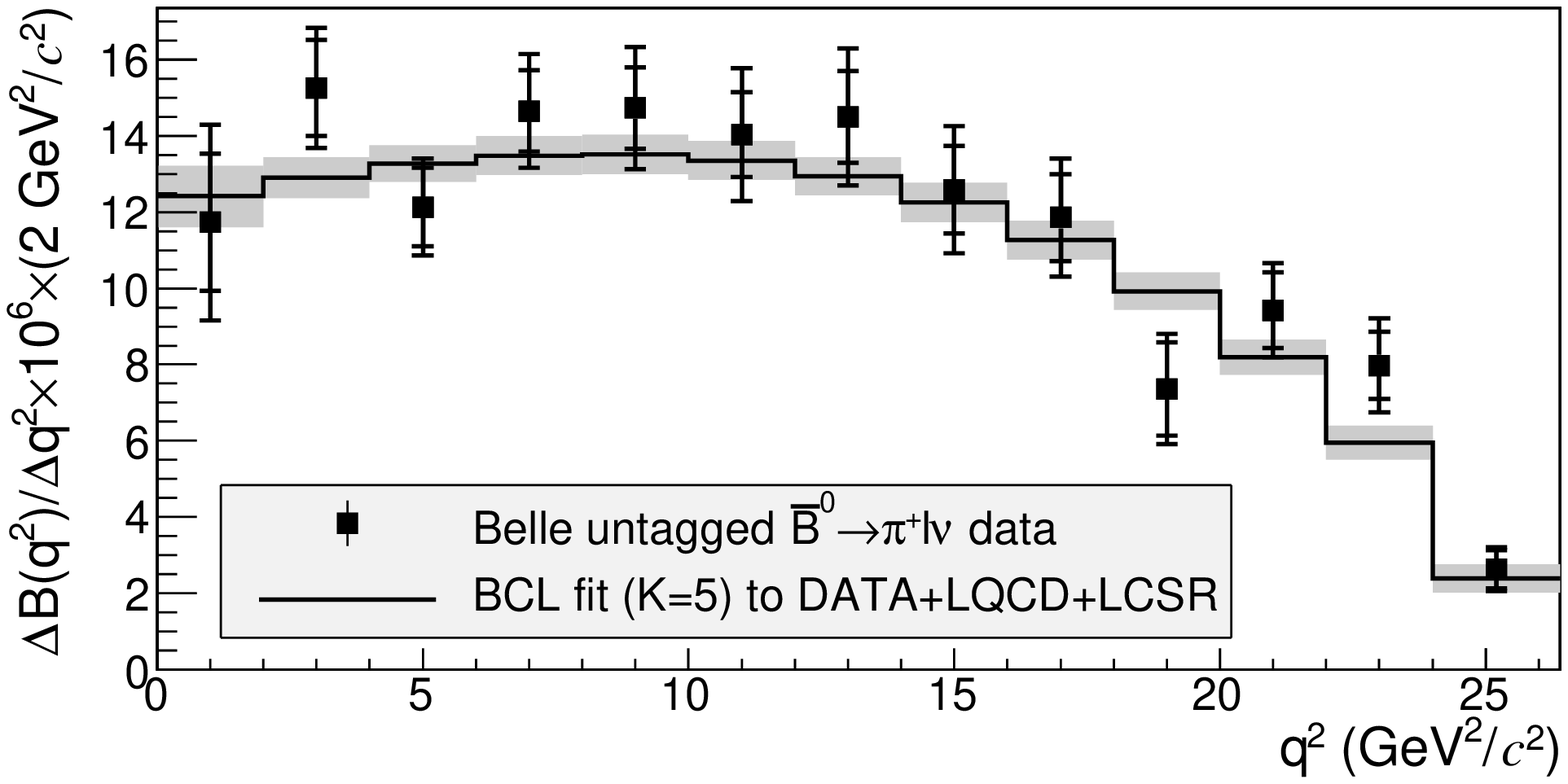}
  \includegraphics[width=0.495\textwidth]{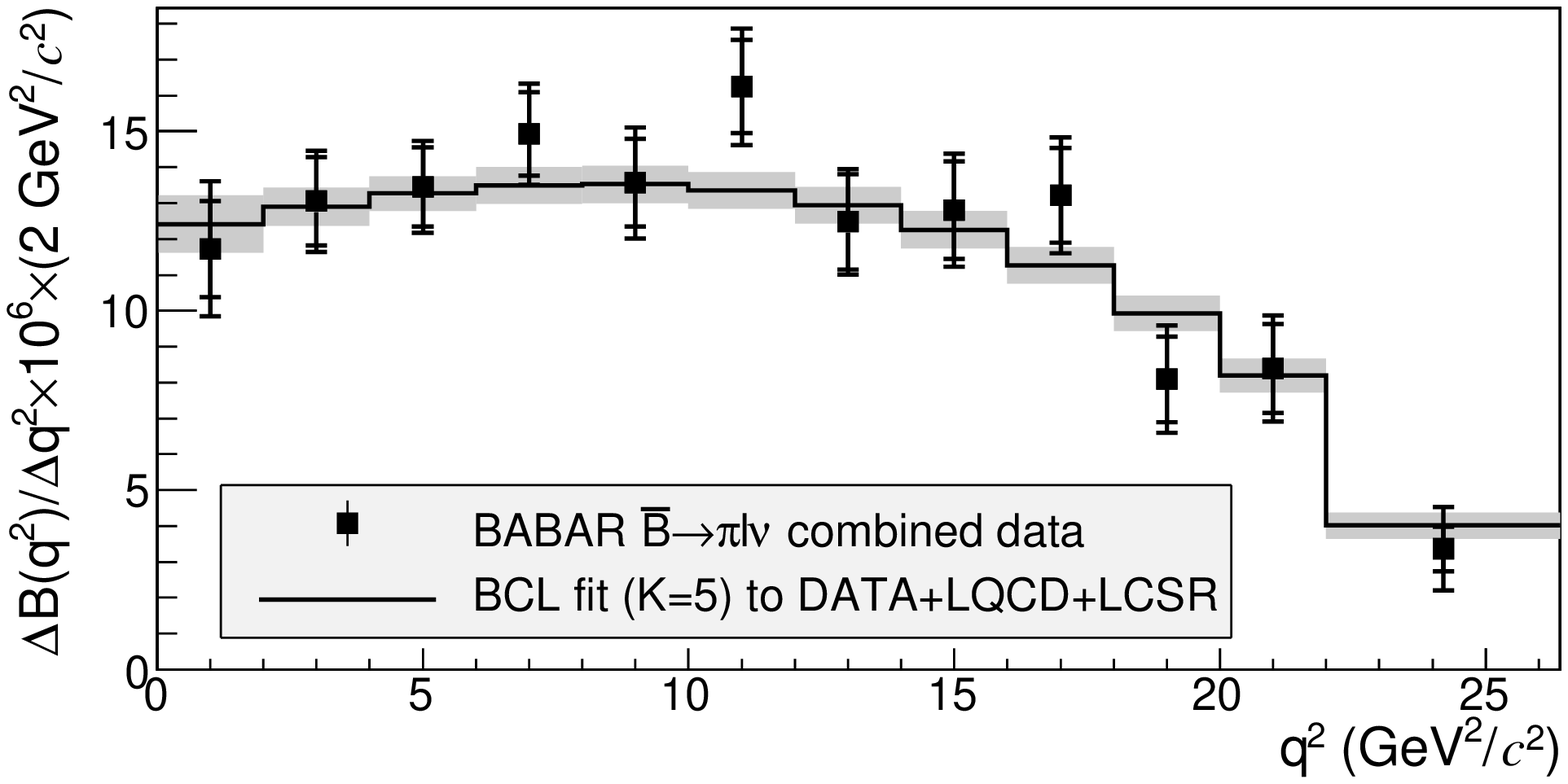}
  \includegraphics[width=0.495\textwidth]{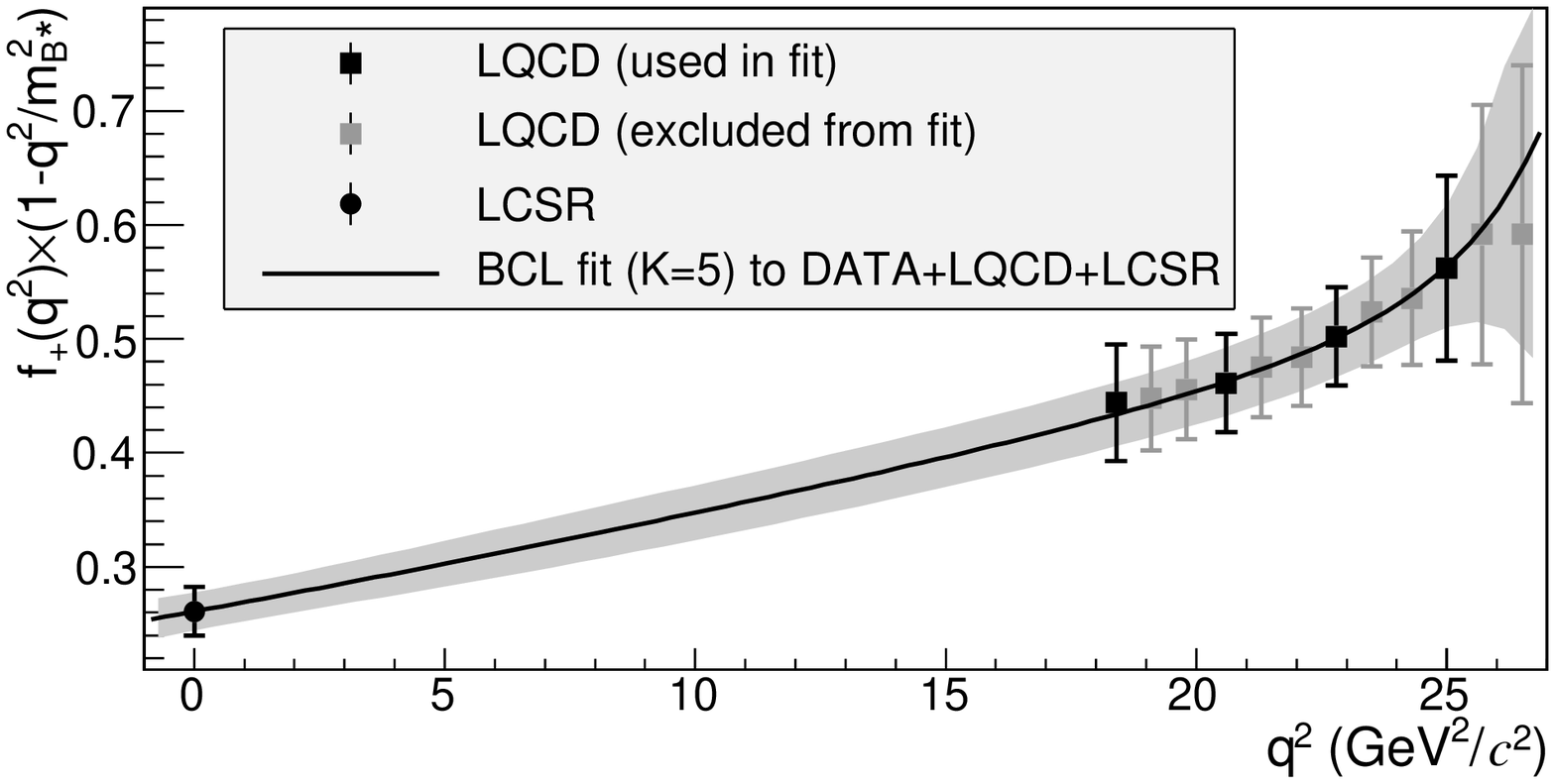}
  \caption{\label{fig:vubq2fitall}The combined fit to the most recent data
  and theory inputs using the BCL parametrization with the number of free
  parameters $N=6$ using untagged Belle~\cite{BelleUntaggedPi} and
  \babar~\cite{BaBarUntaggedPiOmegaEta} and tagged (this study) data. Shaded
  regions represent the uncertainties of the fit.}
\end{figure*}

To estimate the effect of truncating the series in $z$, we use the most recent
untagged Belle~\cite{BelleUntaggedPi} and \babar~\cite{BaBarUntaggedPiOmegaEta} data,
which have better statistical and completely different systematic
uncertainties. An example of the combined fit is shown in
Fig.~\ref{fig:vubq2fitall}, with $N=6$, the largest number of
parameters where the unitarity condition is not saturated. For larger
numbers of parameters, we artificially force the unitarity condition by
adding a component to the $\chi^2$ function that is zero if the
unitarity condition is not saturated and steeply rising to infinity when
approaching the unitarity bound:
\begin{equation}
\chi^2_\mathrm{unitarity} = \dfrac{1}{1-u^{150}} - 1,\ u = \sum\limits_{j,k=0}^{K} B_{jk}b_j b_k. 
\end{equation}
The values of $|V_{ub}|$ extracted using different numbers of terms in
the form factor expansion are shown in Table~\ref{table:vubvsN}. To be
sure that the $\chi^2$ minimum is a true minimum, we repeated the fit many times,
starting with a random initial vector of parameters $\vec{b}$ that satisfies the
unitarity condition. It can be seen from Table~\ref{table:vubvsN} that the value of $|V_{ub}|$
is stable starting from $N=4$; additional parameters only slightly
improve the overall $\chi^2$.  From this, we can conclude that the number of
terms in the expansion, $K = N - 1 =3$, is enough to describe the current data
with a negligibly small $\sim0.5$\% systematic uncertainty due to
unaccounted terms in the expansion.
\begin{table}
\caption{\label{table:vubvsN}Fit results with different number of free
parameters $N$ and forced unitarity bound on the coefficients of the
expansion using untagged Belle~\cite{BelleUntaggedPi},
\babar~\cite{BaBarUntaggedPiOmegaEta} and tagged (this study) data and
the LCSR and LQCD calculations. Note that in this approach the
$|V_{ub}|$ error cannot be reliably estimated for $N>6$ near to the unitarity
bound.}  \vspace*{2mm}
\begin{tabular}{cccc}
  \hline
  N & $|V_{ub}|\times10^3$ & $\sum B_{jk}b_jb_k$ & $\chi^2/\mbox{ndf}$ \\
  \hline
  3&$3.47\pm0.21$&$0.0148$&46.39/47\\ 
  4&$3.41\pm0.22$&$0.0232$&45.37/46\\ 
  5&$3.39\pm0.22$&$0.1073$&44.76/45\\ 
  6&$3.39\pm0.22$&$0.2289$&44.74/44\\ 
  7&$3.39\pm0.20$&$0.9501$&44.65/43\\ 
  8&$3.39\pm0.08$&$0.9503$&44.65/42\\ 
  9&$3.39\pm0.09$&$0.9525$&44.62/41\\ 
  10&$3.39\pm0.09$&$0.9525$&44.62/40\\ 
  11&$3.39\pm0.09$&$0.9527$&44.59/39\\ 
  12&$3.39\pm0.11$&$0.9531$&44.59/38\\ 
  13&$3.39\pm0.09$&$0.9538$&44.58/37\\ 
  14&$3.39\pm0.10$&$0.9539$&44.58/36\\ 
  15&$3.39\pm0.09$&$0.9545$&44.56/35\\ 
  \hline
\end{tabular}
\end{table}

As a result of the model-independent fit of both the \bpiplnu and
\bpizlnu differential branching fractions measured in this analysis,
the LQCD form factor points and LCSR prediction with $N=4$, we quote
$|V_{ub}| = (3.52\pm0.29)\times10^{-3}$. It is difficult to unequivocally
separate the experimental and theoretical uncertainties so we quote
only a total error. Using only the LCSR prediction or LQCD points in
the fit as shown in Table~\ref{table:vubq2fit}, we can conclude that
the LCSR prediction and LQCD points have almost equal contributions to
the total uncertainty.

\begin{figure}[h]
  \includegraphics[width=0.695\textwidth]{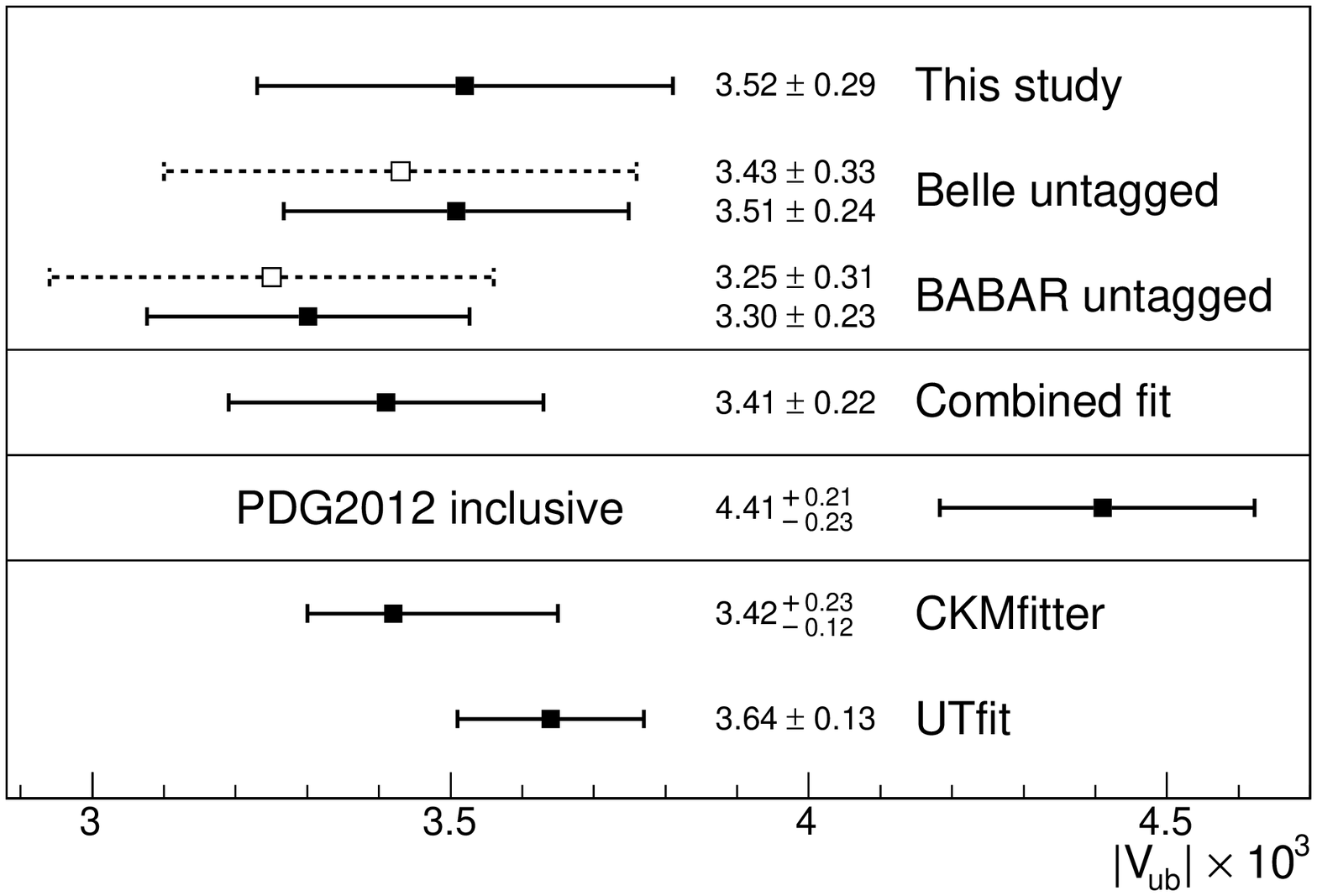}
  \caption{\label{fig:vub_comp}The comparison of \vub values obtained
  with the BCL form factor parameterization with the number of terms
  in expansion $K=3$, using LCSR and LQCD inputs for tagged (this
  analysis), untagged Belle~\cite{BelleUntaggedPi} and
  \babar~\cite{BaBarUntaggedPiOmegaEta} data as well as a combined fit
  with those three data sets. The value of \vub based on inclusive
  semileptonic decays of $B$ mesons is taken from the recent
  PDG~\cite{PDG} review.  The CKMfitter~\cite{ckmfitter} and
  UTfit~\cite{utfit} estimates of \vub are from global fits where
  \vub related inputs are excluded.  The dashed lines represent \vub
  values quoted in the original papers.}
\end{figure}

In Fig.~\ref{fig:vub_comp}, we show the value of \vub obtained in this
analysis, compared to values obtained from other recent measurements
and global determinations.  The extracted value of \vub has comparable
precision to, and agrees with, the values obtained from untagged
Belle~\cite{BelleUntaggedPi} and \babar~\cite{BaBarUntaggedPiOmegaEta}
data using the same method of determination as in this analysis.  The
figure shows both the values quoted in the Belle and \babar papers and
the values obtained by refitting using the original data and the method
used in the present analysis. The combined fit shown uses data from
all three analyses. Our value is also in agreement with the results of
global fits performed by the CKMfitter~\cite{ckmfitter} and
UTfit~\cite{utfit} groups, where they excluded \vub related inputs from
the fits.  The tension between the value of \vub extracted from \bpilnu
decays and that measured in inclusive semileptonic decays of $B$
mesons, represented in the figure by the latest PDG~\cite{PDG} value, remains
significant ($\sim3\sigma$).

We note that all theoretical predictions used in $|V_{ub}|$ extraction
procedures described above assume no photon emission in the final
state. For future experiments with much larger data samples, it will
become crucial for theory to take into account radiative effects to
describe high precision experimental data.

\section{Conclusion}
We measure the total branching fractions for \bpizlnu, \bpiplnu,
\brhozlnu, \brhoplnu and \bomegalnu decays using fully
reconstructed hadronic $B$ decays as a tag. This technique 
provides exceptionally clean signal samples and thus low systematic
uncertainty in the final results. The \bpizlnu and \bpiplnu branching
fractions are in a good agreement with the previous Belle
result~\cite{BelleUntaggedPi} using an untagged reconstruction technique
and with the recent \babar
measurement~\cite{BaBarUntaggedPiOmegaEta}, and also with the isospin symmetry
relation. The \brhozlnu and \brhoplnu branching fractions are also in good
agreement with the isospin symmetry relation. The \brholnu
branching fraction is 43\% (2.7$\sigma$) higher than the PDG value and
its precision is almost a factor of two better. This raises the probability that,
in the previous analyses, backgrounds to \brholnu decays may have been
considerably overestimated.

For the first time, we have an indication of neutral charmless hadronic
states above 1~\GeVcc in invariant mass in semileptonic decays of $B$
mesons.  The broad peak observed in the $\pi^+\pi^-$ invariant mass
distribution around 1.3~\GeVcc is dominated in our fit by the
\bftwolnu decay where $f_2\to \pi^+\pi^-$.  The fitted yield is 2-3
times higher than expected from the ISGW2 model and, assuming the
absence of the non-resonant \bpipilnu decay, it has high
statistical significance.  A dedicated study is needed to fully
explore the region above 1~\GeVcc.

From the studied decays, we extract values of $|V_{ub}|$ in various
$q^2$ regions where the theoretical predictions of the hadronic form factors
are valid. The resulting values of $|V_{ub}|$ are in a good agreement with
each other. We also measure the $q^2$ dependence of the partial
branching fractions, which can be used to test the theoretical
predictions for the hadronic form factors.

For the \bpilnu decay, we extract a value of $|V_{ub}|=
(3.52\pm0.29)\times10^{-3}$ using our measured partial branching
fractions, a recent LCSR calculation, LQCD points and a
model-independent description of the $f_{+}(q^2)$ hadronic form
factor. We also present values of $|V_{ub}|$ obtained from fits where
we do not assume that the theoretical inputs from LCSR and LQCD can be
included in the same fit. Within the BCL parametrization, we have
shown that 3 terms in the $z$ expansion are enough to extract a value
of $|V_{ub}|$ with negligibly small systematic uncertainty due to
truncating the expansion.

\begin{acknowledgments}
We thank the KEKB group for the excellent operation of the
accelerator; the KEK cryogenics group for the efficient
operation of the solenoid; and the KEK computer group,
the National Institute of Informatics, and the 
PNNL/EMSL computing group for valuable computing
and SINET4 network support.  We acknowledge support from
the Ministry of Education, Culture, Sports, Science, and
Technology (MEXT) of Japan, the Japan Society for the 
Promotion of Science (JSPS), and the Tau-Lepton Physics 
Research Center of Nagoya University; 
the Australian Research Council and the Australian 
Department of Industry, Innovation, Science and Research;
Austrian Science Fund under Grant No. P 22742-N16;
the National Natural Science Foundation of China under
contract No.~10575109, 10775142, 10875115 and 10825524; 
the Ministry of Education, Youth and Sports of the Czech 
Republic under contract No.~MSM0021620859;
the Carl Zeiss Foundation, the Deutsche Forschungsgemeinschaft
and the VolkswagenStiftung;
the Department of Science and Technology of India; 
the Istituto Nazionale di Fisica Nucleare of Italy; 
The BK21 and WCU program of the Ministry Education Science and
Technology, National Research Foundation of Korea Grant No.\ 
2010-0021174, 2011-0029457, 2012-0008143, 2012R1A1A2008330,
BRL program under NRF Grant No. KRF-2011-0020333,
and GSDC of the Korea Institute of Science and Technology Information;
the Polish Ministry of Science and Higher Education and 
the National Science Center;
the Ministry of Education and Science of the Russian
Federation and the Russian Federal Agency for Atomic Energy;
the Slovenian Research Agency;
the Basque Foundation for Science (IKERBASQUE) and the UPV/EHU under 
program UFI 11/55;
the Swiss National Science Foundation; the National Science Council
and the Ministry of Education of Taiwan; and the U.S.\
Department of Energy and the National Science Foundation.
This work is supported by a Grant-in-Aid from MEXT for 
Science Research in a Priority Area (``New Development of 
Flavor Physics''), and from JSPS for Creative Scientific 
Research (``Evolution of Tau-lepton Physics'').
\end{acknowledgments}

\section*{APPENDIX}
In this Appendix, we present the fitted event yields, unfolded yields,
efficiencies and corresponding partial branching fractions for the
decays investigated in this study, in bins of $q^2$, in
Tables~\ref{table:bpiclnuvsq2}, \ref{table:bpi0lnuvsq2}, 
\ref{table:brhoclnuvsq2}, \ref{table:brho0lnuvsq2}
and \ref{table:bomega3plnuvsq2}. We also give the statistical
correlations between $q^2$ bins in Tables~\ref{table:bpiclnucrr},
\ref{table:bpi0lnucrr}, \ref{table:brhoclnucrr},
\ref{table:brho0lnucrr} and \ref{table:bomega3plnucrr}.

\begin{table}
  \caption{\label{table:bpiclnuvsq2}Raw \bpiplnu yields obtained from
    the two-dimensional fit, unfolded yields, efficiencies and partial
    branching fractions in bins of $q^2$.} \vspace*{2mm}
  \begin{tabular}{
      @{\hspace{0.3cm}}r@{$-$}l
      r@{$\pm$}l
      r@{$\pm$}l
      r@{$\pm$}l
      r@{$\pm$}l@{\hspace{0.1cm}}
    }
    \hline
    \multicolumn{2}{c}{$\Delta q^2$} &
    \multicolumn{2}{c}{$N_\mathrm{fit}$} &
    \multicolumn{2}{c}{$N_\mathrm{fit}^\mathrm{unfolded}$} &
    \multicolumn{2}{c}{$\varepsilon$} &
    \multicolumn{2}{c}{$\Delta\Br$} \\
    \multicolumn{2}{c}{\GeVGeVcc} & \multicolumn{2}{c}{} &
    \multicolumn{2}{c}{} & \multicolumn{2}{c}{$10^{-3}$} &
    \multicolumn{2}{c}{$10^{-6}$} \\
    \hline
    
 0& 2& 53.9&8.6 &  55.5&9.2 & 1.90&0.07 &  19.5&3.2 \\
 2& 4& 35.4&7.5 &  33.0&8.4 & 2.07&0.07 &  10.6&2.7 \\
 4& 6& 42.5&7.3 &  44.5&8.2 & 1.96&0.06 &  15.1&2.8 \\
 6& 8& 30.5&6.6 &  29.8&7.2 & 2.05&0.06 &   9.7&2.3 \\
 8&10& 27.2&6.4 &  25.1&6.9 & 2.14&0.06 &   7.8&2.2 \\
10&12& 48.9&8.2 &  50.7&9.0 & 2.13&0.06 &  15.9&2.8 \\
12&14& 43.0&7.8 &  43.0&8.5 & 2.13&0.06 &  13.5&2.7 \\
14&16& 40.7&7.9 &  41.2&8.5 & 2.02&0.06 &  13.6&2.8 \\
16&18& 34.0&7.5 &  34.6&8.0 & 2.16&0.07 &  10.7&2.5 \\
18&20& 39.7&8.2 &  40.1&8.7 & 2.31&0.09 &  11.6&2.5 \\
20&22& 35.6&8.0 &  36.4&8.6 & 2.06&0.12 &  11.8&2.8 \\
22&24& 21.6&6.3 &  21.5&6.8 & 2.14&0.21 &   6.7&2.1 \\
24&26&  8.0&6.3 &   5.6&6.3 & 1.35&0.39 &   2.8&3.1 \\
\hline
\multicolumn{2}{c}{Full range} & 461.1&27.4 & 461.1&27.4 & 2.07&0.02 & 149.4&9.1 \\

    \hline
  \end{tabular}
\end{table}

\begin{table*}
  \caption{\label{table:bpiclnucrr}Normalised statistical correlation
  matrix in percent for the \bpiplnu partial branching fractions.} \vspace*{2mm}
  \begin{tabular}{c*{13}{D{.}{.}{4.1}}}
  \hline
    
$\Delta q^2$, \GeVGeVcc & \multicolumn{1}{r}{0-2}& \multicolumn{1}{r}{2-4}& \multicolumn{1}{r}{4-6}& \multicolumn{1}{r}{6-8}& \multicolumn{1}{r}{8-10}& \multicolumn{1}{r}{10-12}& \multicolumn{1}{r}{12-14}& \multicolumn{1}{r}{14-16}& \multicolumn{1}{r}{16-18}& \multicolumn{1}{r}{18-20}& \multicolumn{1}{r}{20-22}& \multicolumn{1}{r}{22-24}& \multicolumn{1}{r}{24-26}\\ \hline
0-2 & 100.0 & -14.5 &  1.0 & -0.1 &  0.0 &  0.0 &  0.0 &  0.0 &  0.0 & -0.0 & -0.0 &  0.1 &  0.1 \\
2-4 & -14.5 & 100.0 & -10.0 &  0.8 & -0.0 &  0.1 & -0.2 & -0.1 & -0.1 &  0.1 & -0.1 & -0.0 &  0.0 \\
4-6 &  1.0 & -10.0 & 100.0 & -9.4 &  0.3 & -0.1 &  0.0 &  0.1 &  0.1 & -0.2 & -0.2 &  0.0 &  0.4 \\
6-8 & -0.1 &  0.8 & -9.4 & 100.0 & -7.8 &  0.5 &  0.0 &  0.0 &  0.1 & -0.1 &  0.2 &  0.2 &  0.5 \\
8-10 &  0.0 & -0.0 &  0.3 & -7.8 & 100.0 & -9.7 &  0.3 &  0.1 &  0.1 &  0.0 &  0.4 &  0.3 &  0.9 \\
10-12 &  0.0 &  0.1 & -0.1 &  0.5 & -9.7 & 100.0 & -7.2 &  0.5 &  0.1 &  0.3 &  0.5 &  0.4 &  1.3 \\
12-14 &  0.0 & -0.2 &  0.0 &  0.0 &  0.3 & -7.2 & 100.0 & -6.2 &  0.4 &  0.2 &  0.1 &  0.2 &  1.0 \\
14-16 &  0.0 & -0.1 &  0.1 &  0.0 &  0.1 &  0.5 & -6.2 & 100.0 & -5.9 &  0.2 &  0.3 &  0.2 &  1.1 \\
16-18 &  0.0 & -0.1 &  0.1 &  0.1 &  0.1 &  0.1 &  0.4 & -5.9 & 100.0 & -4.7 &  0.1 & -0.0 &  0.9 \\
18-20 & -0.0 &  0.1 & -0.2 & -0.1 &  0.0 &  0.3 &  0.2 &  0.2 & -4.7 & 100.0 & -1.8 &  0.2 &  0.5 \\
20-22 & -0.0 & -0.1 & -0.2 &  0.2 &  0.4 &  0.5 &  0.1 &  0.3 &  0.1 & -1.8 & 100.0 & -5.5 & -2.0 \\
22-24 &  0.1 & -0.0 &  0.0 &  0.2 &  0.3 &  0.4 &  0.2 &  0.2 & -0.0 &  0.2 & -5.5 & 100.0 & -1.2 \\
24-26 &  0.1 &  0.0 &  0.4 &  0.5 &  0.9 &  1.3 &  1.0 &  1.1 &  0.9 &  0.5 & -2.0 & -1.2 & 100.0 \\

    \hline
  \end{tabular}
\end{table*}

\begin{table}
  \caption{\label{table:bpi0lnuvsq2}Raw \bpizlnu yields obtained from
    the two-dimensional fit, unfolded yields, efficiencies and partial
    branching fractions in bins of $q^2$.} \vspace*{2mm}
  \begin{tabular}{
      @{\hspace{0.3cm}}r@{$-$}l
      r@{$\pm$}l
      r@{$\pm$}l
      r@{$\pm$}l
      r@{$\pm$}l@{\hspace{0.1cm}}
    }
    \hline
    \multicolumn{2}{c}{$\Delta q^2$} &
    \multicolumn{2}{c}{$N_\mathrm{fit}$} &
    \multicolumn{2}{c}{$N_\mathrm{fit}^\mathrm{unfolded}$} &
    \multicolumn{2}{c}{$\varepsilon$} & \multicolumn{2}{c}{$\Delta\Br$} \\
    \multicolumn{2}{c}{\GeVGeVcc} & \multicolumn{2}{c}{} &
    \multicolumn{2}{c}{} & \multicolumn{2}{c}{$10^{-3}$} &
    \multicolumn{2}{c}{$10^{-6}$} \\
    \hline
    
 0& 4& 45.2&7.8 &  50.2&8.9 & 1.76&0.06 &  18.1&3.2 \\
 4& 8& 45.9&8.2 &  44.7&9.6 & 1.80&0.06 &  15.7&3.4 \\
 8&12& 35.1&7.2 &  33.8&8.4 & 1.99&0.06 &  10.8&2.7 \\
12&16& 51.7&8.7 &  54.2&10.0 & 1.86&0.06 &  18.5&3.4 \\
16&20& 33.1&7.3 &  32.3&8.5 & 1.93&0.07 &  10.7&2.8 \\
20&24& 16.3&6.0 &  13.5&6.8 & 1.69&0.14 &   5.1&2.6 \\
24&28&  6.1&4.4 &   4.5&5.0 & 1.24&0.51 &   2.3&2.5 \\
\hline
\multicolumn{2}{c}{Full range} & 233.3&20.6 & 233.3&20.6 & 1.83&0.03 &  81.2&7.4 \\

    \hline
  \end{tabular}
  
\end{table}

\begin{table}
  \caption{\label{table:bpi0lnucrr}Normalised statistical correlation
  matrix in percent for the \bpizlnu partial branching fractions.} \vspace*{2mm}
  \begin{tabular}{c*{7}{D{.}{.}{4.1}}}
    \hline
    
$\Delta q^2$, \GeVGeVcc & \multicolumn{1}{r}{0-4}& \multicolumn{1}{r}{4-8}& \multicolumn{1}{r}{8-12}& \multicolumn{1}{r}{12-16}& \multicolumn{1}{r}{16-20}& \multicolumn{1}{r}{20-24}& \multicolumn{1}{r}{24-28}\\ \hline
0-4 & 100.0 & -15.9 &  3.5 &  1.2 &  1.3 &  1.6 &  2.1 \\
4-8 & -15.9 & 100.0 & -14.9 &  3.0 &  1.3 &  1.5 &  2.5 \\
8-12 &  3.5 & -14.9 & 100.0 & -9.8 &  3.4 &  2.8 &  4.2 \\
12-16 &  1.2 &  3.0 & -9.8 & 100.0 & -11.2 &  2.8 &  3.5 \\
16-20 &  1.3 &  1.3 &  3.4 & -11.2 & 100.0 & -11.5 &  2.6 \\
20-24 &  1.6 &  1.5 &  2.8 &  2.8 & -11.5 & 100.0 & -13.1 \\
24-28 &  2.1 &  2.5 &  4.2 &  3.5 &  2.6 & -13.1 & 100.0 \\

    \hline
  \end{tabular}
\end{table}

\begin{table}
  \caption{\label{table:brhoclnuvsq2}Raw \brhoplnu yields obtained
    from the two-dimensional fit, unfolded yields, efficiencies and
    partial branching fractions in bins of $q^2$.  } \vspace*{2mm}
  \begin{tabular}{
      @{\hspace{0.3cm}}r@{$-$}l
      r@{$\pm$}l
      r@{$\pm$}l
      r@{$\pm$}l
      r@{$\pm$}c@{\hspace{0.1cm}
      }
    } \hline
    \multicolumn{2}{c}{$\Delta q^2$} &
    \multicolumn{2}{c}{$N_\mathrm{fit}$} &
    \multicolumn{2}{c}{$N_\mathrm{fit}^\mathrm{unfolded}$} &
    \multicolumn{2}{c}{$\varepsilon$} & \multicolumn{2}{c}{$\Delta\Br$} \\
    \multicolumn{2}{c}{\GeVGeVcc} & \multicolumn{2}{c}{} &
    \multicolumn{2}{c}{} & \multicolumn{2}{c}{$10^{-4}$} &
    \multicolumn{2}{c}{$10^{-6}$} \\
    \hline
    
 0& 4& 35.5&9.3 &  37.2&10.6 & 6.76&0.19 &  37.3&10.6 \\
 4& 8& 72.1&10.5 &  76.2&12.3 & 7.20&0.18 &  71.8&11.6 \\
 8&12& 88.1&11.9 &  90.0&13.7 & 7.58&0.17 &  80.6&12.3 \\
12&16& 80.2&11.9 &  78.0&13.5 & 7.32&0.18 &  72.3&12.5 \\
16&20& 69.4&11.2 &  67.6&12.4 & 7.33&0.22 &  62.6&11.5 \\
20&24&  4.9&4.7 &   1.1&5.2 & 4.43&0.62 &   1.7&7.9 \\
\hline
\multicolumn{2}{c}{Full range} & 350.2&27.3 & 350.2&27.3 & 7.22&0.09 & 326.2&26.3 \\

    \hline
  \end{tabular}
\end{table}

\begin{table}
  \caption{\label{table:brhoclnucrr}Normalised statistical correlation
  matrix in percent for the \brhoplnu partial branching fractions.} \vspace*{2mm}
  \begin{tabular}{c*{6}{D{.}{.}{4.1}}}
    \hline
    
$\Delta q^2$, \GeVGeVcc & \multicolumn{1}{r}{0-4}& \multicolumn{1}{r}{4-8}& \multicolumn{1}{r}{8-12}& \multicolumn{1}{r}{12-16}& \multicolumn{1}{r}{16-20}& \multicolumn{1}{r}{20-24}\\ \hline
0-4 & 100.0 & -14.1 &  2.4 &  0.9 &  1.1 &  1.6 \\
4-8 & -14.1 & 100.0 & -11.2 &  3.0 &  2.4 &  2.5 \\
8-12 &  2.4 & -11.2 & 100.0 & -7.5 &  5.0 &  5.8 \\
12-16 &  0.9 &  3.0 & -7.5 & 100.0 & -8.8 &  6.1 \\
16-20 &  1.1 &  2.4 &  5.0 & -8.8 & 100.0 & -10.2 \\
20-24 &  1.6 &  2.5 &  5.8 &  6.1 & -10.2 & 100.0 \\

    \hline
  \end{tabular}
\end{table}

\begin{table}
  \caption{\label{table:brho0lnuvsq2}Raw \brhozlnu yields obtained
    from the two-dimensional fit, unfolded yields, efficiencies and
    partial branching fractions in bins of $q^2$.} \vspace*{2mm}
  \begin{tabular}{
      @{\hspace{0.3cm}}r@{$-$}l
      r@{$\pm$}l
      r@{$\pm$}l
      r@{$\pm$}l
      r@{$\pm$}l@{\hspace{0.1cm}}
    }
    \hline
    \multicolumn{2}{c}{$\Delta q^2$} &
    \multicolumn{2}{c}{$N_\mathrm{fit}$} &
    \multicolumn{2}{c}{$N_\mathrm{fit}^\mathrm{unfolded}$} &
    \multicolumn{2}{c}{$\varepsilon$} & \multicolumn{2}{c}{$\Delta\Br$} \\
    \multicolumn{2}{c}{\GeVGeVcc} & \multicolumn{2}{c}{} &
    \multicolumn{2}{c}{} & \multicolumn{2}{c}{$10^{-3}$} &
    \multicolumn{2}{c}{$10^{-6}$} \\
    \hline
    
 0& 2& 35.7&9.2 &  36.6&10.0 & 1.85&0.06 &  12.4&3.4 \\
 2& 4& 52.6&9.4 &  52.1&10.4 & 2.14&0.06 &  15.3&3.1 \\
 4& 6& 55.5&9.3 &  55.4&10.5 & 1.95&0.06 &  17.9&3.4 \\
 6& 8& 70.2&10.2 &  72.4&11.3 & 2.10&0.06 &  21.7&3.4 \\
 8&10& 52.6&9.2 &  50.7&10.3 & 2.23&0.06 &  14.3&2.9 \\
10&12& 66.7&10.4 &  67.0&11.6 & 2.12&0.06 &  19.9&3.4 \\
12&14& 75.3&10.6 &  77.9&11.8 & 2.18&0.06 &  22.4&3.4 \\
14&16& 77.4&11.1 &  78.7&12.1 & 2.28&0.06 &  21.7&3.3 \\
16&18& 64.8&10.7 &  63.1&11.4 & 2.27&0.07 &  17.5&3.2 \\
18&20& 52.3&9.5 &  51.7&10.2 & 2.27&0.09 &  14.3&2.8 \\
20&22& 18.9&6.9 &  16.3&7.1 & 1.74&0.17 &   5.9&2.6 \\
\hline
\multicolumn{2}{c}{Full range} & 621.9&34.8 & 621.9&34.8 & 2.13&0.02 & 183.5&10.4 \\

    \hline
  \end{tabular}
\end{table}

\begin{table*}
  \caption{\label{table:brho0lnucrr}Normalised statistical correlation
  matrix in percent for the \brhozlnu partial branching fractions. } \vspace*{2mm}
  \begin{tabular}{c*{11}{D{.}{.}{4.1}}}
    \hline
    
$\Delta q^2$, \GeVGeVcc & \multicolumn{1}{r}{0-2}& \multicolumn{1}{r}{2-4}& \multicolumn{1}{r}{4-6}& \multicolumn{1}{r}{6-8}& \multicolumn{1}{r}{8-10}& \multicolumn{1}{r}{10-12}& \multicolumn{1}{r}{12-14}& \multicolumn{1}{r}{14-16}& \multicolumn{1}{r}{16-18}& \multicolumn{1}{r}{18-20}& \multicolumn{1}{r}{20-22}\\ \hline
0-2 & 100.0 & -12.3 &  0.8 &  0.0 &  0.2 &  0.2 &  0.2 &  0.3 &  0.3 &  0.3 &  0.3 \\
2-4 & -12.3 & 100.0 & -10.9 &  0.8 &  0.5 &  0.5 &  0.6 &  0.7 &  0.9 &  0.8 &  0.8 \\
4-6 &  0.8 & -10.9 & 100.0 & -9.7 &  0.9 &  0.9 &  0.8 &  0.9 &  1.3 &  1.1 &  1.1 \\
6-8 &  0.0 &  0.8 & -9.7 & 100.0 & -8.8 &  1.4 &  1.2 &  1.5 &  1.8 &  1.4 &  1.6 \\
8-10 &  0.2 &  0.5 &  0.9 & -8.8 & 100.0 & -9.3 &  1.5 &  1.6 &  2.1 &  1.7 &  1.8 \\
10-12 &  0.2 &  0.5 &  0.9 &  1.4 & -9.3 & 100.0 & -6.6 &  1.9 &  2.7 &  2.3 &  2.4 \\
12-14 &  0.2 &  0.6 &  0.8 &  1.2 &  1.5 & -6.6 & 100.0 & -4.8 &  1.8 &  1.8 &  2.4 \\
14-16 &  0.3 &  0.7 &  0.9 &  1.5 &  1.6 &  1.9 & -4.8 & 100.0 & -3.9 &  2.3 &  2.5 \\
16-18 &  0.3 &  0.9 &  1.3 &  1.8 &  2.1 &  2.7 &  1.8 & -3.9 & 100.0 & -3.8 &  2.7 \\
18-20 &  0.3 &  0.8 &  1.1 &  1.4 &  1.7 &  2.3 &  1.8 &  2.3 & -3.8 & 100.0 & -5.2 \\
20-22 &  0.3 &  0.8 &  1.1 &  1.6 &  1.8 &  2.4 &  2.4 &  2.5 &  2.7 & -5.2 & 100.0 \\

    \hline
  \end{tabular}
\end{table*}

\begin{table}
  \caption{\label{table:bomega3plnuvsq2}Raw \bomegappplnu obtained
    from the two-dimensional fit, unfolded yields, efficiencies and
    partial branching fractions in bins of $q^2$.} \vspace*{2mm}
  \begin{tabular}{
      @{\hspace{0.3cm}}r@{$-$}l
      r@{$\pm$}l
      r@{$\pm$}l
      r@{$\pm$}l
      r@{$\pm$}l@{\hspace{0.1cm}}
    }
    \hline
    \multicolumn{2}{c}{$\Delta q^2$} &
    \multicolumn{2}{c}{$N_\mathrm{fit}$} &
    \multicolumn{2}{c}{$N_\mathrm{fit}^\mathrm{unfolded}$} &
    \multicolumn{2}{c}{$\varepsilon$} & \multicolumn{2}{c}{$\Delta\Br$} \\
    \multicolumn{2}{c}{\GeVGeVcc} & \multicolumn{2}{c}{} &
    \multicolumn{2}{c}{} & \multicolumn{2}{c}{$10^{-4}$} &
    \multicolumn{2}{c}{$10^{-6}$} \\
    \hline
    
 0& 7& 23.7&6.3 &  24.4&6.6 & 7.59&0.24 &  22.8&6.2 \\
 7&14& 50.7&9.2 &  51.5&9.7 & 6.48&0.20 &  56.5&10.6 \\
14&21& 24.6&7.8 &  23.1&7.8 & 4.84&0.22 &  33.9&11.5 \\
\hline
\multicolumn{2}{c}{Full range} &  99.0&15.0 &  99.0&15.0 & 6.42&0.14 & 113.3&18.0 \\

    \hline
  \end{tabular}
\end{table}

\begin{table}
  \caption{\label{table:bomega3plnucrr}Normalised statistical correlation
  matrix in percent for the \bomegappplnu partial branching fractions.} \vspace*{2mm}
  \begin{tabular}{c*{3}{D{.}{.}{4.1}}}
    \hline
    
$\Delta q^2$, \GeVGeVcc & \multicolumn{1}{r}{0-7}& \multicolumn{1}{r}{7-14}& \multicolumn{1}{r}{14-21}\\ \hline
0-7 & 100.0 &  1.5 & 10.3 \\
7-14 &  1.5 & 100.0 &  9.1 \\
14-21 & 10.3 &  9.1 & 100.0 \\

    \hline
  \end{tabular}
\end{table}


\begin{thebibliography}{99}
\bibitem{beta}
The angle $\phi_1$ is also referred to by the symbol $\beta$
in the literature; similarly, $\phi_2\equiv\alpha$ and $\phi_3\equiv\gamma$.
\bibitem{conj}
Charge-conjugate decays are implied throughout this paper.
\end{thebibliography}
\end{document}